\pdfoutput=1
\documentclass[sn-nature,iicol]{sn-jnl}
\usepackage{graphicx}%
\usepackage{multirow}%
\usepackage{amsmath,amssymb,amsfonts}%
\usepackage{amsthm}%
\usepackage{mathrsfs}%
\usepackage[title]{appendix}%
\usepackage{xcolor}%
\usepackage{textcomp}%
\usepackage{manyfoot}%
\usepackage{booktabs}%
\usepackage{algorithm}%
\usepackage{algorithmicx}%
\usepackage{algpseudocode}%
\usepackage{listings}%
\usepackage[utf8]{inputenc}
\usepackage[T1]{fontenc}
\usepackage{mathptmx}
\usepackage{etoolbox}
\usepackage{braket}
\usepackage{subcaption}
\usepackage{caption}
\usepackage{bm}
\usepackage{hyperref}
\usepackage{natbib}
\usepackage{float}
\usepackage{comment}
\usepackage{diagbox}

\begin{document}

\title[Article Title]{Information-entropy-driven generation of material-agnostic datasets for machine-learning interatomic potentials}

\author*[1]{\fnm{Aparna} \sur{P. A. Subramanyam}}\email{apasubramanyam@lanl.gov} 
\author*[1]{\fnm{Danny} \sur{Perez}}\email{danny\_perez@lanl.gov}

\affil[1]{\orgdiv{Theoretical Division T-1}, \orgname{Los Alamos National Laboratory}, \orgaddress{\city{Los Alamos}, \postcode{87545}, \state{New Mexico}, \country{USA}}}

\abstract{In contrast to their empirical counterparts, machine-learning interatomic potentials (MLIAPs) promise to deliver near-quantum accuracy over broad regions of configuration space. However, due to their generic functional forms and extreme flexibility, they can catastrophically fail to capture the properties of novel, out-of-sample configurations, making the quality of the training set a determining factor, especially when investigating materials under extreme conditions. 
We propose a novel automated dataset generation method based on the maximization of the information entropy of the feature distribution, aiming at an extremely broad coverage of the configuration space in a way that is agnostic to the  properties of specific target materials. 
The ability of the dataset to capture unique material properties is demonstrated on a range of unary materials, including elements with the fcc (Al), bcc (W), hcp (Be, Re and Os), graphite (C), and trigonal (Sb, Te) ground states. MLIAPs trained to this dataset are shown to be accurate over a range of application-relevant metrics, as well as extremely robust over very broad swaths of configurations space, even without dataset fine-tuning or hyper-parameter optimization, making the approach extremely attractive to rapidly and autonomously develop general-purpose MLIAPs suitable for simulations in extreme conditions.
}


\maketitle

\section{Introduction}\label{sec:Introduction}

The high computational cost of {\em ab initio} electronic structure methods such as density functional theory (DFT) severely constrains the time and length-scales accessible to direct simulations. This has for decades motivated the development of computationally-efficient interatomic potentials that act as surrogates to quantum calculations. Early potentials relied on physically-motivated functional forms that involved a few empirical parameters that could be determined using only a small amount of training data (e.g., elastic constants, cohesive energy, lattice constants, etc.). Such simple functional forms rooted in physics typically resulted in robust potentials, at the cost of a generally significant compromise in accuracy \cite{PhysRevMaterials.8.013803}. In contrast, modern machine learning interatomic potentials (MLIAPs) instead invoke generic features (also known as descriptors) and/or flexible ML architectures to describe local atomic environments and capture their energetic contributions \cite{PhysRevLett.98.146401, 10.1063/1.4712397, 10.1063/1.3553717, PhysRevB.87.184115, PhysRevLett.104.136403, THOMPSON2015316, 10.1063/1.5017641, doi:10.1137/15M1054183, PhysRevB.99.014104}. 
The high complexity and flexibility of these potentials dramatically increases the number of adjustable parameters compared to earlier generations of potentials, enabling unprecedented accuracy at the cost of a very significant increase in the amount of required training data. Crucially, the lack of rigid physics-based underpinnings potentially can lead to poor extrapolative capabilities, making MLIAPs susceptible to catastrophic failures in regimes where training data is either scarce or missing \cite{fu2023forces, Stocker_2022}, even when test errors on hold-out data suggests that the potentials should be very accurate \cite{Stocker_2022}. Observations suggest that assembling diverse datasets with high energy structures that are far from equilibrium significantly improves the stability of simulations \cite{Stocker_2022}. A similar conclusion was reached in Ref.\ \cite{NPJ_Montes}, where serious limitations in the robustness of models trained to hand-curated and narrowly tailored datasets were observed. This highlights the critical importance of training MLIAPs to very diverse data, especially in applications to extreme conditions where it is very difficult to {\em a priori} predict the regions of configuration space that will be visited by simulations. This recognition has led to the development of a number of approaches aiming at dramatically improving dataset curation compared to the manual domain-expertise-driven approaches that were previously common.
\begin{figure}[p]
\begin{subfigure}[b]{0.45\columnwidth}
\includegraphics[width=\textwidth]{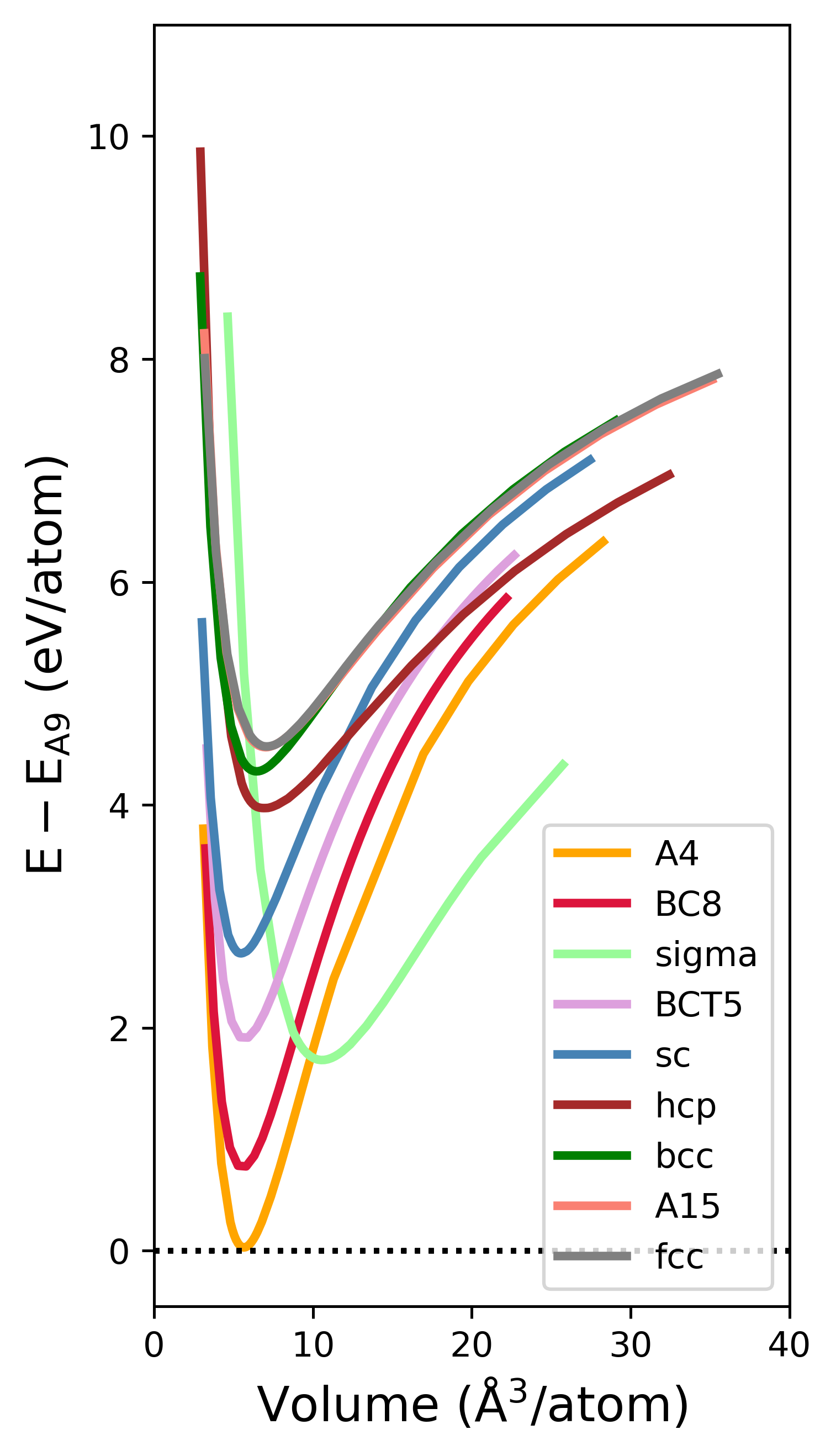}
\caption{DFT}
\label{fig:C_DFT_EV}
\end{subfigure}\qquad
\begin{subfigure}[b]{0.45\columnwidth}
\includegraphics[width=\textwidth]{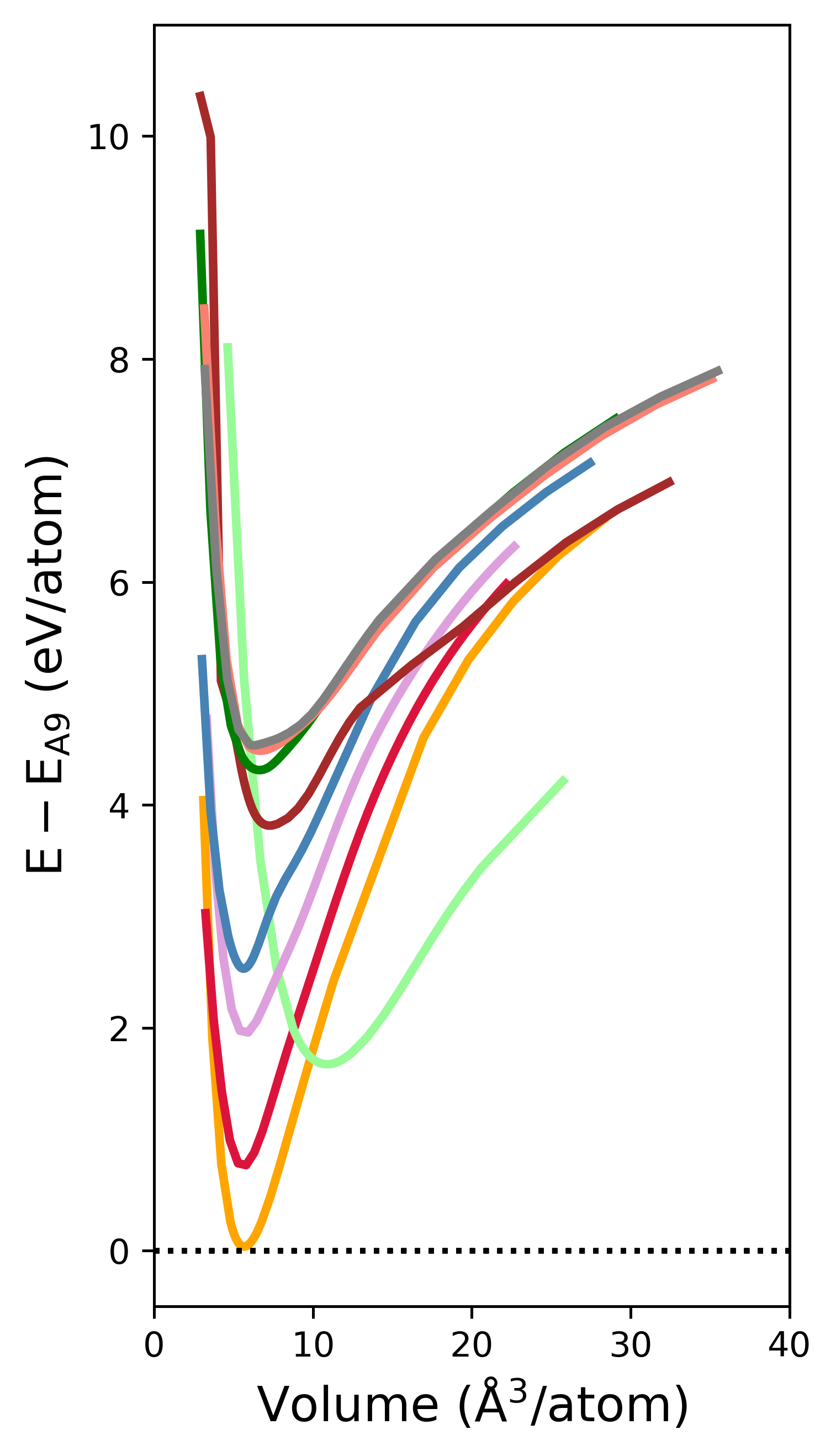}
\caption{ACE \cite{Qamar2023}}
\label{fig:C_Qamar_EV}
\end{subfigure}\qquad
\begin{subfigure}[b]{0.45\columnwidth}
\includegraphics[width=\textwidth]{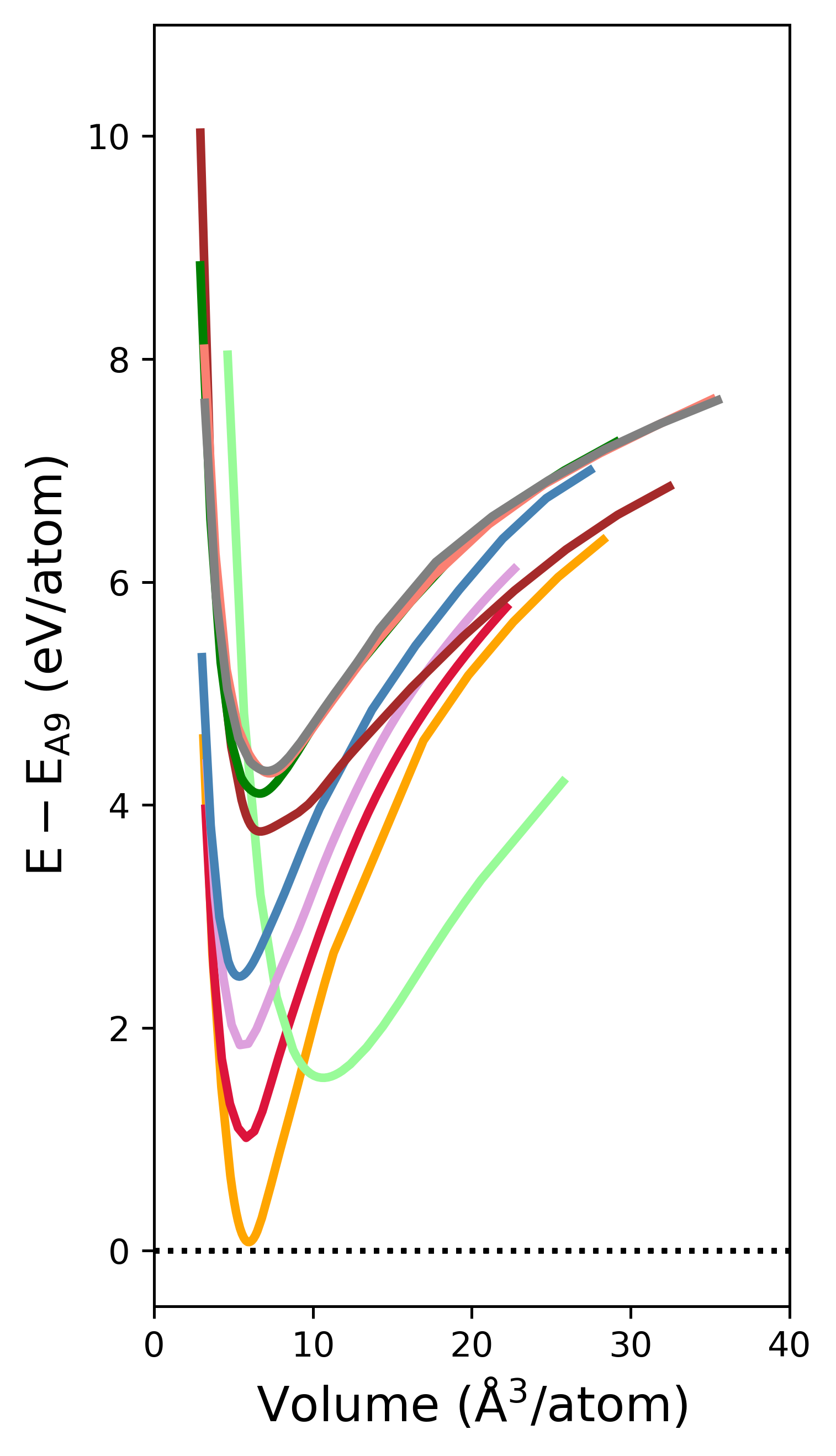}
\caption{ACE (this work)}
\label{fig:C_ACE-3}
\end{subfigure}\qquad
\begin{subfigure}[b]{0.45\columnwidth}
\includegraphics[width=\textwidth]{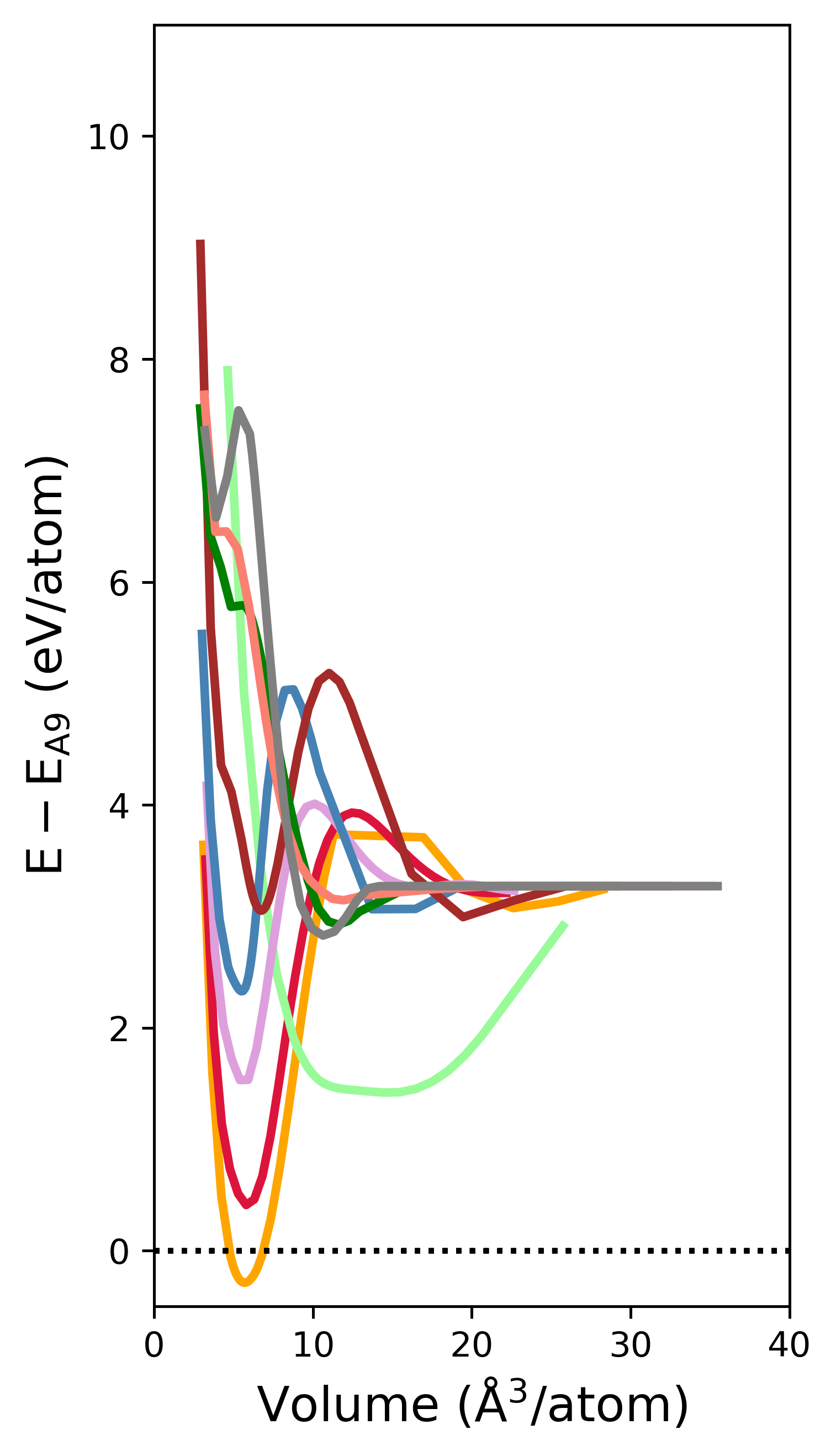}
\caption{SNAP \cite{PhysRevB.106.L180101}}
\label{fig:C_SNAP_EV}
\end{subfigure}\qquad
\begin{subfigure}[b]{0.45\columnwidth}
\includegraphics[width=\textwidth]{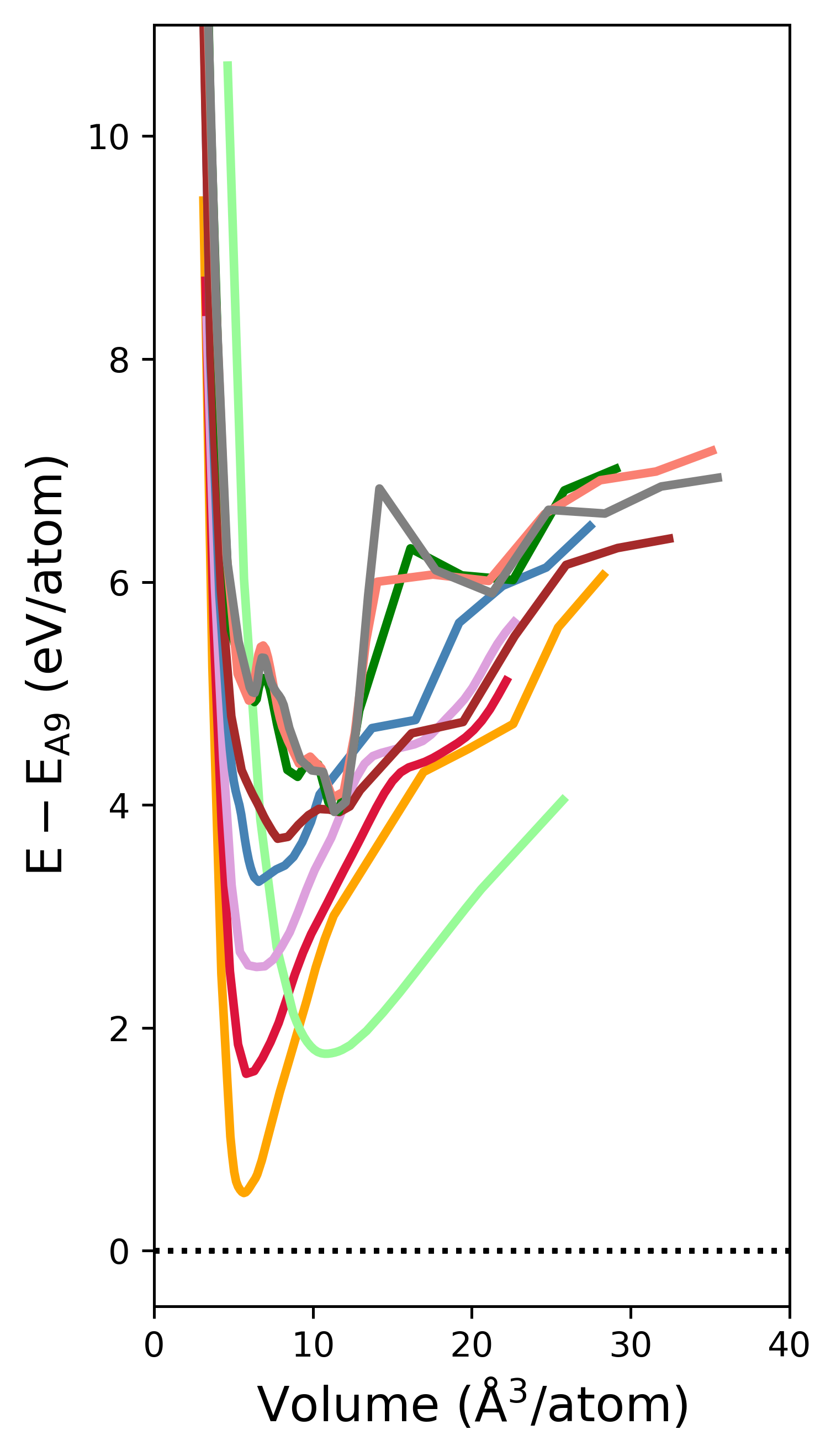}
\caption{ANI-1xnr \cite{Zhang2024}}
\label{fig:C_ANI1xnr_EV}
\end{subfigure}\qquad
\begin{subfigure}[b]{0.45\columnwidth}
\includegraphics[width=\textwidth]{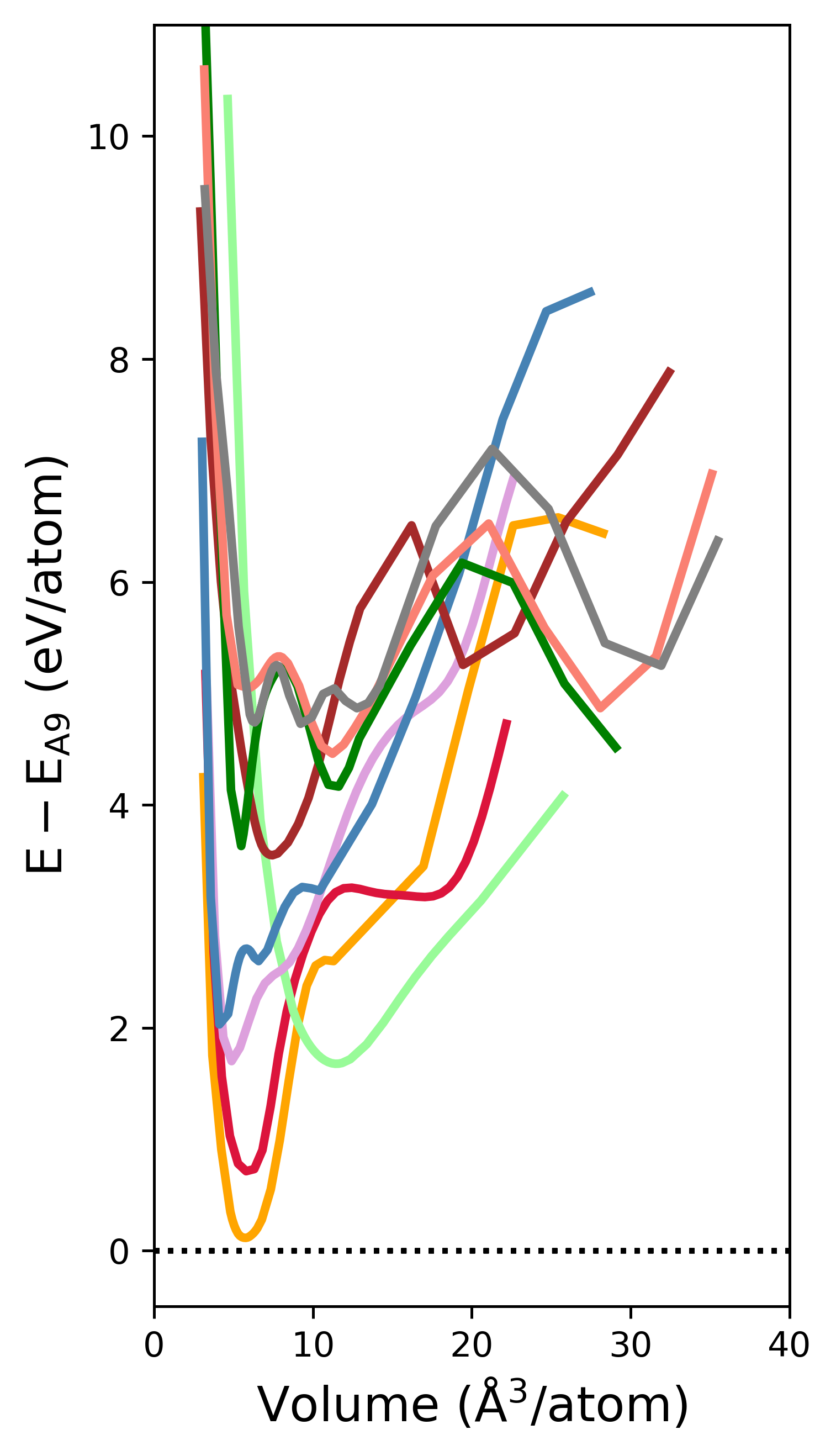}
\caption{GAP \cite{10.1063/5.0091698}}
\label{fig:C_GAP-U_EV}
\end{subfigure}\qquad
\caption{Transferability of different MLIAPs for C to nine crystal structures.}
\label{fig:MLIAPs_to_C_EV}
\end{figure}

Perhaps the most well-known approach in this class is active learning, where iteratively-refined potentials are used to drive further exploration of the  configuration space, repeating the training/exploration cycle multiple times. These techniques rely on a combination of diverse initialization and/or efficient sampling algorithms 
(e.g., molecular dynamics \cite{ANI_Al, PhysRevMaterials.3.023804, PhysRevB.100.014105}, metadynamics \cite{10.1063/1.5020067}, uncertainty-biased MD \cite{vanderOord2023, Kulichenko2023}, evolutionary structural searches \cite{PhysRevB.99.064114, Bernstein2019}, local structure optimization and diversity based selection \cite{Bernstein2019}, clustering \cite{Sivaraman2020}, dimensionality reduction \cite{Qi2024}, etc.) to ensure broad coverage of configuration space. Methods in this class often also rely on uncertainty quantification (e.g., using extrapolative score, d-optimality \cite{PODRYABINKIN2017171, 10.1063/1.5005095, GUBAEV2019148, PhysRevMaterials.7.043801, PhysRevB.99.064114}, Bayesian regression \cite{PhysRevB.100.014105, vanderOord2023, Vandermause2020, Xie2023}, etc.) to flag configurations where the current potential is deemed unreliable for inclusion in the next iteration of the training set. These approaches can be a powerful way to enrich datasets with targeted information directly relevant to certain conditions or applications. However, given the rugged nature of actual energy surfaces, the diversity of the dataset produced by active learning can be expected to be significantly affected by the details of the initialization and sampling procedures. Care must therefore be taken if one aims at a very broad exploration as energetically or entropically hard-to-reach regions can be expected to be the norm rather than the exception.

Active-learning approaches are material-specific, in that preliminary approximations of a specific potential energy surface drives the data generation process. They provide a natural way to inject material specific information in the training sets, but can also significantly increases the time required to obtain models due to the need for iterative refinement. Other approaches instead opt for material-agnostic strategies. E.g., datasets assembled from extensive sampling of distorted, deformed, and randomized crystal structures (even those that were not expected to be physically-relevant for the target materials) have performed extremely well when tested on defects structures that were not explicitly introduced in the training data \cite{PhysRevB.107.104103}. This approach has the advantage of being completely material-agnostic and easily automatable, removing the time-consuming step of iterative potential refinement.  The present work generalizes another material-agnostic approach that was introduced in Refs.\ \cite{10.1063/5.0013059,NPJ_Montes}. Here, dataset curation is recast as an optimization problem where the objective function is the information entropy of the dataset, as measured in feature space. Since uniform distributions maximize the information entropy, this approach explicitly favors a broad coverage in feature space instead of specifically focusing on regions that are deemed important {\em a priori}. This method is hence also material-agnostic in that datasets are assembled strictly based on structural characteristics, independently of the projected thermodynamic relevance of the corresponding configurations. This approach was shown to confer MLIAPs extremely good transferability, in contrast to MLIAPs trained on a traditional hand-curated dataset which are locally accurate, but whose fidelity dramatically degrades for out-of-sample configurations \cite{NPJ_Montes}. This study has also shown a pronounced but favorable accuracy/robustness tradeoff for potentials trained to entropy-maximized datasets: while training MLIAPs to extremely diverse datasets was observed to incur an accuracy penalty on low energy structures, the transferability of the potentials to a broad range of configurations is dramatically improved compared to targeted approaches. Transferability is paramount in many applications, especially when investigating the behavior of materials in extreme conditions --- such as temperature, pressure, irradiation, etc. --- where simulations should remain well behaved in conditions that are very far from thermodynamic equilibrium. However, in applications where the relevant region of configuration space is small and easily delineated, more targeted approaches to dataset curation could prove to be more appropriate.

To exemplify these transferability challenges, we tested several popular MLIAPs developed for C using a range of different data generation strategies and base ML architectures. Energy-volume curves for different crystal structures are reported in Fig.~\ref{fig:MLIAPs_to_C_EV}. We emphasize that the developers of these models (Spectral Neighbor Analysis Potential (SNAP) \cite{PhysRevB.106.L180101}, ANI-1xnr \cite{Zhang2024} and Gaussian Approximation Potential (GAP) \cite{10.1063/5.0091698}) in general made no claims on accuracy of the models for these phases, except in the case of the Atomic Cluster Expansion (ACE) model from Ref. \cite{Qamar2023} (c.f., Fig.~\ref{fig:MLIAPs_to_C_EV}(b)) where such data was actually used in training. The SNAP, ANI-1xnr and GAP models show numerous spurious features in their energy landscapes. This indicates that even with expressive ML architectures and careful data curation by domain experts, ML potentials cannot be expected to capture properties that they were not explicitly trained to and are hence prone to unphysical behaviors away from regions covered by training data. In contrast, as will be shown below, the potentials trained in this work (c.f., Fig.~\ref{fig:MLIAPs_to_C_EV}(c)) captures the behavior of a broad range of phases that were not explicitly injected in the training set. This indicates that one-shot diverse-by-design data generation approaches can in fact efficiently constrain the behavior of flexible ML potentials over broad swaths of the feature space without requiring external guidance from domain experts or iterative improvement with active learning, opening the door to the straightforward and principled generation of training sets that confer extreme robustness to ML potentials.

In the following, we generalize the aforementioned entropy optimization approach to generate a compact, yet ultra-diverse, material-agnostic dataset that can be applied to train robust unary MLIAPs across the periodic table. The power of this approach is demonstrated by training potentials in the family of ACE \cite{PhysRevB.99.014104} to a diverse set of elements (Be, C, Al, Sb, Te, W, Re, Os), thoroughly validating the results using physical properties that are not explicitly targeted in the training data. The proposed approach provides a systematic, simple, and fast approach to MLIAP development that alleviates the need for fine-tuned material-specific datasets, while retaining the type of exceptional transferability and predictive capabilities that is required to investigate materials in broad range of conditions. Although not pursued here, this method can easily be combined with traditional active learning techniques to further fine-tune the models in applications-specific regions of configuration space. 

The manuscript is organized as follows. In Sec.\ \ref{sec:Results}, we explain the workflow of the entropy maximization approach followed by a comparison of the diversity of the resulting dataset to those from literature in Sec.\ \ref{sec:diversity}. In Sec.\ \ref{sec:W}, we extensively validate the performance of an ACE potential for W that is fitted to the newly generated dataset 
and analyze the accuracy/robustness tradeoff of the resulting W ACE models 
in Sec.\ \ref{sec:tradeoff}. The material agnostic nature of the current dataset 
is demonstrated in Sec.\ \ref{sec:agnostic} by validating ACE models for Be, C, Al, Sb, Te, Re, and Os. The exceptional robustness of the ACE models for Al, W, and C is further corroborated by cross-validating on multiple datasets available from the literature in Sec.~\ref{sec:comparison_MLIAPs}. Details of the computational procedure are given in Sec. \ref{sec:Methods}.

\section{Results}\label{sec:Results}

\begin{figure*}[h]%
\centering
\includegraphics[width=0.9\textwidth]{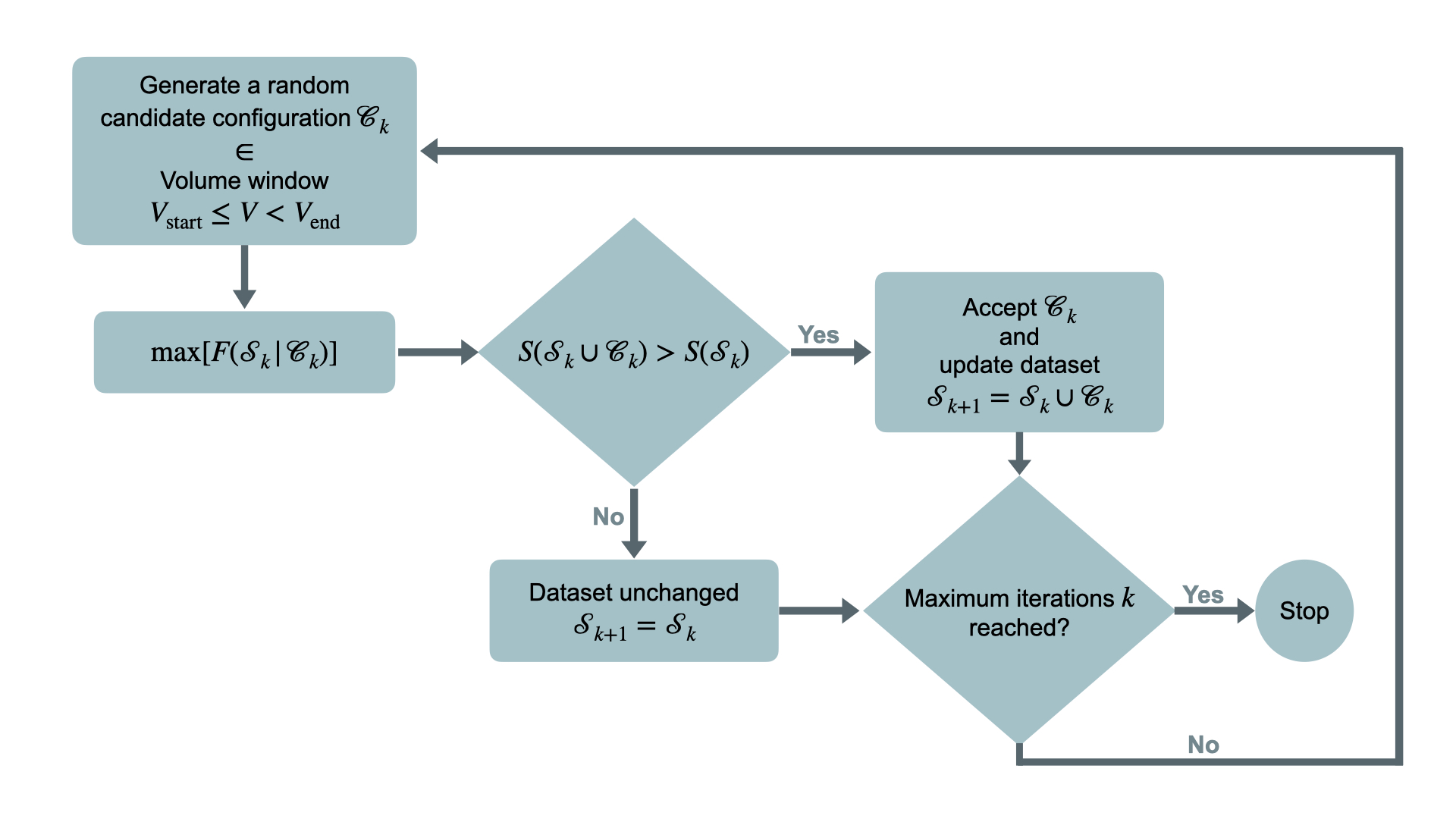}
\caption{Schematics of the greedy entropy-maximization workflow. See text for details.}
\label{fig:workflow}
\end{figure*}

\subsection{Entropy maximization approach}\label{sec:Rev-EM}

In order to systematize the curation of training sets for MLIAPs, the problem is first reformulated in terms of an optimization procedure where the objective function is a measure of the information encoded in the dataset. 
Qualitatively, this objective function should be such that redundant information is penalized and that information diversity is promoted. While not unique, one such natural objective function is the {\em information entropy} of the whole dataset, which is defined as 

\begin{equation}
    S = -\int{p\left(B\right)\log{p\left(B\right)dB}}
    \label{eq:entropy}
\end{equation}
where $B$ is an $m$-dimensional feature vector and $p(B)$ is the probability distribution function of features characterizing the atomistic environments contained in the dataset. Unless otherwise constrained, maximizing the information entropy would lead to a {\em uniform} $p(B)$, hence maximizing coverage and minimizing redundancy. In practice, a uniform distribution might not be achievable due to geometric constraints (e.g., some feature vectors might not be realizable for actual atomistic configurations due to correlations between features). A natural choice of $B$ is to rely on the same type of features used to parameterize local atomic environments in MLIAPs, as they are by construction deemed suitable to capture their energy-determining characteristics and often obey the translation/rotation/permutation symmetries that are inherent to physical interactions. 

A first key numerical challenge is to efficiently estimate the high dimensional integral in Eq.\ \ref{eq:entropy}. In the first incarnation of the entropy optimization method \cite{10.1063/5.0013059}, a non-parametric estimator was employed where the density was estimated based on first-neighbor distances in feature space \cite{beirlant1997nonparametric}. While simple, this approach is not scalable and better suited to low dimensions, and so was restricted to the optimization of the feature entropy {\em within} a given configuration containing a few tens of atoms, and not to the optimization of the entropy of the {\em entire} dataset. As a pragmatic alternative, we instead introduce an approximation of $p(B)$ in terms of a high-dimensional normal distribution 
\begin{equation}
p(B) \simeq N(\mu, \Sigma)  
\end{equation}
with mean $\mu$ and covariance matrix $\Sigma$. In this case, the information entropy of the feature distribution is given analytically as 
\begin{equation}
S=\frac{m}{2}\log(2\pi e) + \frac{1}{2}\log \det(\Sigma),
\label{eq:entropy_normal}
\end{equation}
which, up to irrelevant additive and multiplicative constants, is simply proportional to the logarithm of the determinant of the feature covariance matrix; in an abuse of language, we will refer to $\log \det(\Sigma)$ as the information entropy in the following. This quantity has a simple interpretation in terms of the sum of the logarithm of the variance of the feature distribution along all principal directions: high feature entropies hence corresponds to broad distributions in feature space. 
Note that within this approximation, maximizing the dataset entropy is akin to a d-optimality experiment design procedure, which is known to optimize the differential Shannon information content of the parameter estimates. This correspondence with d-optimality suggests that whether or not the normal approximation is an accurate representation of the actual feature distribution, maximizing Eq.\ \ref{eq:entropy_normal} will nonetheless maximize a well-defined measure of dataset diversity. This is an important point given that, as will be shown below in Fig.\ \ref{fig:bispectrum_scatter_plots}, certain feature distributions show clear signs of non-normality.
D-optimality was introduced in the context of active learning of MLIAPs by Podryabinkin and Shapeev \cite{PODRYABINKIN2017171}, where the determinant of the feature covariance matrix was used as a scoring function for candidate configurations. The original "passive" approach is here generalized to an "active" variant where atomic configurations are explicitly generated so as to maximize this objective function.

In practice, $\Sigma$ is given by the empirical covariance of the features over a given dataset. In the following, we consider two flavors, which are referred to as per-atom and per-configuration respectively. In the per-atom case, each {\em atom} $i$ contributes a feature vector $B_i$, while in the per-configuration variant, each {\em configuration} $\alpha$ contributes a feature vector $\hat{B}_\alpha$ which is the arithmetic mean of the atomic $B_i$ within the configuration. This distinction is made by analogy to the training of an hypothetical MLIAP which would be linear in the $B_i$. In this case, the predicted energy per atom would be proportional to the {\em mean} of the $B_i$ over all atoms in a given configuration. We empirically observe that maximizing the entropy of per-atom features tends to generate internally diverse configurations, while maximizing the entropy of per-configuration features tends to, perhaps surprisingly, generate internally ordered configurations. This can be understood qualitatively in terms of self-averaging where, by definition, the mean of per-atom features must lie within the envelope of the individual per-atom features. Generating internally diverse configurations is therefore in tension with the generation of extreme values of the mean, which is comparatively easier to achieve when all per-atom features take on similar values, hence resulting in internally ordered configurations. We however note that any other, potentially non-linear, combination of atomic features could in principle be used as a basis for the entropy optimization procedure.

A second key numerical challenge is to generate an entropy-maximizing dataset in practice. The approach, which follows that introduced in Ref.\ \cite{10.1063/5.0013059}, is based on a greedy iterative procedure where candidate configurations are sequentially considered for inclusion in the dataset. The greedy optimization workflow is illustrated in Fig.\ \ref{fig:workflow}. Denote the existing dataset at iteration $k$ by $\mathcal{S}_k$. At each iteration $k$, a candidate configuration $\mathcal{C}_k$ is initialized at random, including the cell size, shape, number of atoms, and atomic positions. The feature entropy of $\mathcal{S}_k \cup \mathcal{C}_k$ is then locally maximized with respect to the atomic positions of the atoms in $\mathcal{C}_k$ leaving the existing $\mathcal{S}_k$ fixed, using, e.g., a steepest gradient algorithm. If, upon convergence to a local maximum of the entropy, the information entropy of the dataset is increased by the addition of  $\mathcal{C}_k$, i.e., if $S(\mathcal{S}_k \cup \mathcal{C}_k) > S(\mathcal{S}_k)$, it is added to the dataset --- $\mathcal{S}_{k+1}=\mathcal{S}_k \cup \mathcal{C}_k$ --- otherwise the dataset remains unchanged and $\mathcal{S}_{k+1}=\mathcal{S}_k$.

Since many atomic features used to parameterize MLIAPs vary very rapidly at short distances (which is desirable, since the energy also varies very rapidly in this case), an unconstrained maximization of the feature entropy would lead to a large number of unphysical configurations with overlapping atoms, potentially posing serious convergence issues in electronic structure calculations, in addition to generating mainly configurations of extremely high energies. In order to avoid this, a core repulsion term $V_\mathrm{core}$ is added to the entropy to penalize close approaches between atoms, i.e., as part of the workflow loop, the effective objective function $F$ that is actually maximized is given by:
\begin{equation}
    F(\mathcal{C}_k | \mathcal{S}_k) = K*\log \det [\Sigma(\mathcal{S}_k \cup \mathcal{C}_k)]  - V_\mathrm{core}(\mathcal{C}_k)
    \label{eq:objective}
\end{equation}
where $K$ is a scalar parameter that sets the relative strength of the entropic and core repulsion terms. While this approach introduces an {\em ad hoc} term, it allows for the use of arbitrary features without modification. 

In principle, any differentiable atomic features can be employed; in the following, we consider a feature space that consists of the first 55 so-called bispectrum components of the local atomic density as introduced in the GAP formalism and later leveraged by the SNAP family of MLIAPs \cite{PhysRevLett.104.136403, THOMPSON2015316}. Additional technical details on the optimization procedure are provided in Sec.\ \ref{sec:Methods}.

\begin{figure*}[!h]%
\centering
\includegraphics[width=0.9\textwidth]{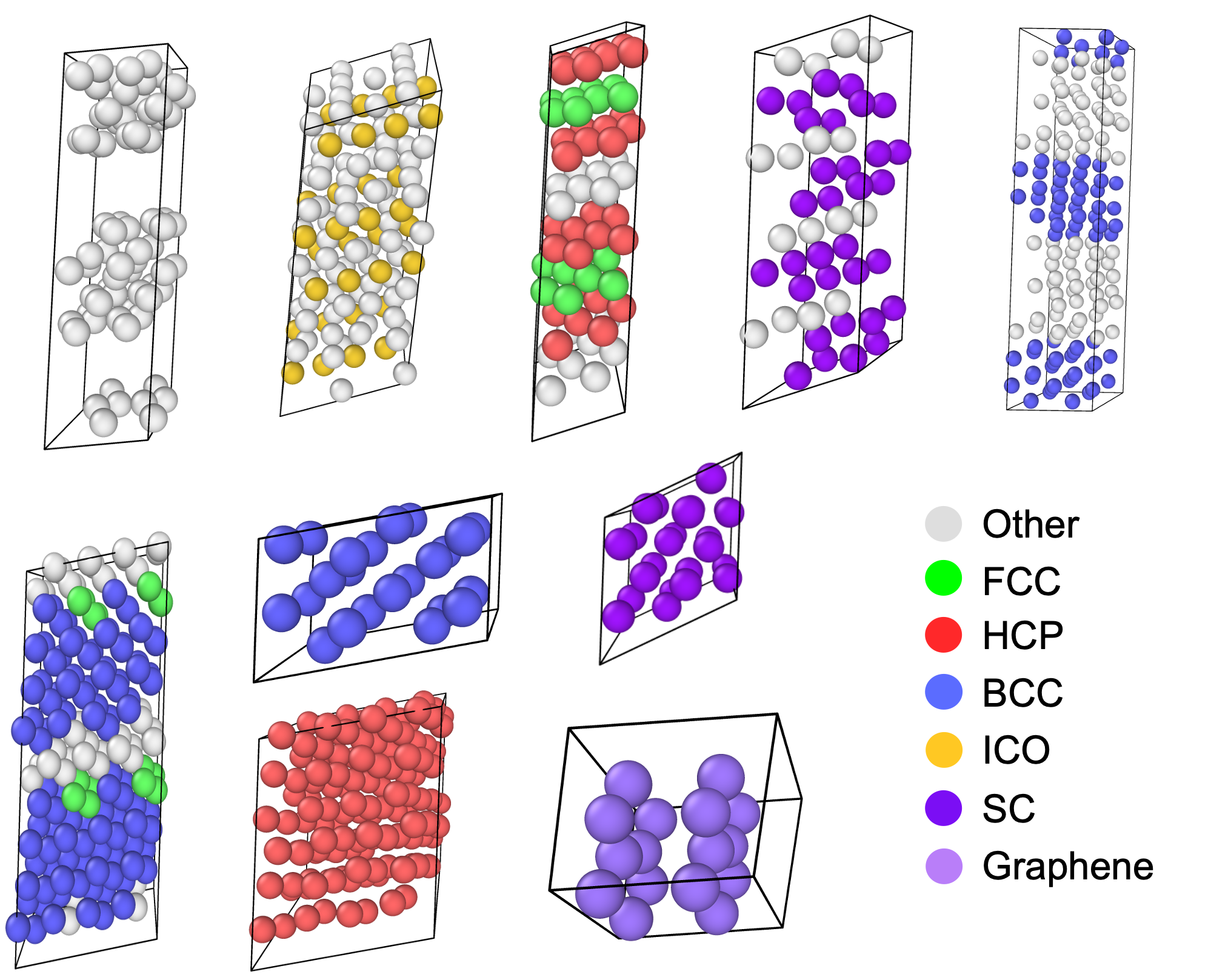}
\caption{Example structures from Rev-EM. Atoms are colored using the polyhedral template matching algorithm \cite{Larsen_2016} implemented in OVITO \cite{Stukowski_2010}. The structures show a wide diversity of local atomic environments including simple cubic (SC), icosahedral (ICO) and Graphene. Note that the unit cells are replicated twice in each direction for ease of visualization.}
\label{fig:example_structures}
\end{figure*}

\subsection{Dataset diversity}\label{sec:diversity}

The workflow introduced in Sec. \ref{sec:Rev-EM} was used to generate a dataset consisting of 30,349 structures with a total of 179,367 atoms, which we refer to as the revised-entropy maximized (Rev-EM) dataset. Roughly half of the configurations were generated with the "per-atom" mode, and half with the "per-configuration" mode.
As shown in Fig.\ \textcolor{blue}{S1}, this dataset is predominantly composed of small unit cells with less than 10 atoms, but contains configurations up to 25 atoms. The figure also shows a very broad distribution of Voronoi volumes per atom, of first neighbor distances, and of coordination numbers, hinting at the large structural diversity of this dataset. Some representative Rev-EM configurations are shown in Fig.\ \ref{fig:example_structures}. 
Although Rev-EM is designed to jointly maximize feature diversity, many configurations contain regions analogous to a broad range of different crystalline and quasi-crystalline orders, from close-packed FCC and HCP, to icosahedral, to simple cubic SC and graphene. These configurations also contain nanostructures that are often important in applications, including FCC/HCP stacking faults, free surfaces, grain boundaries, crystalline/amorphous interfaces, etc.  Partially ordered configurations predominantly result from maximizing the per-configuration feature entropy, while internally-disordered configurations are often produced by maximizing the per-atom feature entropy (see Sec.\ \ref{sec:Methods}). All configurations were randomly seeded, so that specific orderings naturally emerged from the feature-entropy maximization process.

\begin{figure}[h]
\includegraphics[width=\columnwidth]{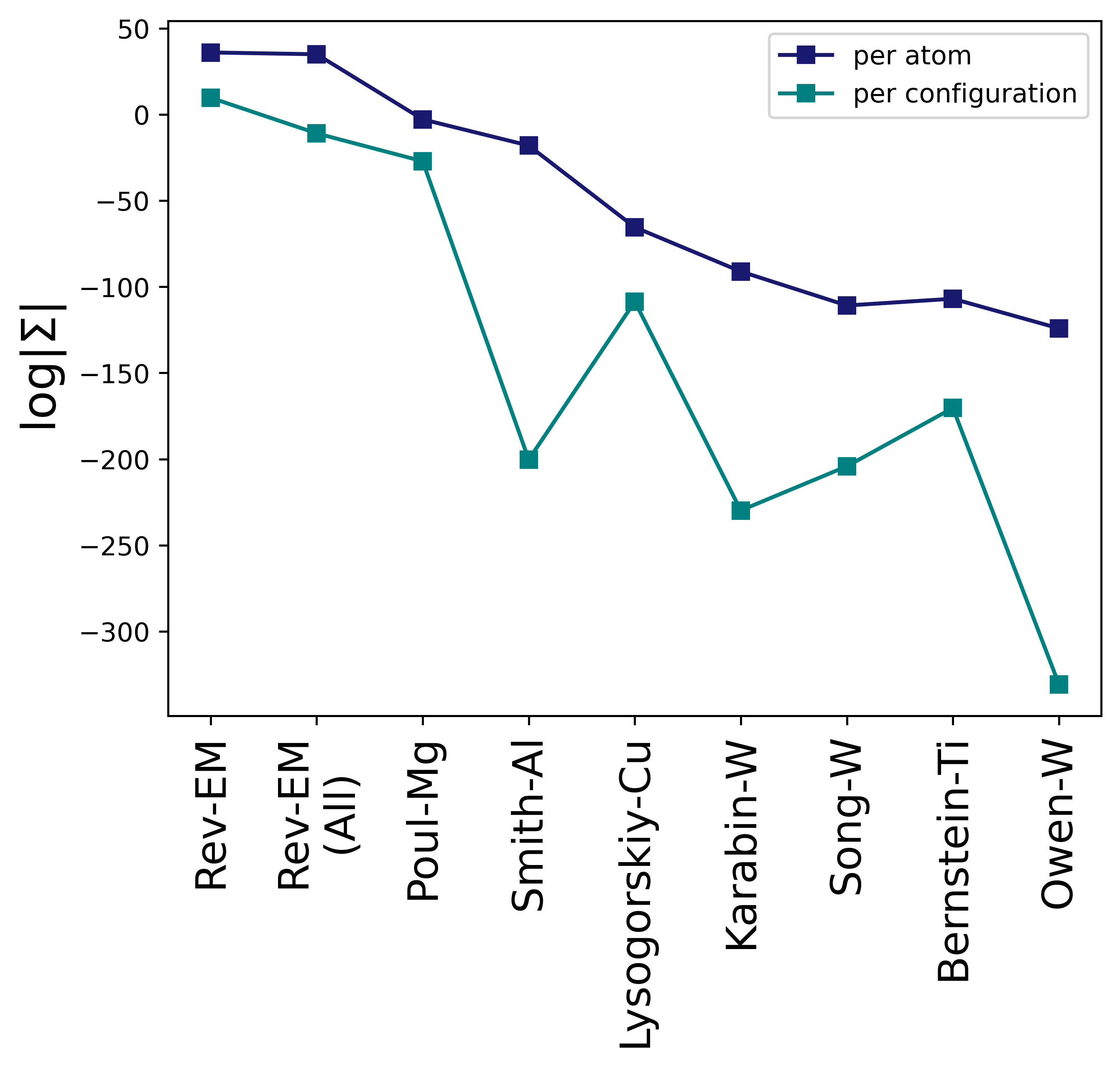}
\caption{Comparison of the information entropy of the configuration averaged and per atom distribution of the bispectrum components for various datasets. The Rev-EM dataset comprises structures generated in the corresponding mode while the Rev-EM (All) dataset is a combination of structures from both modes.}
\label{fig:det_selected}
\end{figure}

In order to quantitatively compare the diversity of Rev-EM with other curation approaches, the bispectrum component feature entropy of a number of publicly available datasets was estimated after volumetrically rescaling all configurations to a common (average) reference density. These datasets include Poul-Mg \cite{PhysRevB.107.104103}, Bernstein-Ti \cite{Bernstein2019}, Smith-Al \cite{ANI_Al}, Lysogorskiy-Cu \cite{PACE_implementation}, Song-W \cite{Song2024}, Owen-W \cite{Owen2024} and Karabin-W, a subset of 20,000 configuration from Ref. \cite{10.1063/5.0013059}. See Sec.\ \ref{sec:Rev-EM} for more details on the different datasets. Given that the bispectrum components exhibit a strong volume dependence, only configurations within $\pm$25\% of the equilibrium density were considered to avoid the specific range of densities in each set dominating the comparison. Note that this does not fully correct for this effect, as the distribution of volumes within each dataset will still differ. The results are reported in Fig.\ \ref{fig:det_selected}.
A comparison of the original un-truncated datasets is presented in Fig.\ \textcolor{blue}{S2}, which shows that the extremely high density configurations included in Poul-Mg confer it a very high information entropy. In addition to recognizing the caveats above, we emphasize that the information entropy should not be construed as an absolute measure of the quality of the datasets and of the MLIAPs derived from them, as each set was created with different applications, purposes, and constraints in mind. Instead, it should be simply taken as a reflection of the diversity of the configurations included in each dataset, which can be beneficial but also potentially detrimental, depending on the application.

\begin{figure*}[h]
\begin{subfigure}[b]{\columnwidth}
\includegraphics[width=\textwidth]{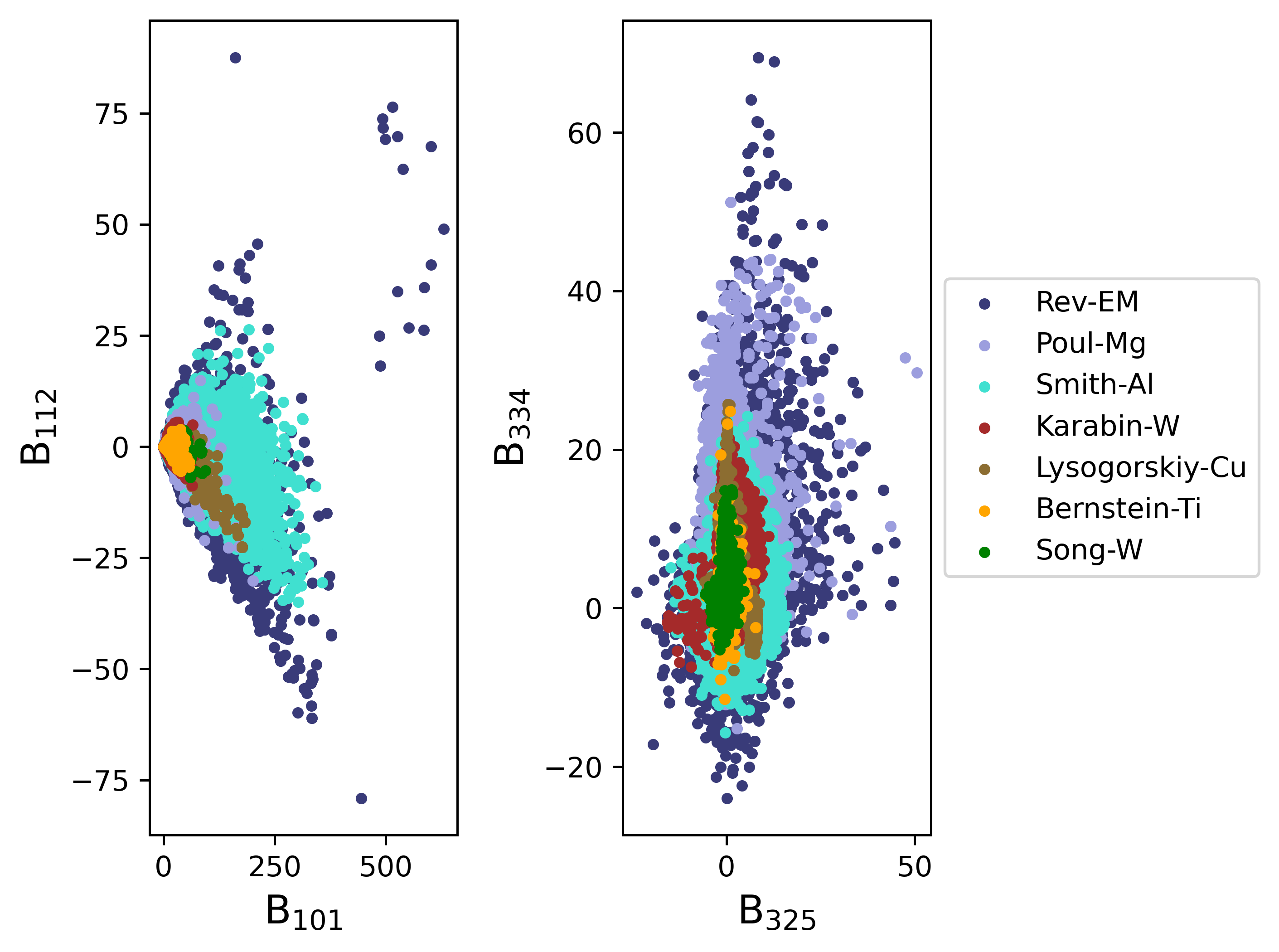}
\label{fig:bispectrum_FM}
\end{subfigure}\qquad
\begin{subfigure}[b]{\columnwidth}
\includegraphics[width=\textwidth]{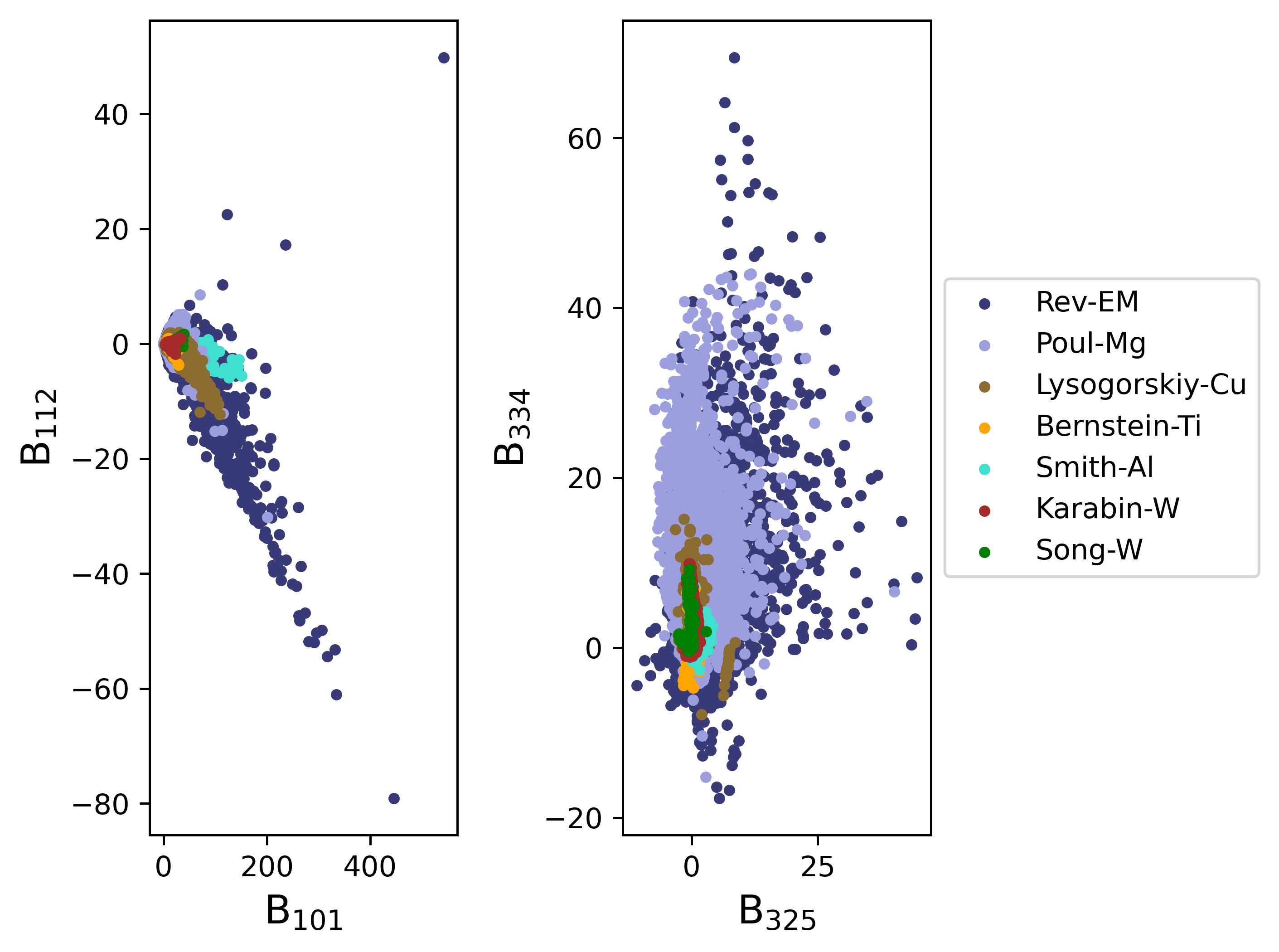}
\label{fig:bispectrum_EM}
\end{subfigure}\qquad
\caption{Comparison of the distribution of per-atom (left) and per-configuration (right) bispectrum components for various datasets over a truncated per-atom volume range (0.75V$_\mathrm{eq}\leq$V$\leq$1.25V$_\mathrm{eq}$).}
\label{fig:bispectrum_scatter_plots}
\end{figure*}

As expected from the fact that the optimization procedure explicitly maximizes the information entropy, Rev-EM shows the highest diversity in both modes. Interestingly, the Poul-Mg dataset \cite{PhysRevB.107.104103}, which was also generated using a materials-agnostic approach aimed at generating broad diversity, also exhibits a very high information entropy. At the other end of the spectrum, the Owen-W dataset predictably shows low information entropy, consistent with its tailor-made nature focusing on snapshots from MD simulations of vacancy-containing crystals and high-temperature liquids. All datasets also show an expected decrease in diversity in configuration-averaged features due to internal self-averaging. The difference is especially significant for 3 datasets: Smith-Al, Karabin-W, and Owen-W. The Karabin-W dataset \cite{10.1063/5.0013059} was generated by maximizing the per-atom entropy of individual configurations with no attempt at generating configurations that differ from each other, leading to considerable self-averaging effects. Similarly, the Owen-W dataset \cite{Owen2024} exhibits low per-configuration entropy, as it is composed of relatively large cells that are topologically very similar to each other. Self-averaging effects are also apparent in the Smith-Al dataset, as shown by comparing the left and right panels of Fig.\ \ref{fig:bispectrum_scatter_plots}.

In this analysis, the information entropy of two datasets whose feature variance varies by a factor of 2 in every direction would differ by a factor of $N_\mathrm{features}*\log(2) = 55*\log(2) \sim 38$, indicating a very significant variability in the feature distribution between these datasets.
This is consistent with visual inspection of the feature distribution along different 2D cuts of feature space, as shown in Fig. \ref{fig:bispectrum_scatter_plots}, which shows significantly more diverse distributions for Rev-EM than for other datasets.

\subsection{Training MLIAPs to Rev-EM}\label{sec:mliap}
We now demonstrate that MLIAPs trained to Rev-EM nonetheless perform extremely well on a broad range of metrics of interest to computational materials scientists. This demonstration proceeds in three steps. First, in Sec.\ \ref{sec:W}, we characterize the performance of a tungsten ACE potential fitted to Rev-EM using an extensive set of common benchmarks. Second, Sec.\ \ref{sec:tradeoff} analyzes the accuracy/robustness tradeoff resulting from this dataset in terms of accuracy on low-energy structures vs transferability to vast regions of configuration space. Third, Sec.\ \ref{sec:agnostic} demonstrates the transferability of Rev-EM across materials, including Al, C, Be, Re, Os, Te, and Sb. 

We emphasize that the goal of the present exercise is to demonstrate the ability of Rev-EM to produce potentials that can exhibit a favorable accuracy/transferability tradeoff in terms of reproducing conventional metrics on low energy structures while retaining the ability to describe the behavior of materials in extreme conditions, not to produce production-quality potentials.
Unless otherwise noted, all potentials share the same hyperparameters, except for the cutoff radius and the energy/force cutoff; reweighting based on energy/force magnitude, and fitting procedure (see Sec.\ \ref{sec:ace} for details of the potential form). These potentials could therefore be significantly improved upon by individually fine-tuning them. Although a systematic exploration is beyond the scope of this paper, evidence is provided below that both the complexity of the potentials and the reweighting scheme are viable mechanisms to alter the accuracy/transferability tradeoff.

Rev-EM was randomly partitioned into around 8000 training configurations and around 22,000 test configurations (the exact number depending on the energy/force cutoffs for each element). More details on the training and test configurations for all elements can be found in Secs.\ \ref{sec:DFT} and \ref{sec:ace}. 
Such a large testing set was purposefully chosen so as to provide a very challenging assessment of the robustness of the potentials on large numbers of ultra-diverse configurations. The reported test errors are therefore expected to constitute a good measure of the transferability of the MLIAPs to genuinely "unseen" and "surprising" configurations, which in general cannot be expected with random hold-out data from low-diversity datasets, where strong correlations between train and test sets would be expected. When training production models, it is advisable to increase the fraction of training to more typical values, which can be expected to further improve accuracy.

\subsection{A case study for W}\label{sec:W}
Rev-EM was first uniformly scaled to correspond to densities typical of W. The distribution of reference DFT energies and forces so obtained are reported in Fig.\ \ref{fig:rescale_datasets}. The distributions are extremely broad (almost 100 eV/atom in energy and $10^4$ eV/\AA~ in forces), but noticeably concentrated at relatively low values which are the most relevant for the majority of applications in materials science in ambient conditions. This indicates that $V_\mathrm{core}$ effectively limits the generation of nonphysical configurations with diverging energies and forces. The core exclusion radius can be adjusted to tailor the energy distributions for different applications.

Three MLIAPs were trained to Rev-EM using different reweighting schemes. This section focuses on ACE-2, which was trained following the standard recipe presented in Sec.\ \ref{sec:ace}. The other two MLIAPs, ACE-1 and ACE-3 will be discussed in Sec.\ \ref{sec:tradeoff}.

\begin{figure}[h]
\includegraphics[width=\columnwidth]{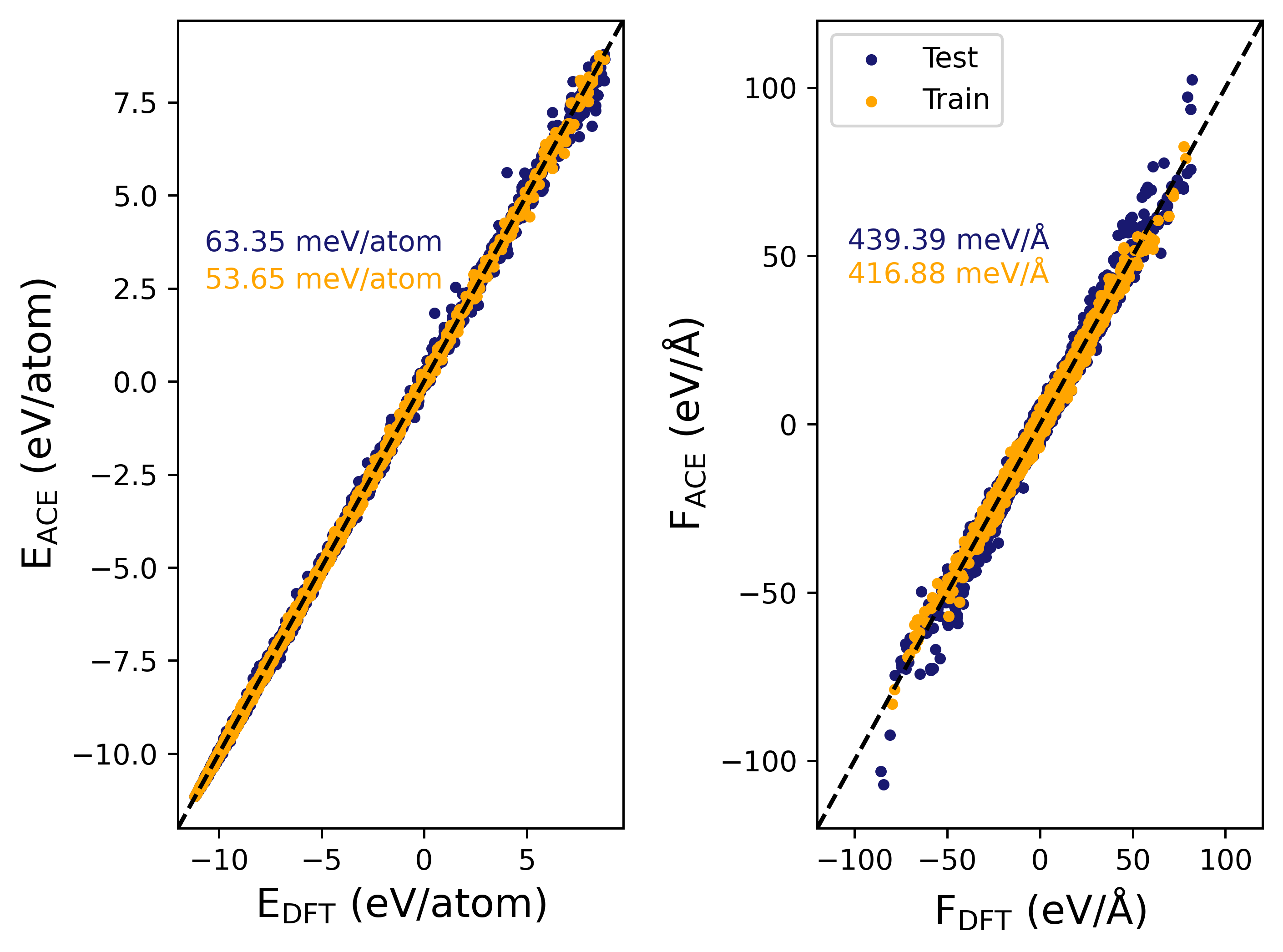}
\caption{Parity plots between the W-ACE predictions and DFT energies and forces for the training dataset (in orange) and the test dataset (in dark blue) comprising of the remaining structures, with the respective RMSEs. The parity plots show only structures lying below the threshold energy and force for W. See text for details.}
\label{fig:ACE_W_entropy_performance}
\end{figure}

The parity plot in Fig. \ref{fig:ACE_W_entropy_performance} shows the train and test performance of ACE-2. A total of 661 structures were excluded due to their energies and/or forces being above threshold values (20 eV/atom and 100 eV/\AA, respectively). The training energy and force RMSEs are 53.65 meV/atom and 416.88 meV/\AA~ respectively and slightly larger at 63.35 meV/atom and 439.39 meV/\AA~ respectively on the much larger test dataset. No catastrophic mispredictions are observed even up to high energies and forces, suggesting that fitting to a small subset of Rev-EM set makes MLIAPs extremely robust even when tested on very large and diverse test sets. 

While the reported errors are relatively large compared to fine-tuned MLIAPs, it is important to note that a reweighting scheme was applied during training to give more importance to low energy and force structures that are usually prioritized in applications (see Sec.\ \ref{sec:ace}); this is reflected by errors being generally lower at low energies and forces. For example, RMSE errors are significantly lower at  23.83 meV/atom and 216.25 meV/\AA~ for energies and forces, respectively, for configurations within 1 eV of the lowest-energy structure. Furthermore, transition metals are known to be particularly difficult to describe with MLIAPs, due to the complexity of their electronic structure around the Fermi level \cite{Owen2024}. It finally should be recognized that reported test errors are intimately related to the diversity of the test sets and to the extent to which training and testing sets are correlated to each other. As such, it is generally impossible to rigorously compare the accuracy of different potentials using test errors measured on different sets. Indeed, as shown in Table\ \ref{tab:ACE_New_to_Old_DE_TM23}, the test errors of ACE-2 on other DFT datasets for W available from the literature show significantly lower errors than on holdout data from Rev-EM, supporting the idea that these are functionally low-energy subsets of Rev-EM. As will be shown below, the reverse is typically observed when testing potentials trained to other datasets to Rev-EM, i.e., they tend to exhibit significantly higher errors than reported on their original test set. Note that the original energies and forces reported by the respective studies were used to evaluate the testing errors; different DFT settings could therefore lead to a overestimation of the errors when mixing training and test sets from different studies.

As we now show, a simple (and unoptimized) energy and force-based reweighting scheme leads to MLIAPs that are accurate at low energies, while retaining exceptional robustness at high energies. To do so, ACE-2 was thoroughly validated using a broad range of low-energy properties of general interest in materials science, including several defects and prototypical crystal structures extracted from the AFLOW database of crystallographic prototypes \cite{MEHL2017S1, HICKS2019S1, HICKS2021110450}. 

\begin{table}[!h]
\begin{tabular}{c | c | c }
\hline\hline
Dataset & Energy & Force \\
& RMSE (meV/atom) & RMSE (meV/\AA) \\
\hline
Rev-EM-W & 63.35 & 439.39 \\
Karabin-W & 15.26 & 313.62 \\
Song-W & 44.37 & 263.54 \\ 
Owen-W & 41.05 & 294.27 \\
\hline\hline
\end{tabular}
\caption{W ACE-2 RMS test errors on various W datasets available from the literature.
}
\label{tab:ACE_New_to_Old_DE_TM23} 
\end{table}

\begin{figure*}[h]
\begin{subfigure}[b]{0.7\columnwidth}
\includegraphics[width=\textwidth]{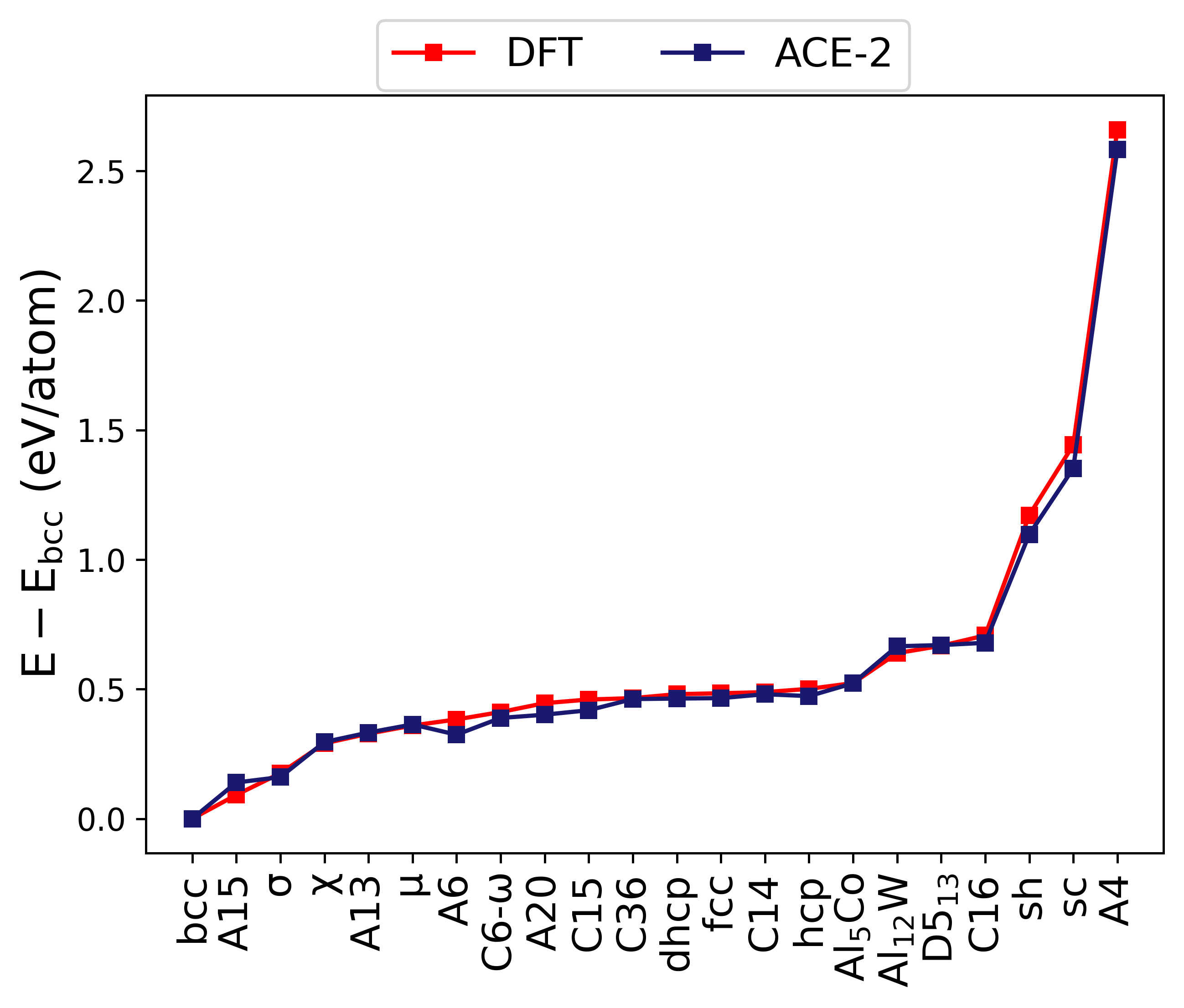}
\caption{Equilibrium formation energy}
\label{fig:W_E}
\end{subfigure}\qquad
\begin{subfigure}[b]{0.7\columnwidth}
\includegraphics[width=\textwidth]{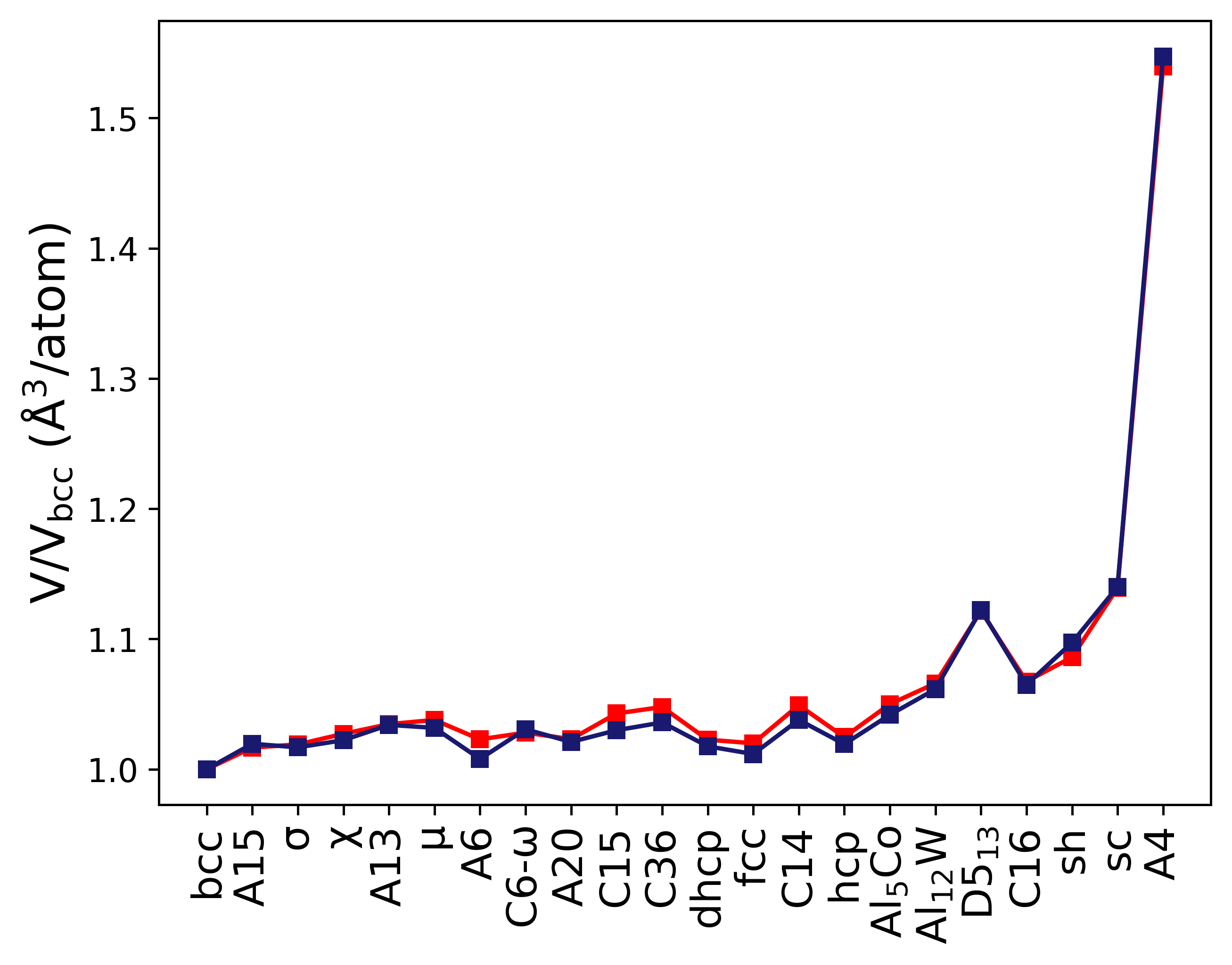}
\caption{Equilibrium volume}
\label{fig:W_V}
\end{subfigure}\qquad
\begin{subfigure}[b]{0.7\columnwidth}
\includegraphics[width=\textwidth]{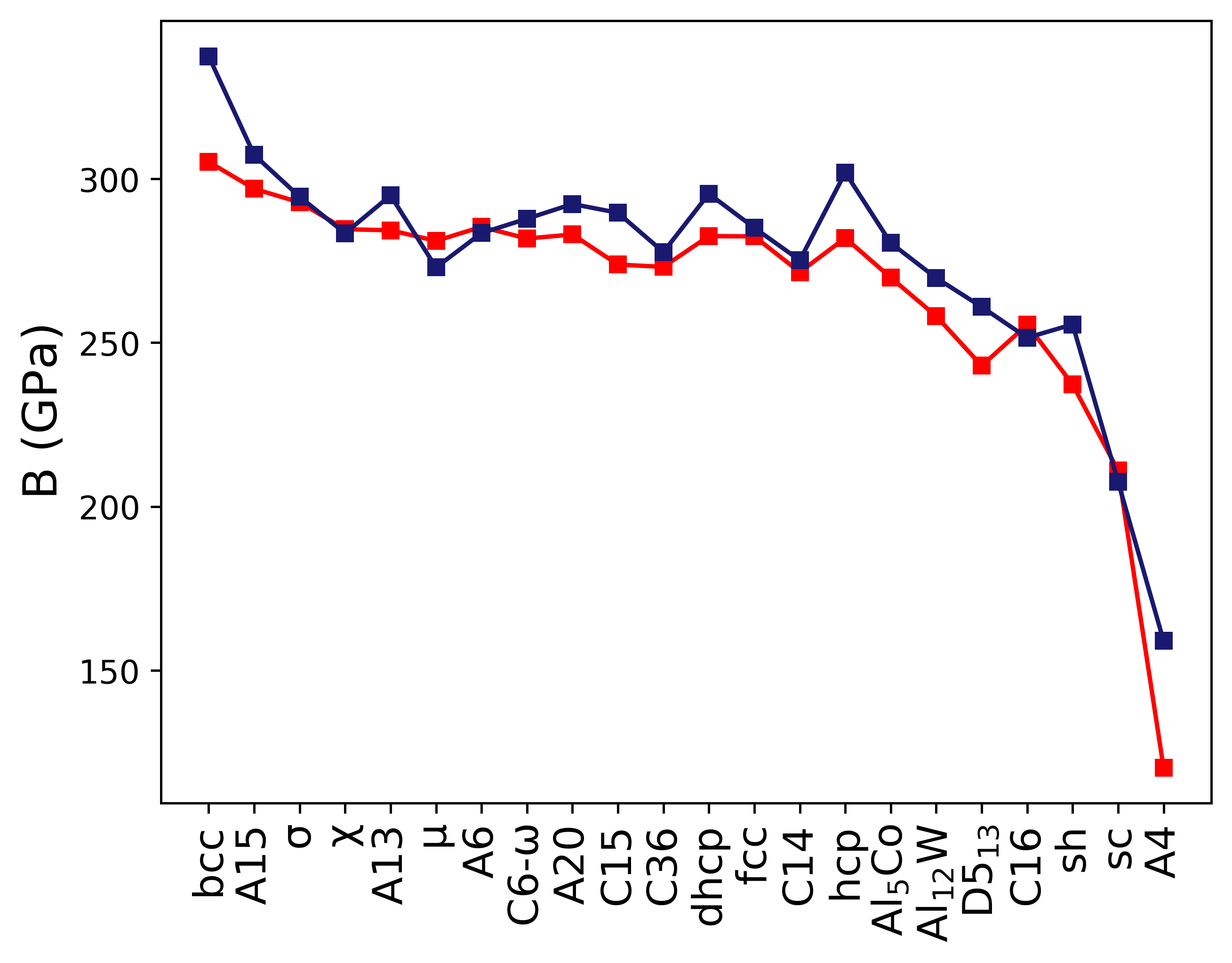}
\caption{Bulk modulus}
\label{fig:W_B}
\end{subfigure}\qquad
\caption{Performance of the Rev-EM ACE-2 model on prototypical crystal structures for W.}
\label{fig:Prototypes_W}
\end{figure*}

\begin{figure}[h]
\includegraphics[width=\columnwidth]{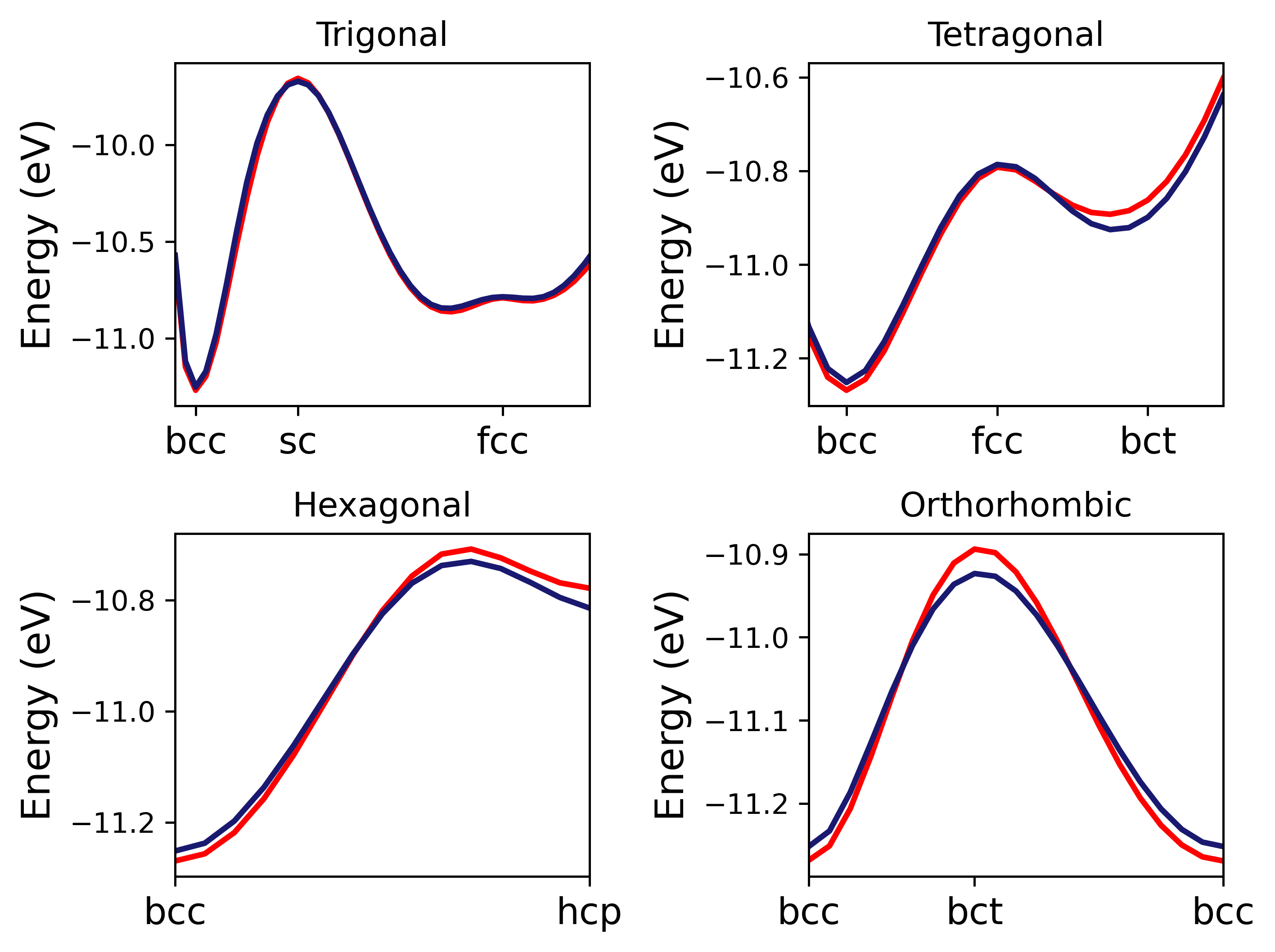}
\caption{Transformation paths between crystal structures for the ACE-2 model (blue) and DFT(red).} 
\label{fig:W_transformation_paths}
\end{figure}

Fig.\ \ref{fig:Prototypes_W} shows that ACE-2 accurately predicts the relative formation energies, formation volumes and bulk moduli of a set of 18 crystal structures ranging from the known BCC ground state to high-energy structures such as A4 (which is about 2.5 eV/atom higher in energy). None of these crystal structures were manually introduced into Rev-EM and that some of them (including TCP and laves phases) have unit cells that are larger than the largest  configuration present in Rev-EM (25 atoms). 
The largest relative errors are observed for the bulk modulus, with errors less than about 5\% on all phases. Agreement with cubic elastic constants is similarly good, as shown in Tab.\ \ref{tab:W_defects}, although an error of about 23\% is observed for $C_{12}$. Transformation paths (generated following the procedure presented in Refs.\ \cite{Cak_2014, Paidar_1999, PhysRevB.66.094110}) between some of these same crystal structures are also extremely well captured, as shown in Fig.\ \ref{fig:W_transformation_paths}, demonstrating that ACE-2 accurately captures the energy landscape between the local minima corresponding to each structure.

\begin{figure}[h]
\includegraphics[width=\columnwidth]{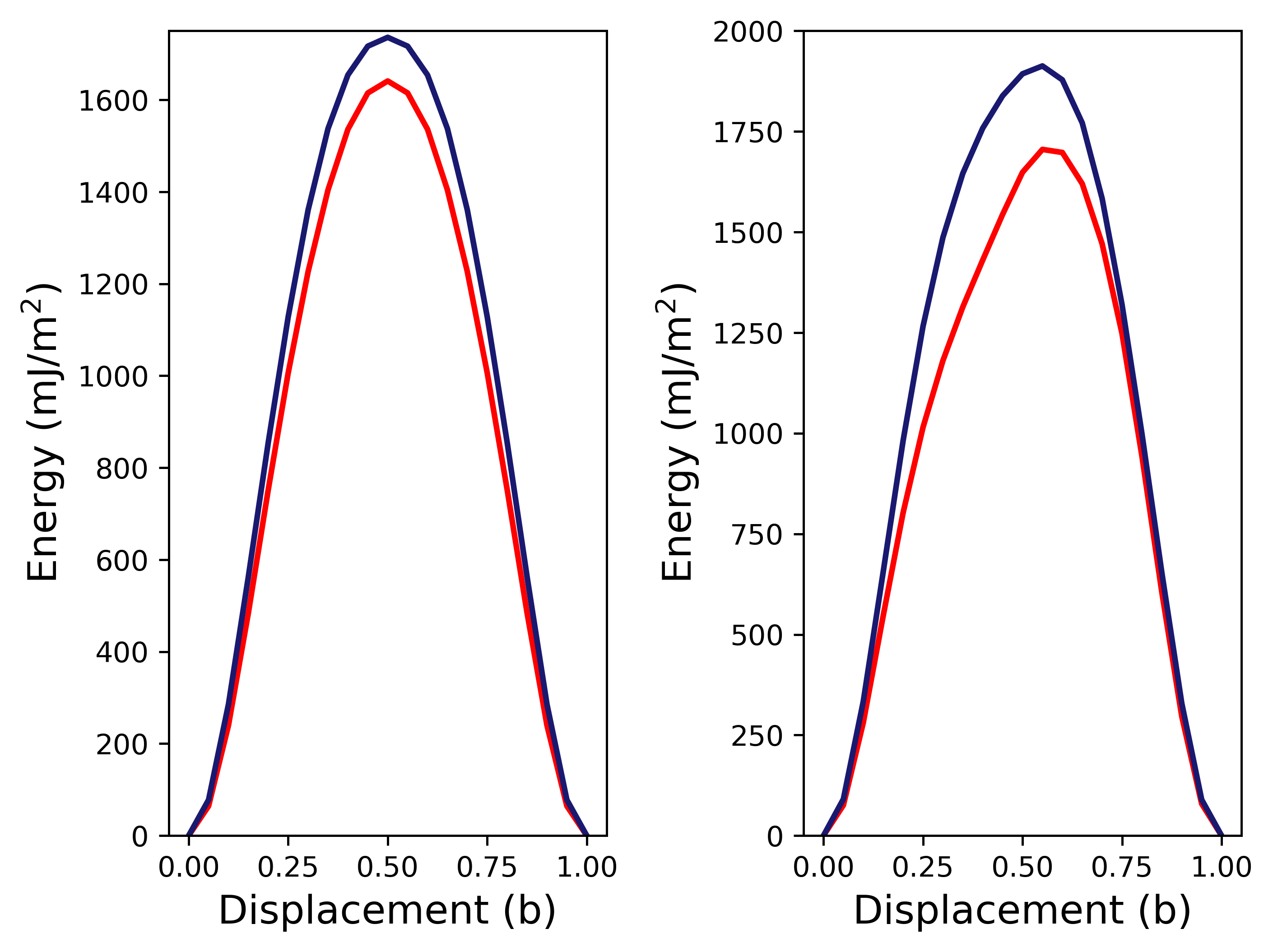}
\caption{Generalized stacking fault energies of the (left) (110) and (right) (112) planes in the [111] direction, where the displacement is shown in units of the Burgers vector (\textbf{b}). DFT results are in red and the ACE results are in blue.} 
\label{fig:W_GSFE}
\end{figure}

ACE-2 also accurately captures the formation energies of a number of surfaces, as reported in Tab.\ \ref{tab:W_defects}. Reference reconstruction-free surface structures are taken from the Crystalium web application at the Materials Virtual Lab website \cite{Tran2016}. Both the absolute values and ordering of the different facets are generally well captured, with the largest discrepancy being observed for the (100) surface (229 mJ/m$^2$, about 5.7\%), with the errors on other orientations being significantly smaller. This agreement extends to other planar defects, e.g., to the generalized stacking fault energy surfaces (GSFE), which is critical for an accurate representation of dislocation core structures. The GSFE for the (110) and (112) planes along the [111] direction are shown in Fig. \ref{fig:W_GSFE}. 
The GSFE is symmetrical in the (110) plane while being asymmetric in the (112) plane, which is reproduced by ACE-2, which slightly overestimates the unstable stacking fault energies.

The energetics of point defects, which are, e.g., critical to describe the behavior of materials under irradiation, is also relatively well captured by the ACE-2 model, as shown by the percentage errors in Tab. \ \ref{tab:W_defects}. The point defects were calculated using a $4\times 4\times 4$ bcc W supercell (with the bulk, mono-vacancy, di-vacancy and SIA supercells containing 128, 127, 126 and 129 atoms respectively). Both vacancy and di-vacancy formation energies are well captured, albeit with a slight overestimation for the di-vacancies. The ordering of the formation energies of the different interstitial configurations is captured (which is critical to predict the relative stability of different interstitial variants), although the absolute formation energies are systematically overestimated by about 1 eV (around 10\%). 
\begin{table}[!h]
\resizebox{\columnwidth}{!}{%
\begin{tabular}{c | c | c }
\hline\hline
 & DFT & ACE-2 \\ 
 & & (E$_\mathrm{min}$+2 eV) \\
\hline
Elastic constants (GPa) & & \\
\hline
C$_\mathrm{11}$ & 526 & $+$9.9\% \\
C$_\mathrm{12}$ & 197 & $+$23.4\% \\
C$_\mathrm{44}$ & 145 & $+$6.4\% \\
\hline
Vacancy formation energies (eV) & & \\
\hline
Mono & 3.25 & $+$0.3\% \\
Di (1NN) & 6.59 & $-$4.5\% \\
Di (2NN) & 6.91 & $-$5.56\% \\
\hline
Self-Interstitial atoms (eV) & & \\
\hline
<111> & 10.04 & $+$11.6\% \\
<110> & 10.45 & $+$9.4\% \\
TS & 11.63 & $+$7.1\% \\
<100> & 12.1  & $+$7.2\% \\
OS & 12.17 & $+$8.0\% \\
\hline
Surface formation energies (mJ/m$^2$) & & \\
\hline
110 & 3292.43 & $+$1.09\% \\
111 & 3515.92 & $+$2.36\% \\
321 & 3523.67 & $-$0.56\% \\
332 & 3604.73 & $-$1.07\% \\
322 & 3614.26 & $-$1.53\% \\
320 & 3625.98 & $-$2.43\% \\
311 & 3702.01 & $-$1.75\% \\
210 & 3745.56 & $-$3.58\% \\
310 & 3769.54 & $-$3.73\% \\
100 & 4023.14 & $-$5.69\% \\
211 & 4038.08 & $-$0.61\% \\
\hline\hline
\end{tabular}}
\caption{Comparison of various properties predicted by ACE-2 and the DFT reference values for BCC W.
}
\label{tab:W_defects}
\end{table}

Overall, given the relative simplicity of the MLIAP considered here (which contains 1094 adjustable parameters, a modest number by modern standards), the lack of hyperparameter tuning, and the material-agnostic nature of the dataset, the ACE-2 potential fitted to the extremely challenging Rev-EM dataset performs remarkably well on a broad range of low-energy metrics traditionally used to evaluate MLIAPs. This demonstrates that information pertaining to physically-relevant low-energy local environments is contained within Rev-EM. At the same time, the MLIAP is extremely robust, providing a good description of an exceptionally challenging and large test set that extends to high energies and forces. 

\subsection{The accuracy/robustness trade-off of Rev-EM}
\label{sec:tradeoff}
The extreme diversity of Rev-EM could be expected to lead to unacceptably high errors on low-energy configurations that are the most thermodynamically-relevant at ambient conditions unless extremely high complexity MLIAPs are used. In order to mitigate this difficulty, ACE-2 was trained using an energy reweighting scheme that assigned 75\% of the total weight to configurations within 2 eV/atom from the lowest energy configuration in the dataset, and 25\% to the rest (see Sec.\ \ref{sec:ace} for the details of the procedure). 
This reweighting exposes a hyperparameter that can be adjusted to tune the accuracy/robustness tradeoff. To demonstrate this, two other ACE potentials, referred to as ACE-1 and ACE-3, were trained with a high/low energy separation set to 1 eV/atom and 3eV/atom respectively (instead of 2 eV/atom for ACE-2). ACE-1 should intuitively lead to more accurate low-energy properties at the expense of higher errors in extreme conditions. In addition, the upper limit on forces has been decreased from 100 eV/\AA~ for the ACE$-2$ model to 50 eV/\AA~ for the ACE$-1$ model, again benefiting near-equilibrium structures, while the upper limit on the forces remains unchanged at 100 eV/\AA~ for ACE-3.
\begin{table*}[!h]
\centering
\begin{tabular}{c | c c | c c | c c}
\hline\hline
Model & \multicolumn{2}{c|}{All} & \multicolumn{2}{c|}{E$_\mathrm{min}$+1eV} & \multicolumn{2}{c}{E$_\mathrm{min}$+2eV} \\
\hline
& Train & Test & Train & Test & Train & Test \\
\hline
ACE-1 & 74.31/509.2 & 83.77/800.47 & 19.84/192.37 & 23.3/224.16 & 38.03/260.05 & 37.94/281.22 \\
ACE-2 & 53.65/416.9 & 63.35/439.39 & 21.88/204.77 & 24.4/219.74 & 29.72/245.05 & 31.24/254.33 \\
ACE-3 & 47.15/377.9 & 52.45/425.61 & 25.59/210.09 & 26.9/225.65 & 32.05/243.98 & 32.29/255.98 \\
\hline\hline
\end{tabular}
\caption{RMS test errors of the different W ACE models for energies and forces, respectively. The columns represent different energy ranges over which averages are taken: {\em All} over all configurations with energies within 20 eV/atom of the minimum energy structure in Rev-EM, {\em E$_\mathrm{min}$+1 eV} over all configurations within 1 eV/atom of the minimum energy (E$_\mathrm{min}$) structure in Rev-EM, and {\em E$_\mathrm{min}$+2 eV} over all configurations within 2 eV/atom of the minimum energy structure in Rev-EM.
}
\label{tab:ACE_models_RMSEs}
\end{table*}

This intuition is supported by comparing the test errors for the three ACE models (see Tab.\ \ref{tab:ACE_models_RMSEs}). ACE-1 shows significantly higher test errors on the whole dataset, compared to ACE-2 and ACE-3. In exchange, this translates to generally lower errors at low energies, although ACE-1 shows signs of overfitting to low energy structures. In contrast, ACE-3 shows significantly lower errors on the whole test set, at the expense of (slightly) higher errors on low energy configurations. This reweighting scheme can be used to adjust the accuracy of target low energy properties. For example, the ACE-1 model shows significant improvement in two of the three elastic constants, the $C_{11}$, $C_{12}$ and $C_{44}$ for the ACE-1 model are 525 GPa, 198 GPa and 167 GPa respectively, against 578 GPa, 243 GPa and 155 GPa respectively for the ACE-2 model. This indicates that developing new reweighting schemes for Rev-EM is likely to provide significant advantages in terms of adapting the performance of the models to different application areas; this aspect will be explored in a follow-on study.

Encouragingly, we note that the performance of ACE-2 fitted to Rev-EM and tested on datasets available from the literature is quantitatively comparable to that of the MLIAPs originally trained on these datasets (see Table\ \ref{tab:ACE_New_to_Old_DE_TM23}), suggesting that the accuracy/robustness tradeoff inherent to training to ultra-diverse datasets can be made very favorable, as was originally observed in Ref.\cite{NPJ_Montes}. For example, the equivariant message-passing neural network NequiP \cite{Batzner2022, Owen2024} that was trained on the Owen-W dataset yields MAEs of 85.13 meV/atom and 69.02 meV/\AA, respectively for energies and forces, compared to MAEs of 37.76 meV/atom and 227.53 meV/\AA~ for ACE-2. While the energy errors for ACE-2 are about twice as small, its force errors are about three times as large, which is likely attributable to the relatively low weight assigned to forces compared to energies (90\% vs 10\%) in the ACE-2 training. Therefore, the much higher diversity of Rev-EM compared to Owen-W (see Fig.\ \ref{fig:det_selected}) does not inherently lead a to significant degradation in performance when training to Rev-EM once the contributions of each structures to the loss function are reweighted.

\subsection{Rev-EM is material agnostic}
\label{sec:agnostic}
As no material specific information is used during the generation of Rev-EM, it should in principle be equally applicable to other unary materials. To demonstrate this, MLIAPs were trained to eight elements with different thermodynamics, namely fcc (Al), bcc (W), hcp (Be, Re and Os), graphite (C), trigonal A7 (Sb) and trigonal A8 (Te) ground states. Except in the case of C (see details below), these datasets were generated through a uniform volume rescaling of all configurations in Rev-EM according to the density in their known ground states. In all cases except for C, MLIAPs were obtained using the ACE-2 settings described in Sec.\ \ref{sec:ace}.

Tab.\ \ref{tab:ACE_New_to_DE_TM23} reports the test RMSE errors observed for these systems. 
Errors are significantly lower for Be, Al, Sb and Te compared to Re and Os, which is consistent with previous observations that the behavior of late transition metals is especially difficult to capture \cite{Owen2024}, and with the expectation that absolute errors should scale roughly with the cohesive energy. Once again, the ACE-2 potential shows good performance when tested on the Owen dataset, which is available for Re and Os, exhibiting significantly lower errors than for testing on holdout from Rev-EM, as expected from the limited diversity of this targeted dataset. 
Similarly, the ACE-2 model tested on the Smith-Al dataset shows comparable performance to the ANI model trained in Ref. \cite{ANI_Al}, albeit with larger force errors due to the relative energy/force importance in the training \footnote{We observed indications that the DFT calculations on two configurations in the Smith-Al dataset potentially did not converge. These were removed from our analysis.}. Note that the comparisons with literature datasets were based on published DFT data; inconsistencies between functionals and convergence parameters could slightly increase the measured test errors. The results once again show that MLIAPs trained to Rev-EM perform very well both on domain-expert and active-learning curated datasets.

\begin{table}[!h]
\resizebox{\columnwidth}{!}{%
\begin{tabular}{c | c | c }
\hline\hline
Element & Energy & Force \\
\hline
& (meV/atom) & (meV/\AA) \\
\hline
Be & 17.47 & 100.42 \\ 
C & 101.35 & 750.67 \\
Al & 9.76 (8.8) & 44.83 (94.42) \\
Sb & 19.02 & 79.99 \\
Te & 17.9 & 92.95\\
Re & 60.86 (68.31) & 458.29 (296.13) \\
Os & 90.05 (18.63) & 607.04 (362.48) \\
\hline\hline
\end{tabular}}
\caption{ACE-2 test RMS errors for the Rev-EM dataset for different elements. The values in parentheses show the RMSEs on the TM23 datasets made available from Ref. \cite{Owen2024} (for Re and Os) and from the Smith-Al dataset provided in Ref. \cite{ANI_Al} (for Al).}
\label{tab:ACE_New_to_DE_TM23} 
\end{table}

\begin{figure*}[p]
\begin{subfigure}[b]{0.7\columnwidth}
\includegraphics[width=\textwidth]{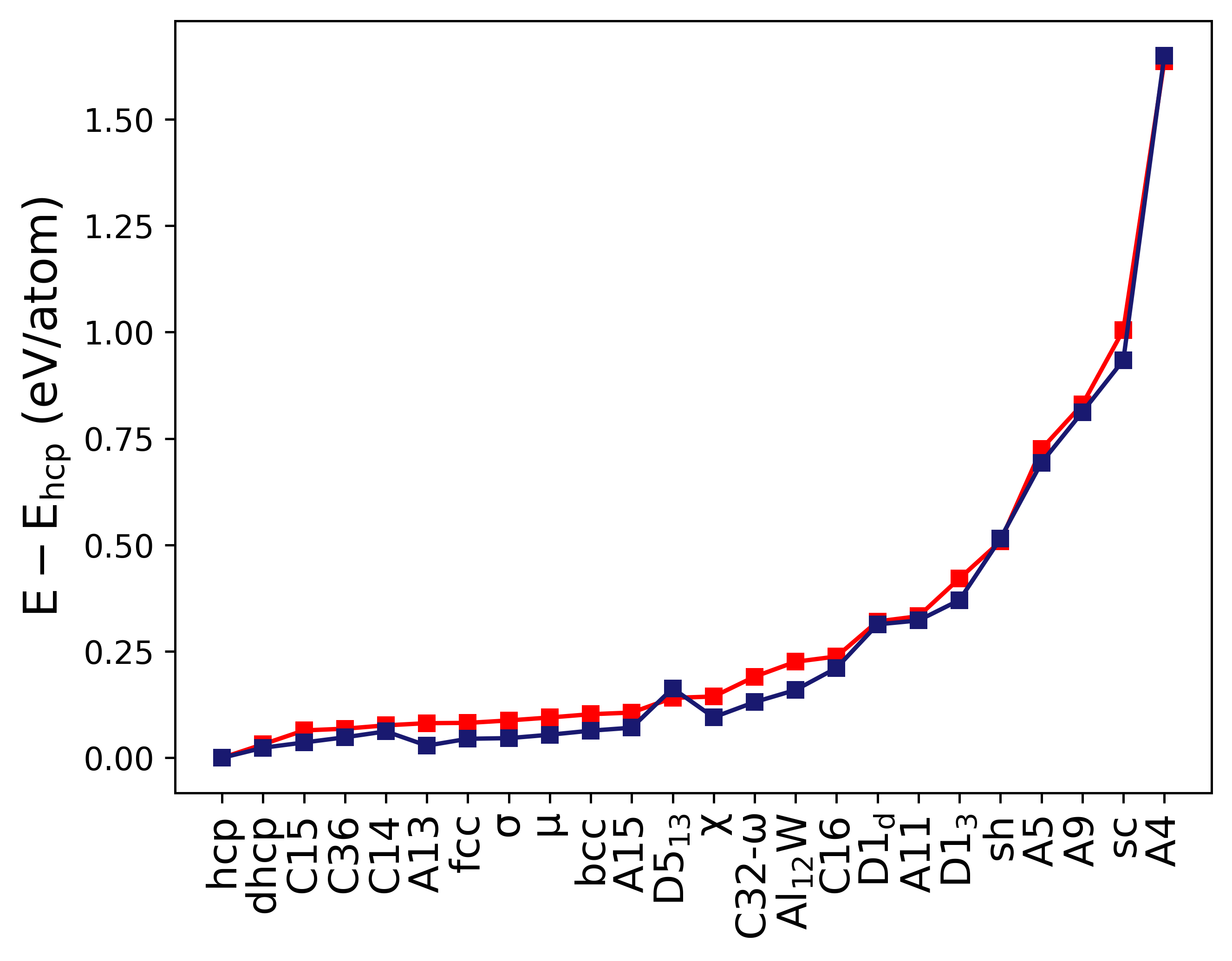}
\caption{Be (Equilibrium energies)}
\label{fig:Be_E}
\end{subfigure}\qquad
\begin{subfigure}[b]{0.7\columnwidth}
\includegraphics[width=\textwidth]{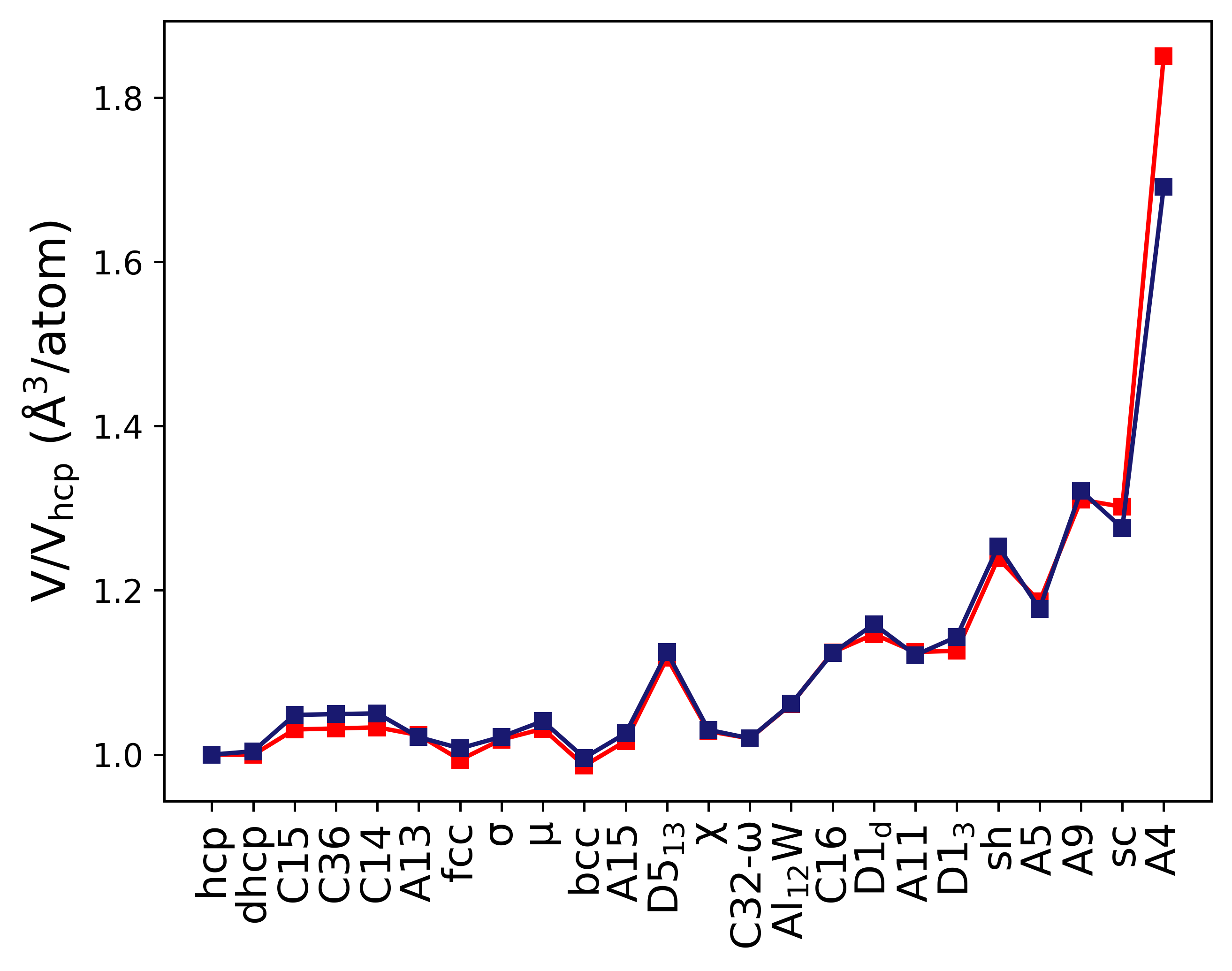}
\caption{Be (Equilibrium volumes)}
\label{fig:Be_V}
\end{subfigure}\qquad
\begin{subfigure}[b]{0.7\columnwidth}
\includegraphics[width=\textwidth]{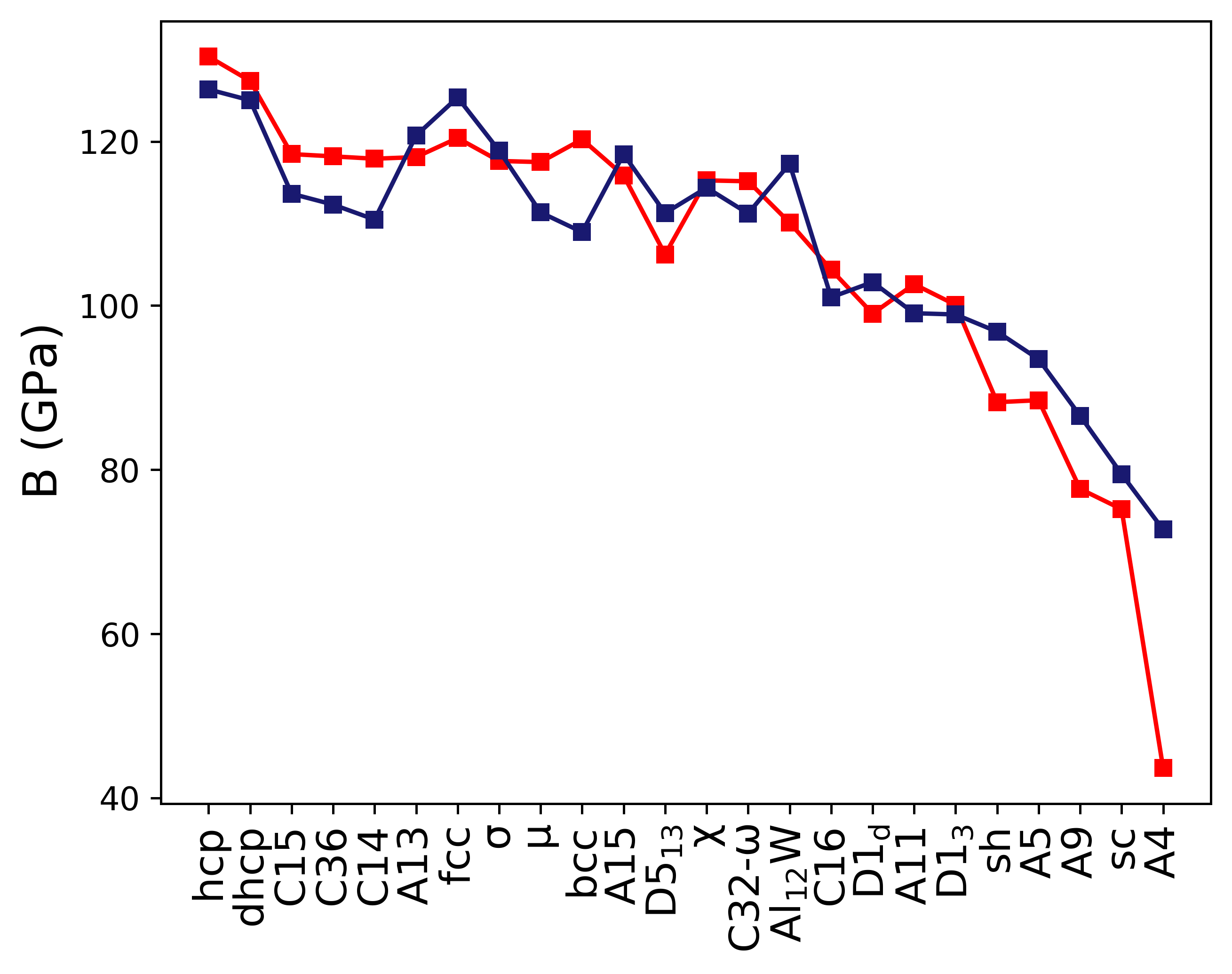}
\caption{Be (Bulk moduli)}
\label{fig:Be_B}
\end{subfigure}\qquad
\begin{subfigure}[b]{0.7\columnwidth}
\includegraphics[width=\textwidth]{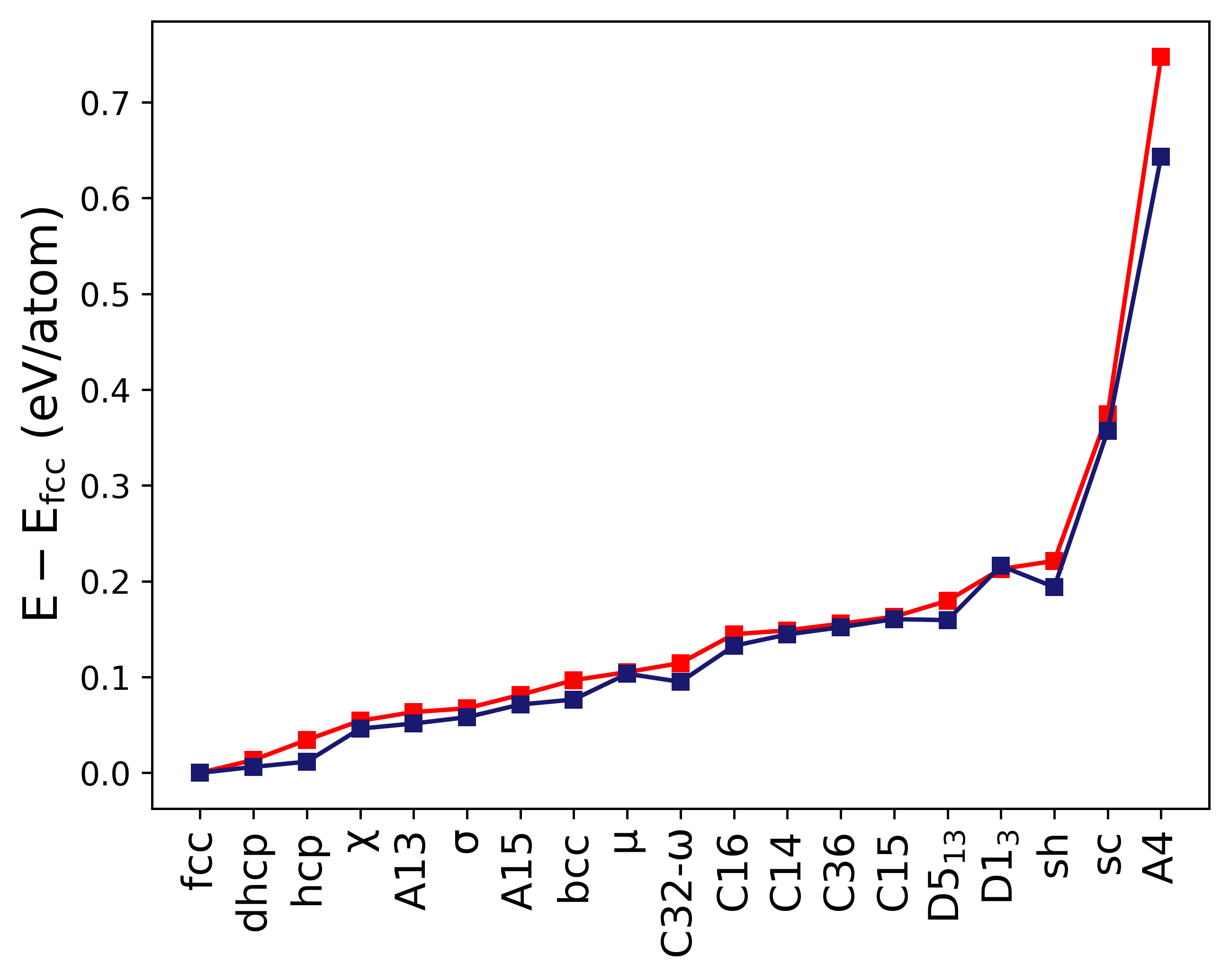}
\caption{Al (Equilibrium energies)}
\label{fig:Al_E}
\end{subfigure}\qquad
\begin{subfigure}[b]{0.7\columnwidth}
\includegraphics[width=\textwidth]{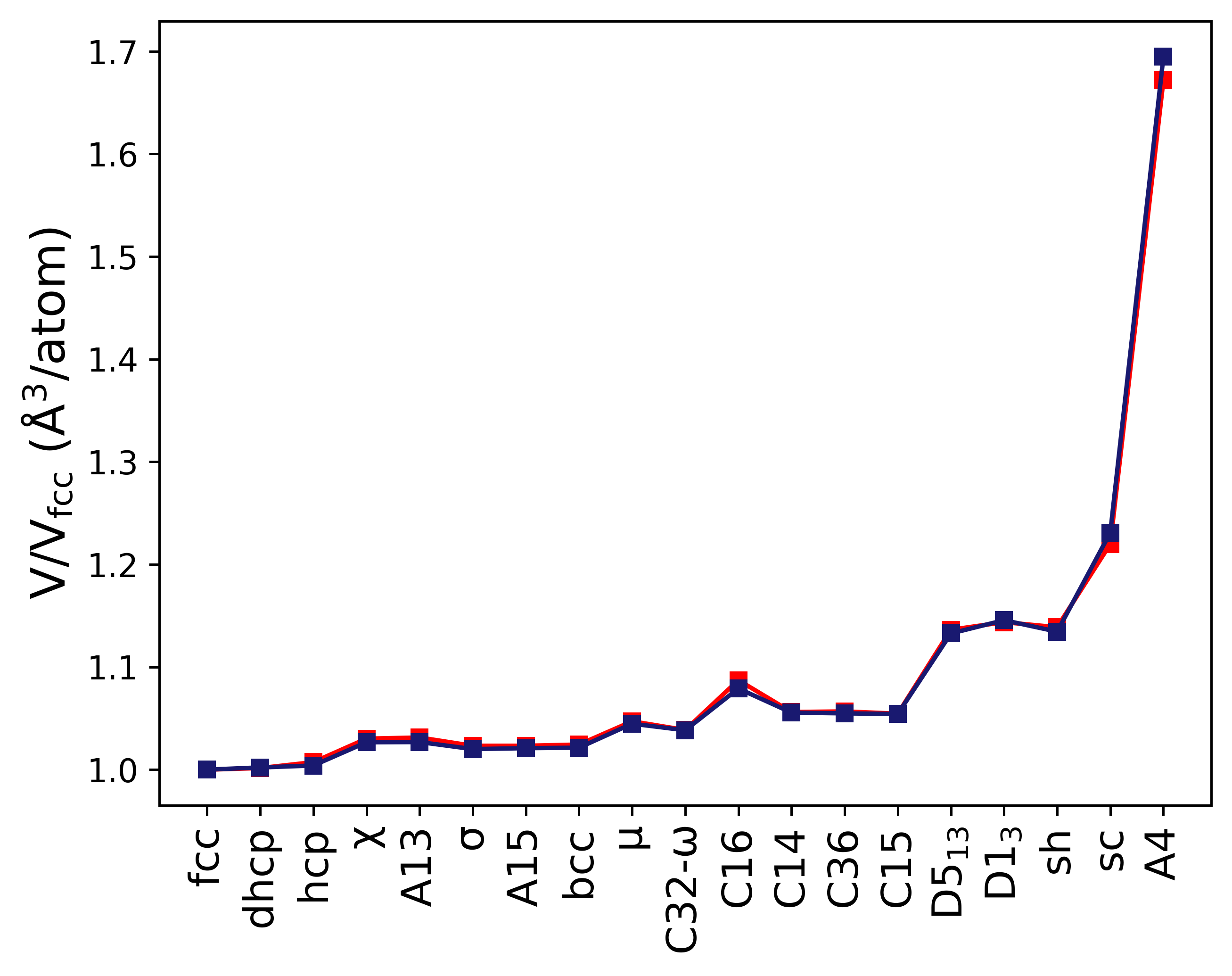}
\caption{Al (Equilibrium volumes)}
\label{fig:Al_V}
\end{subfigure}\qquad
\begin{subfigure}[b]{0.7\columnwidth}
\includegraphics[width=\textwidth]{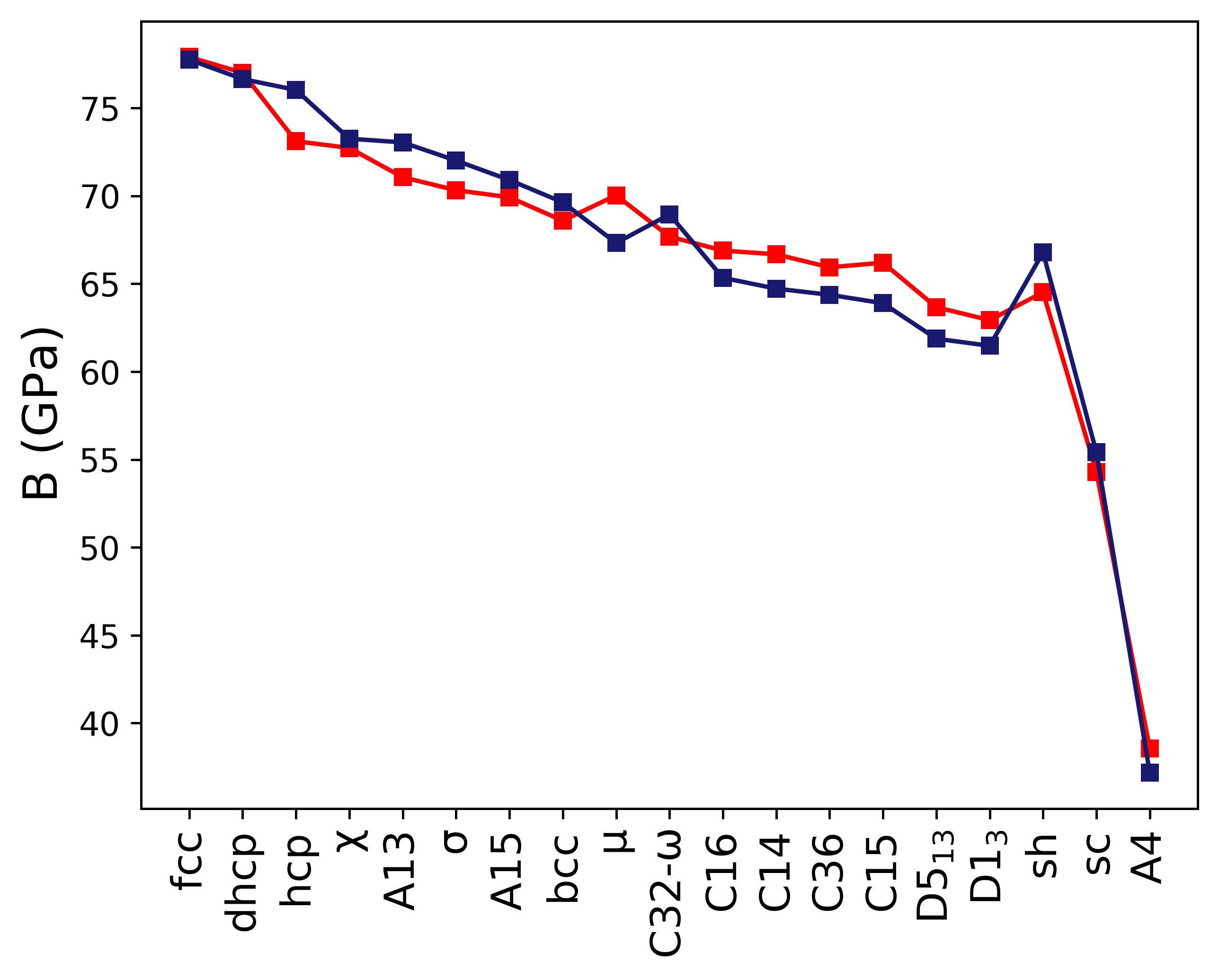}
\caption{Al (Bulk moduli)}
\label{fig:Al_B}
\end{subfigure}\qquad
\begin{subfigure}[b]{0.7\columnwidth}
\includegraphics[width=\textwidth]{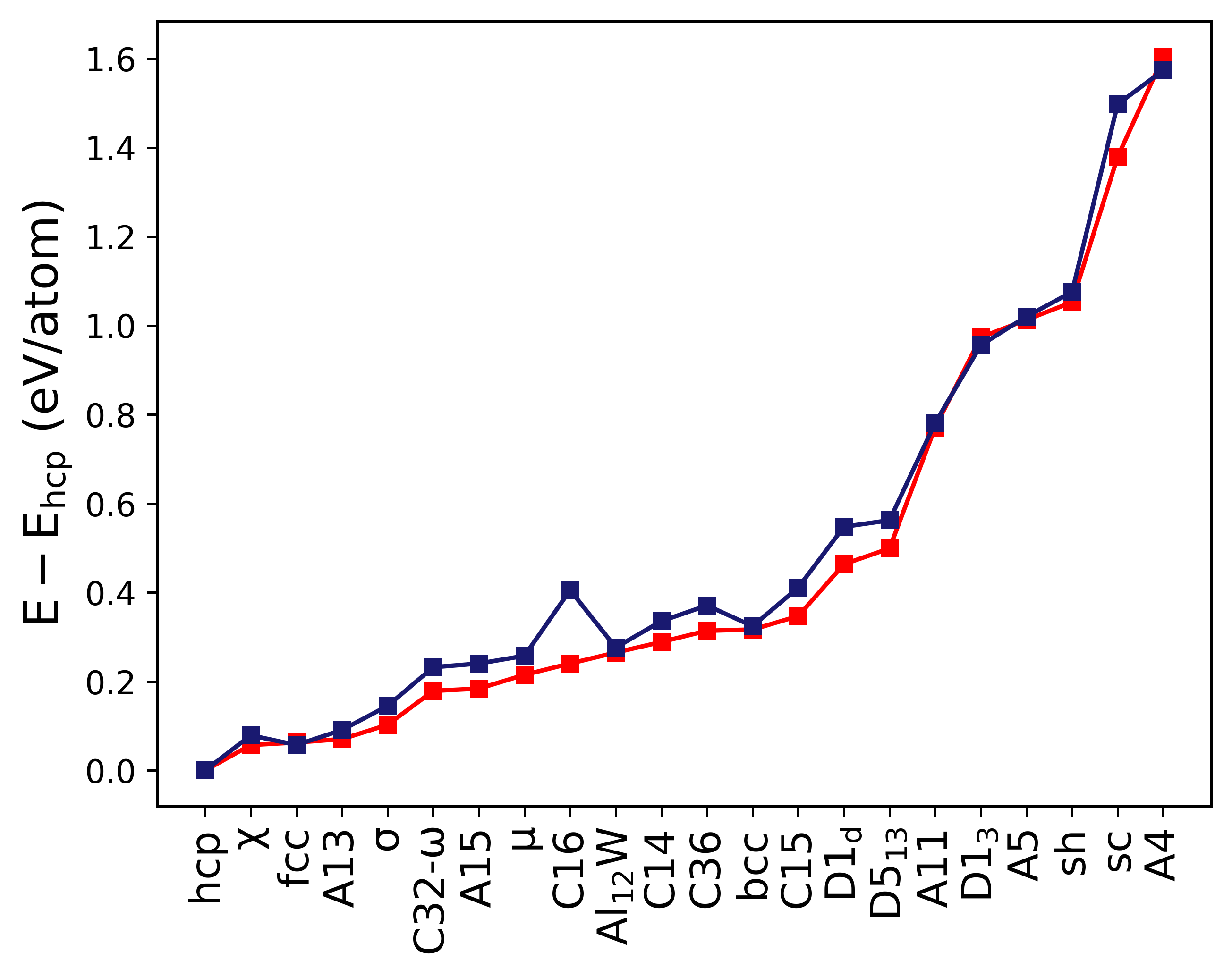}
\caption{Re (Equilibrium energies)}
\label{fig:Re_E}
\end{subfigure}\qquad
\begin{subfigure}[b]{0.7\columnwidth}
\includegraphics[width=\textwidth]{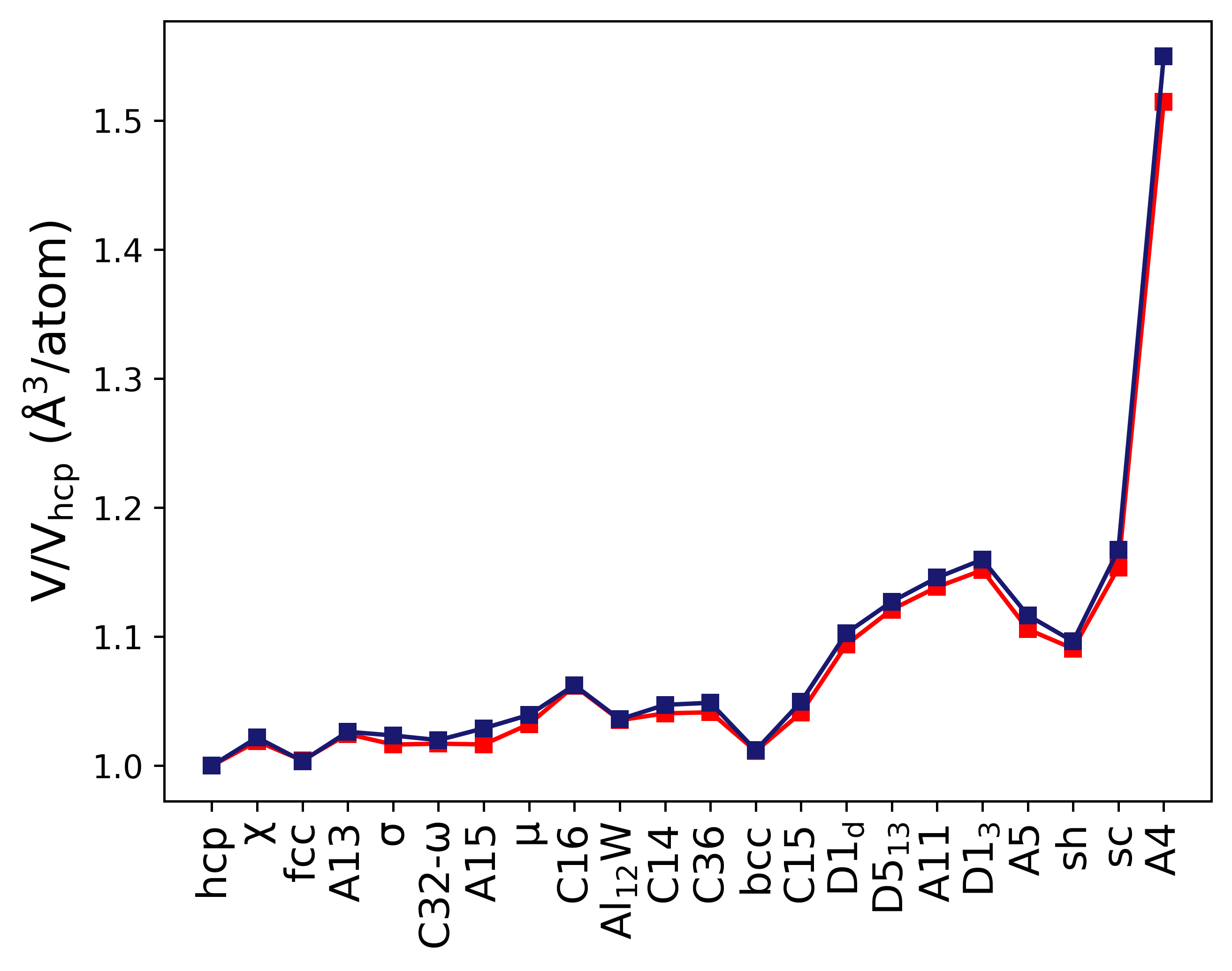}
\caption{Re (Equilibrium volumes)}
\label{fig:Re_C}
\end{subfigure}\qquad
\begin{subfigure}[b]{0.7\columnwidth}
\includegraphics[width=\textwidth]{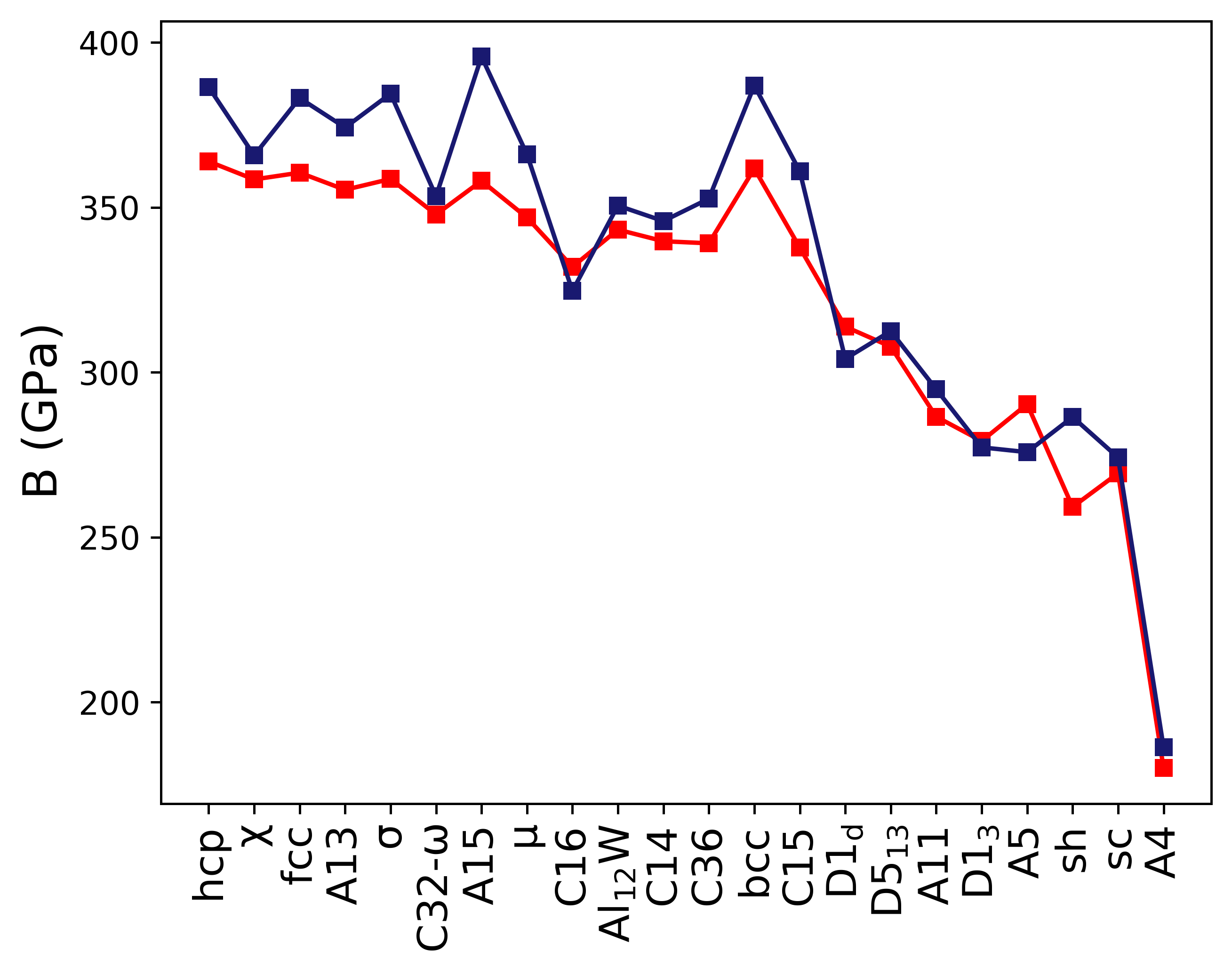}
\caption{Re (Bulk moduli)}
\label{fig:Re_B}
\end{subfigure}\qquad
\begin{subfigure}[b]{0.7\columnwidth}
\includegraphics[width=\textwidth]{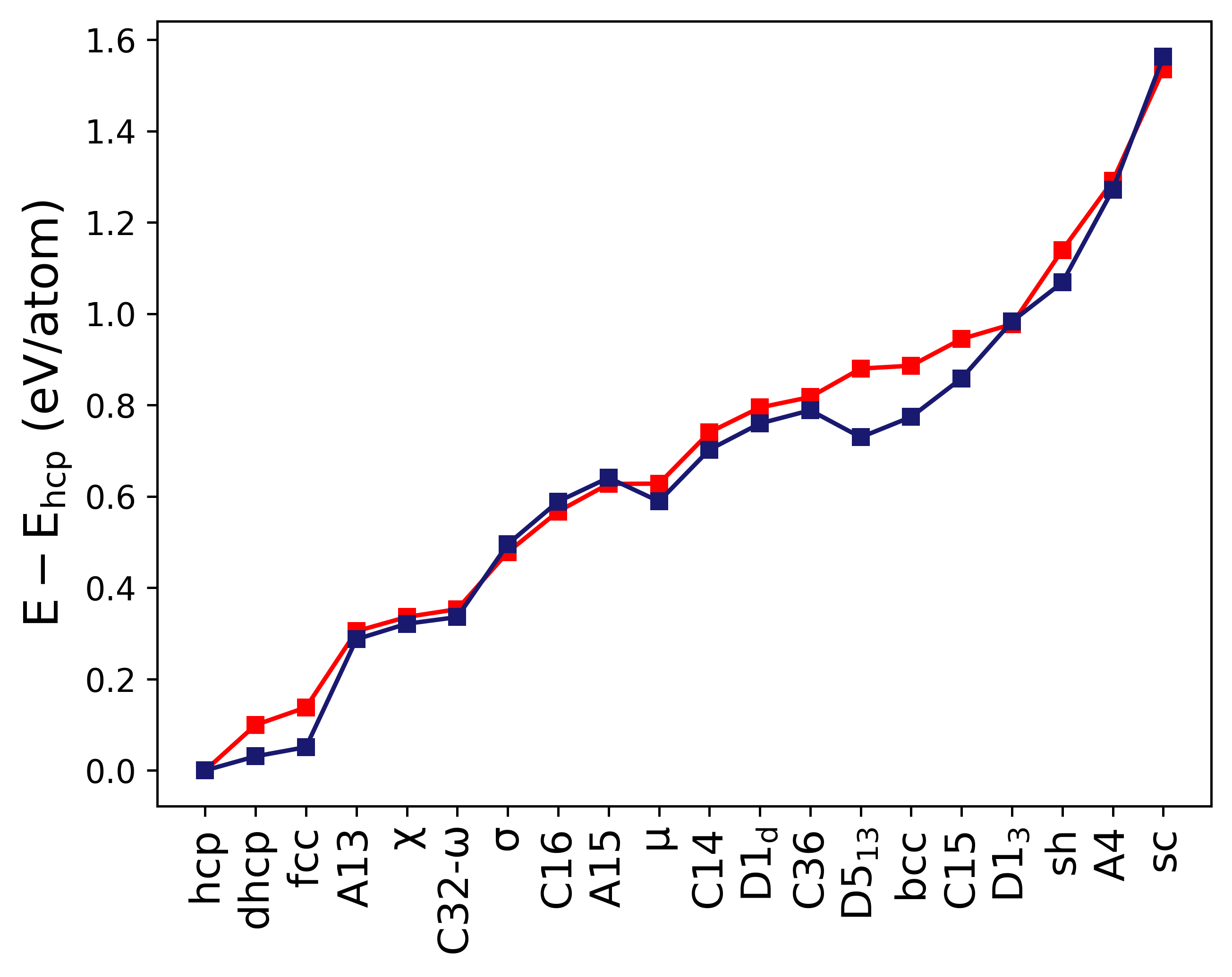}
\caption{Os (Equilibrium energies)}
\label{fig:Os_E}
\end{subfigure}\qquad
\begin{subfigure}[b]{0.7\columnwidth}
\includegraphics[width=\textwidth]{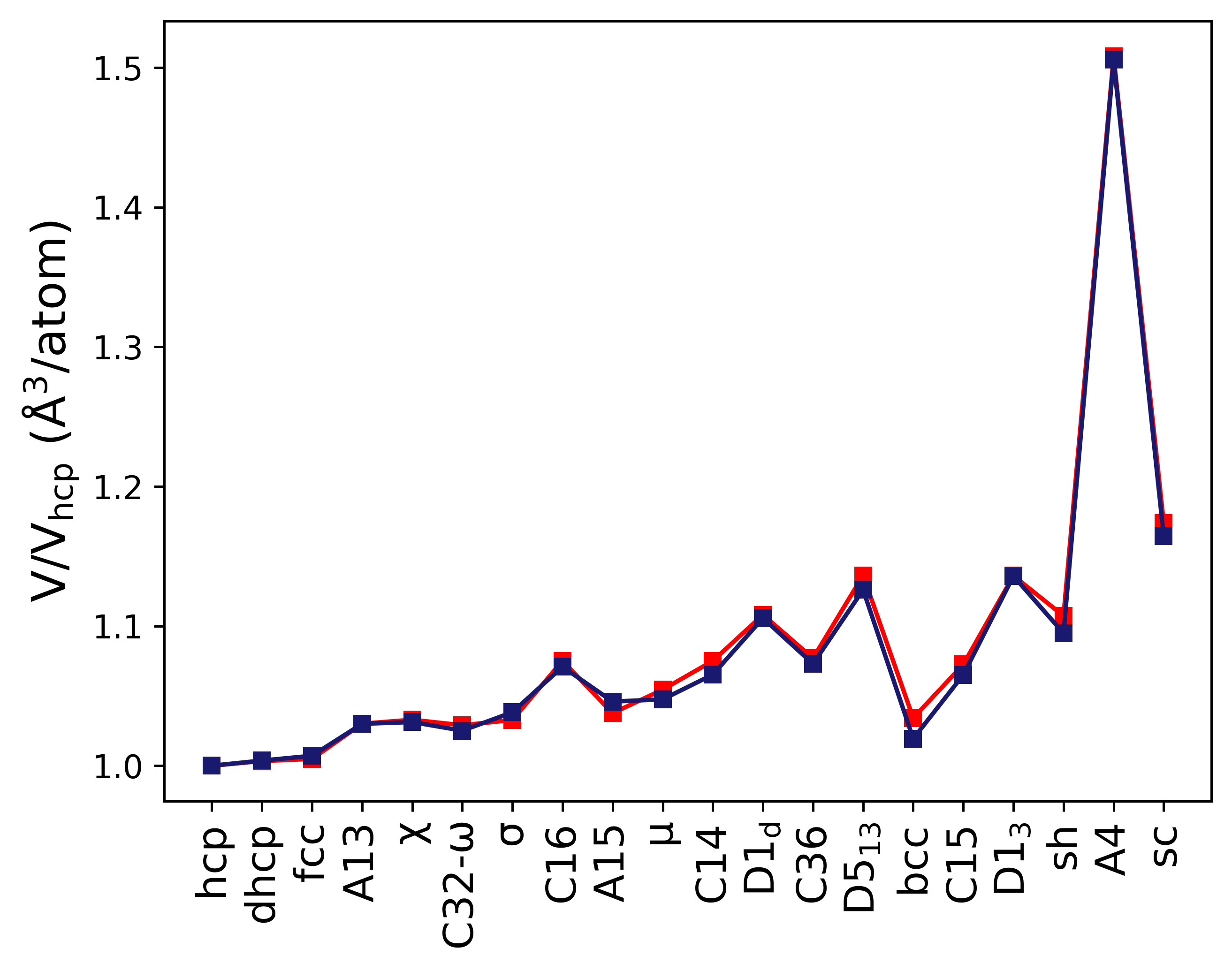}
\caption{Os (Equilibrium volumes)}
\label{fig:Os_V}
\end{subfigure}\qquad
\begin{subfigure}[b]{0.7\columnwidth}
\includegraphics[width=\textwidth]{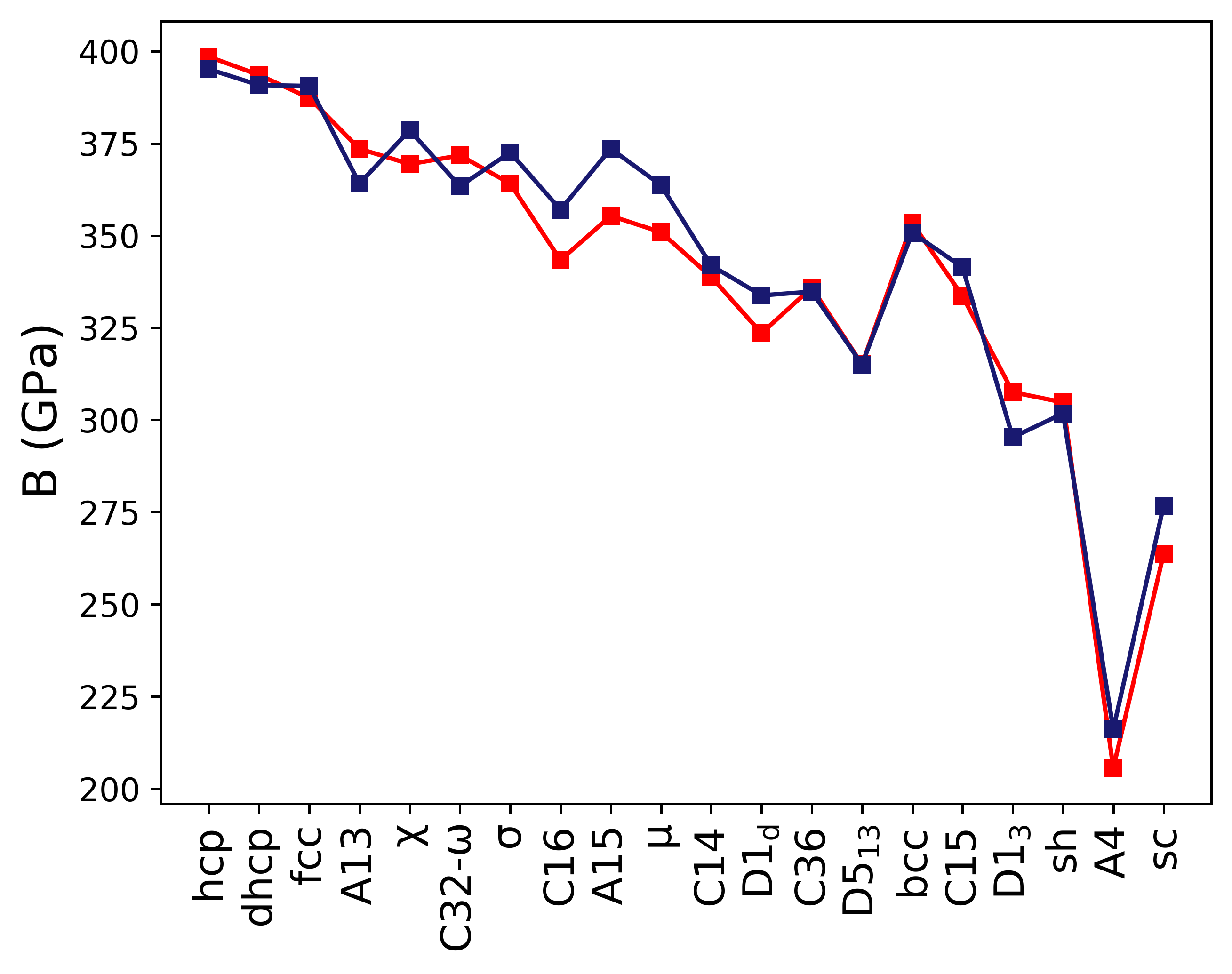}
\caption{Os (Bulk moduli)}
\label{fig:Os_B}
\end{subfigure}\qquad
\caption{Transferability of the trained ACE-2 models (blue) to prototypical structures for Be, Al, Re and Os.}
\label{fig:Prototypes_EVB}
\end{figure*}

Once again, the energy and force reweighting scheme ensures that low energy configurations are given more importance, and hence incur reasonable errors. Figs.\ \ref{fig:Prototypes_EVB} demonstrate that the ACE-2 potentials generally perform extremely well at predicting the properties of a broad range of different crystal structures, both in terms of absolute value and of relative ordering.

\begin{figure*}[h]
\begin{subfigure}[b]{0.7\columnwidth}
\includegraphics[width=\textwidth]{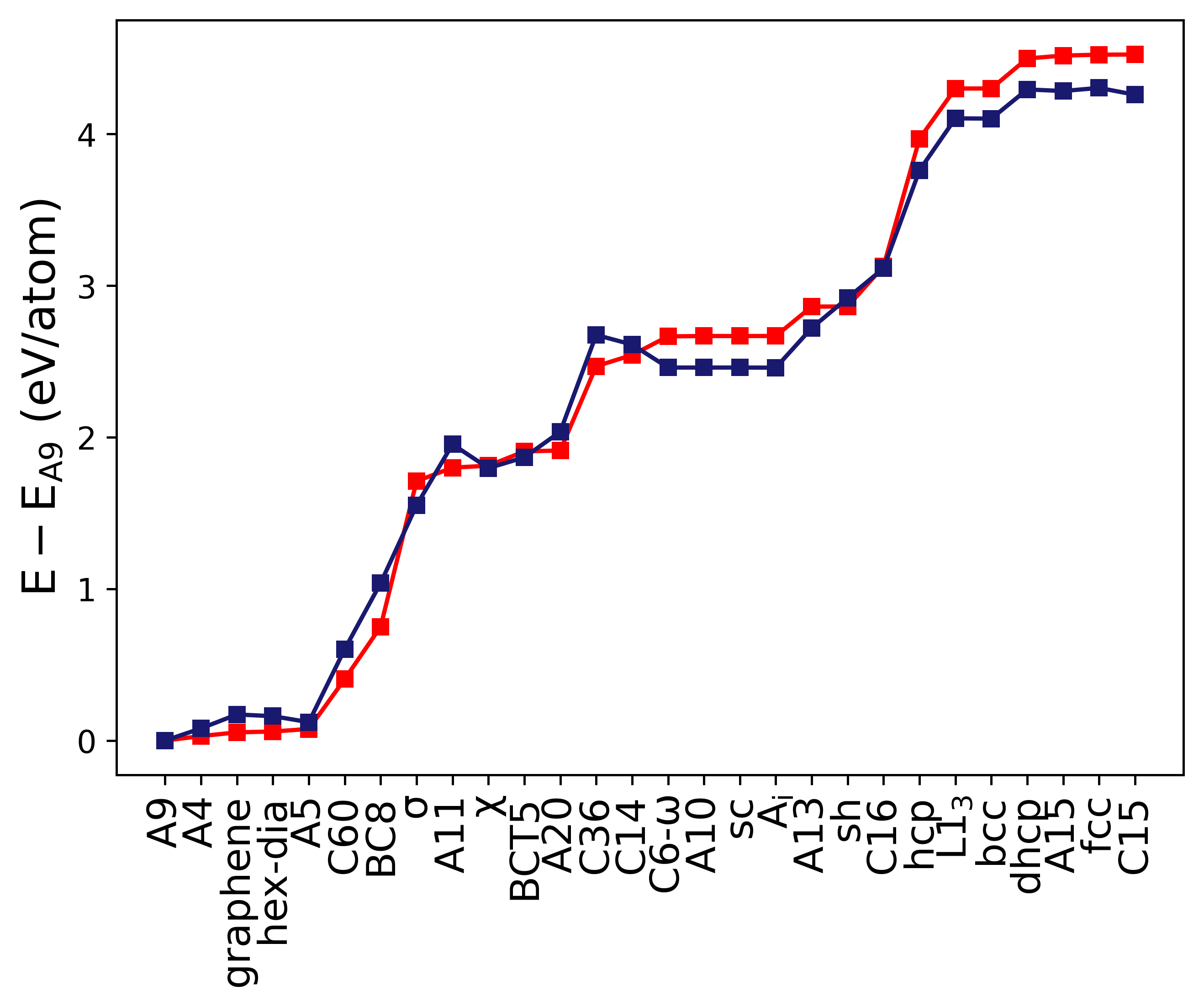}
\caption{C (Equilibrium energies)}
\label{fig:C_E}
\end{subfigure}\qquad
\begin{subfigure}[b]{0.7\columnwidth}
\includegraphics[width=\textwidth]{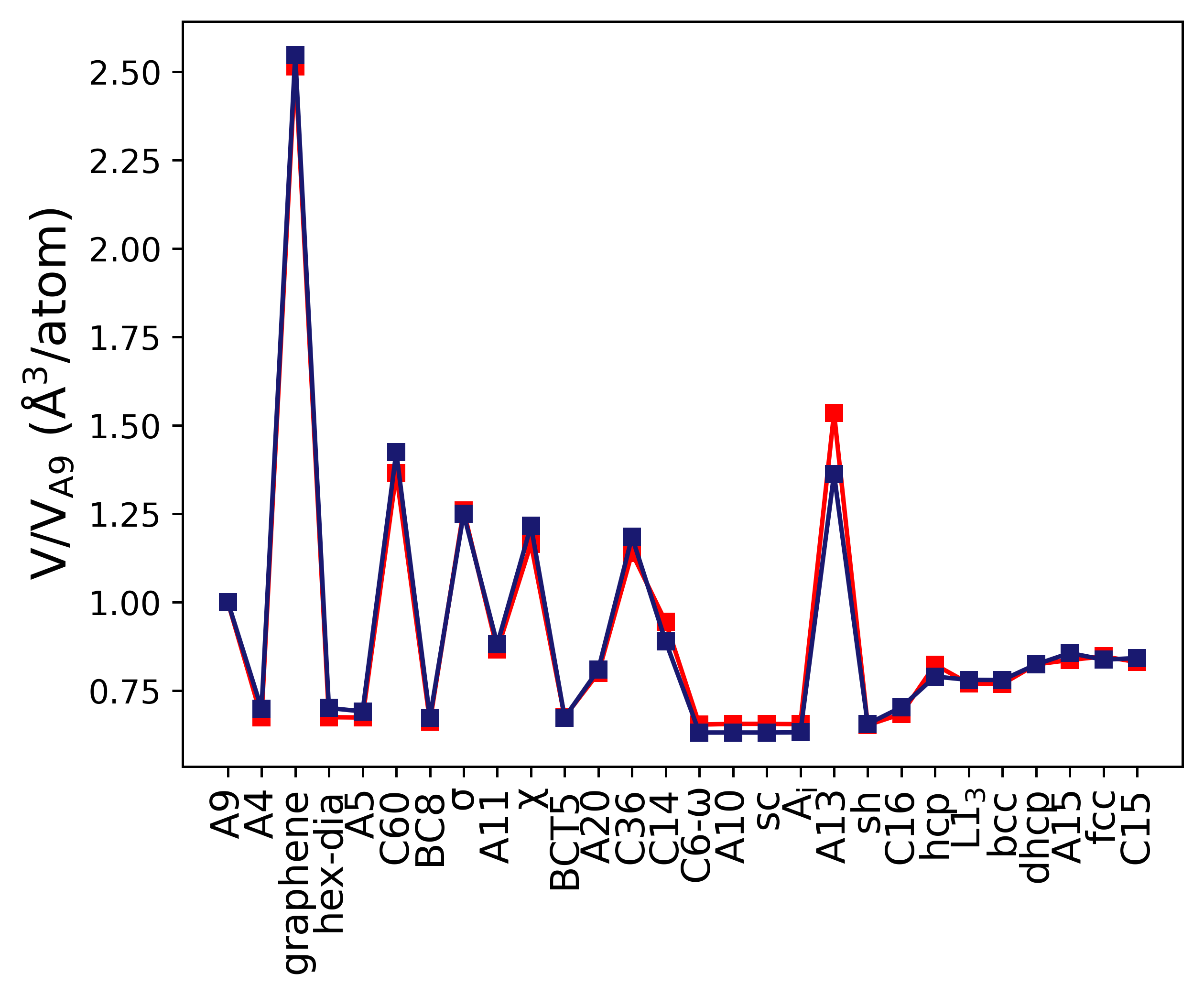}
\caption{C (Equilibrium volumes)}
\label{fig:C_V}
\end{subfigure}\qquad
\begin{subfigure}[b]{0.7\columnwidth}
\includegraphics[width=\textwidth]{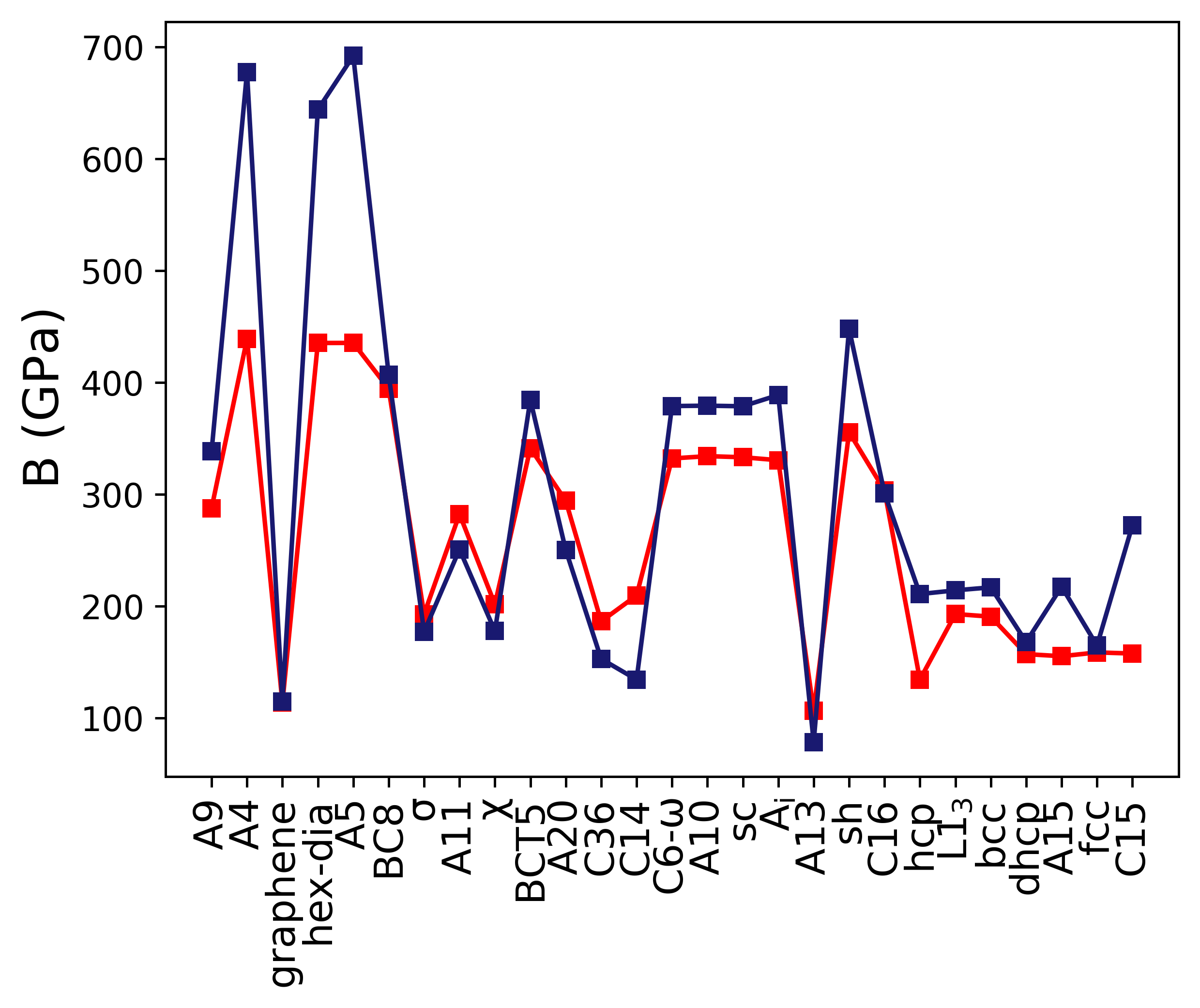}
\caption{C (Bulk moduli)}
\label{fig:C_B}
\end{subfigure}\qquad
\begin{subfigure}[b]{0.7\columnwidth}
\includegraphics[width=\textwidth]{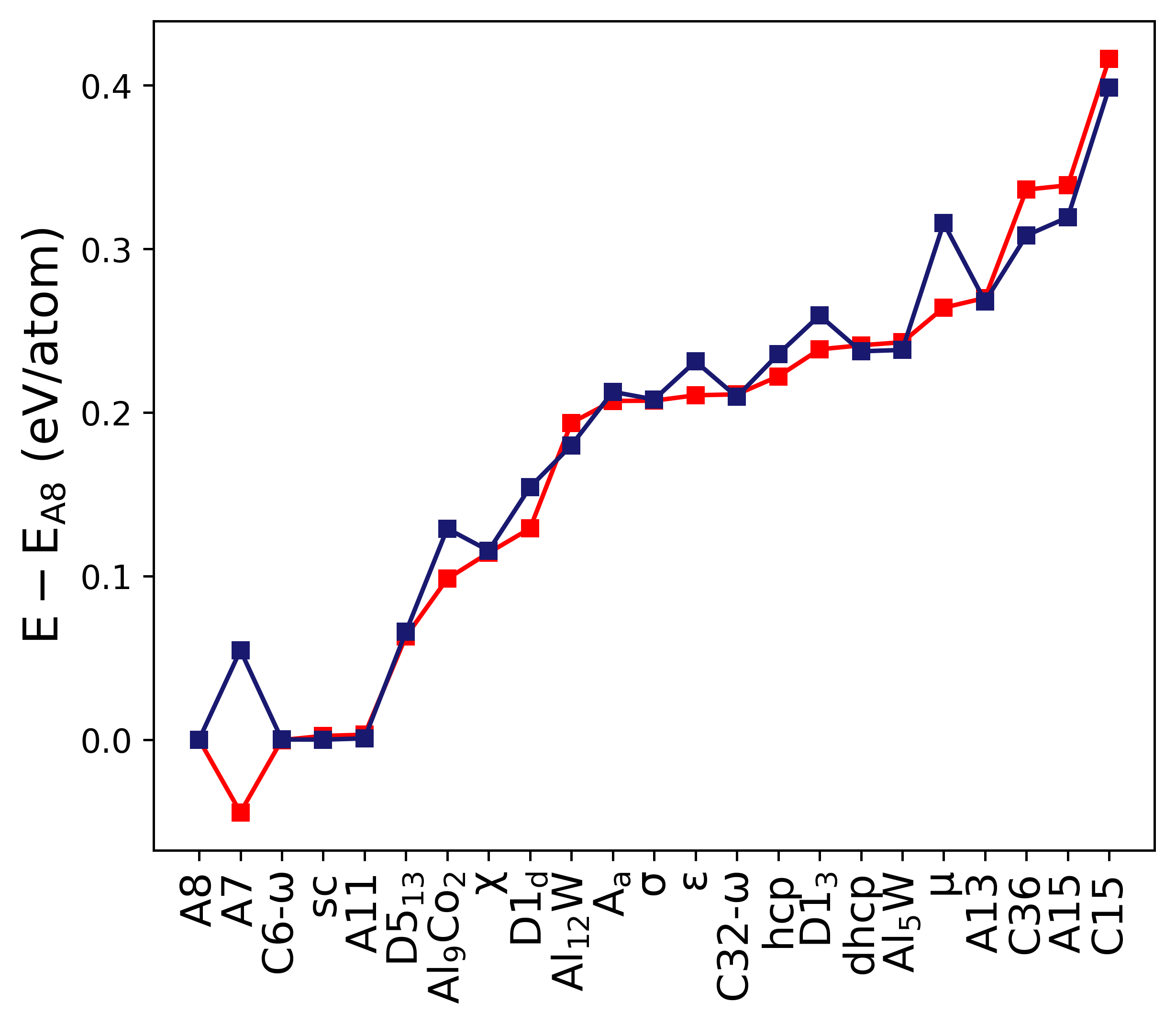}
\caption{Sb (Equilibrium energies)}
\label{fig:Sb_E}
\end{subfigure}\qquad
\begin{subfigure}[b]{0.7\columnwidth}
\includegraphics[width=\textwidth]{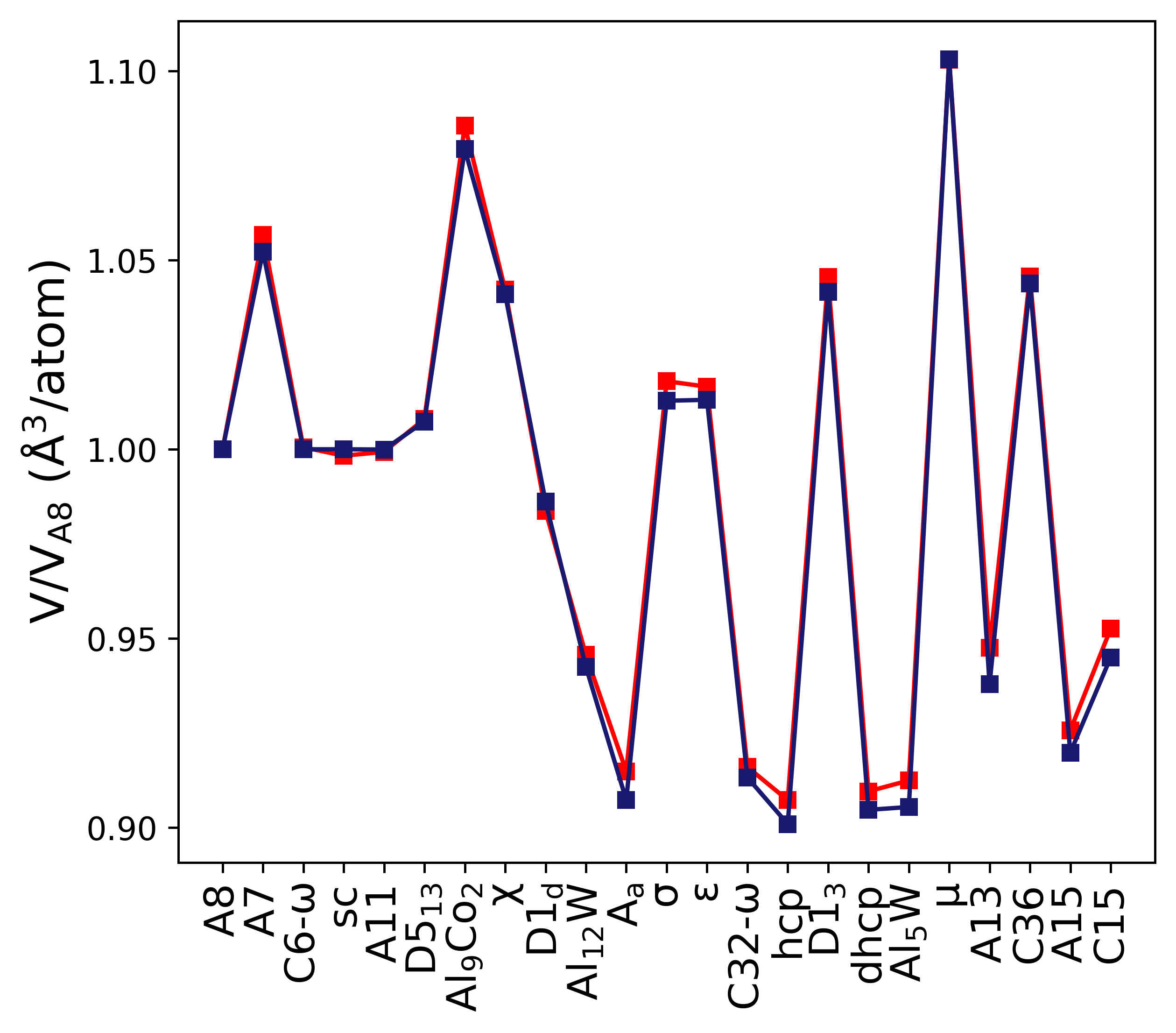}
\caption{Sb (Equilibrium volumes)}
\label{fig:Sb_V}
\end{subfigure}\qquad
\begin{subfigure}[b]{0.7\columnwidth}
\includegraphics[width=\textwidth]{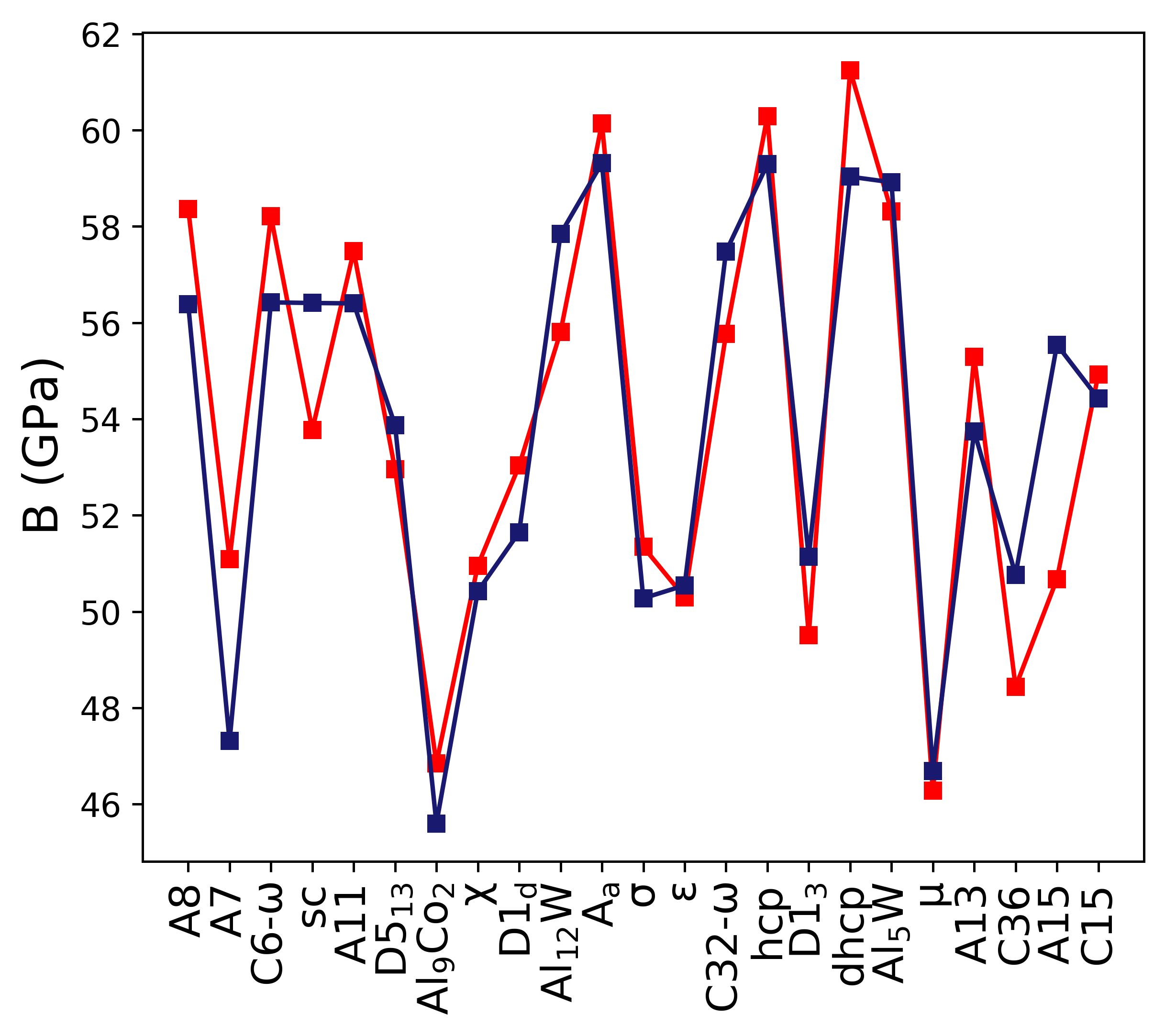}
\caption{Sb (Bulk moduli)}
\label{fig:Sb_B}
\end{subfigure}\qquad
\begin{subfigure}[b]{0.7\columnwidth}
\includegraphics[width=\textwidth]{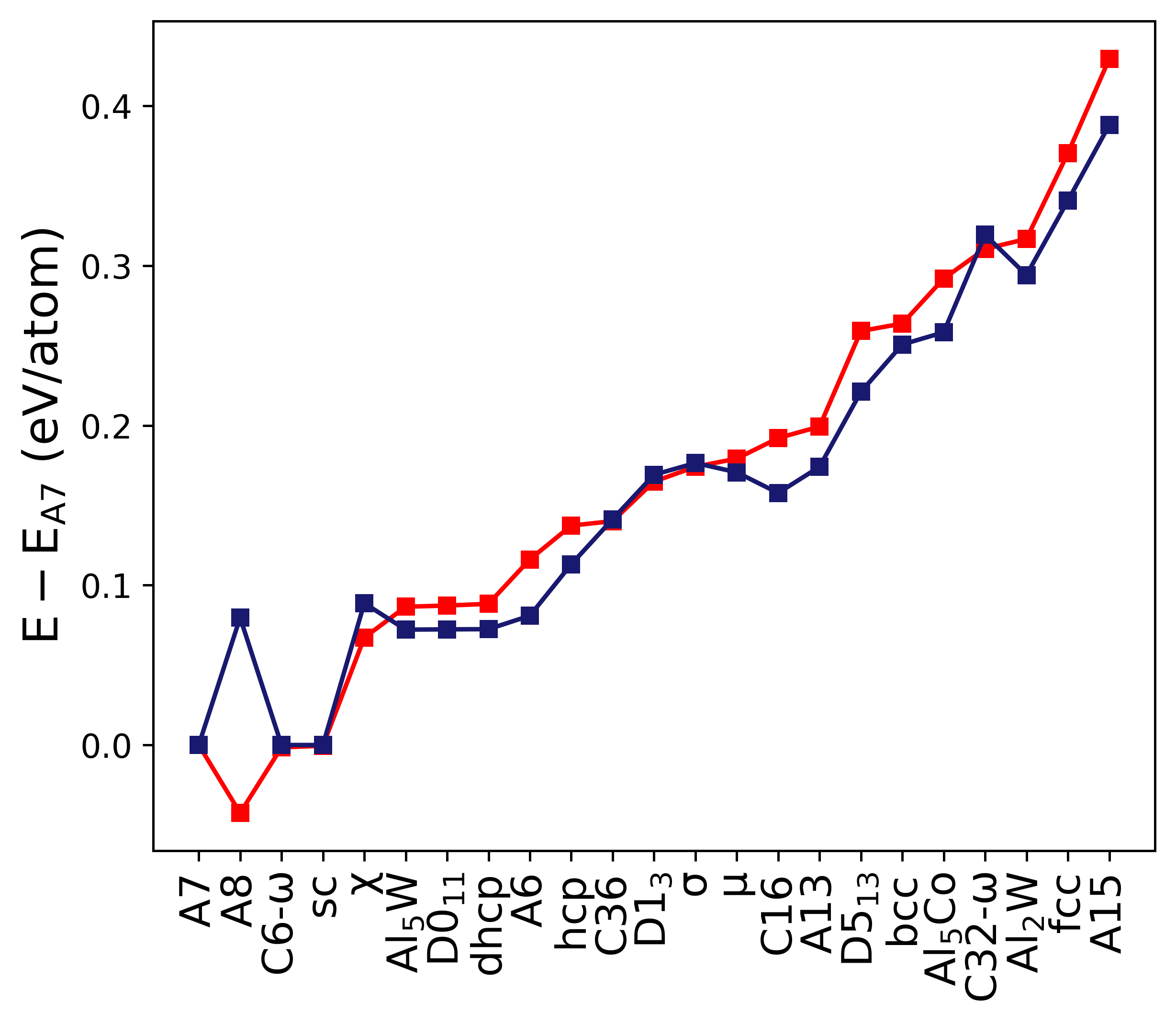}
\caption{Te (Equilibrium energies)}
\label{fig:Te_E}
\end{subfigure}\qquad
\begin{subfigure}[b]{0.7\columnwidth}
\includegraphics[width=\textwidth]{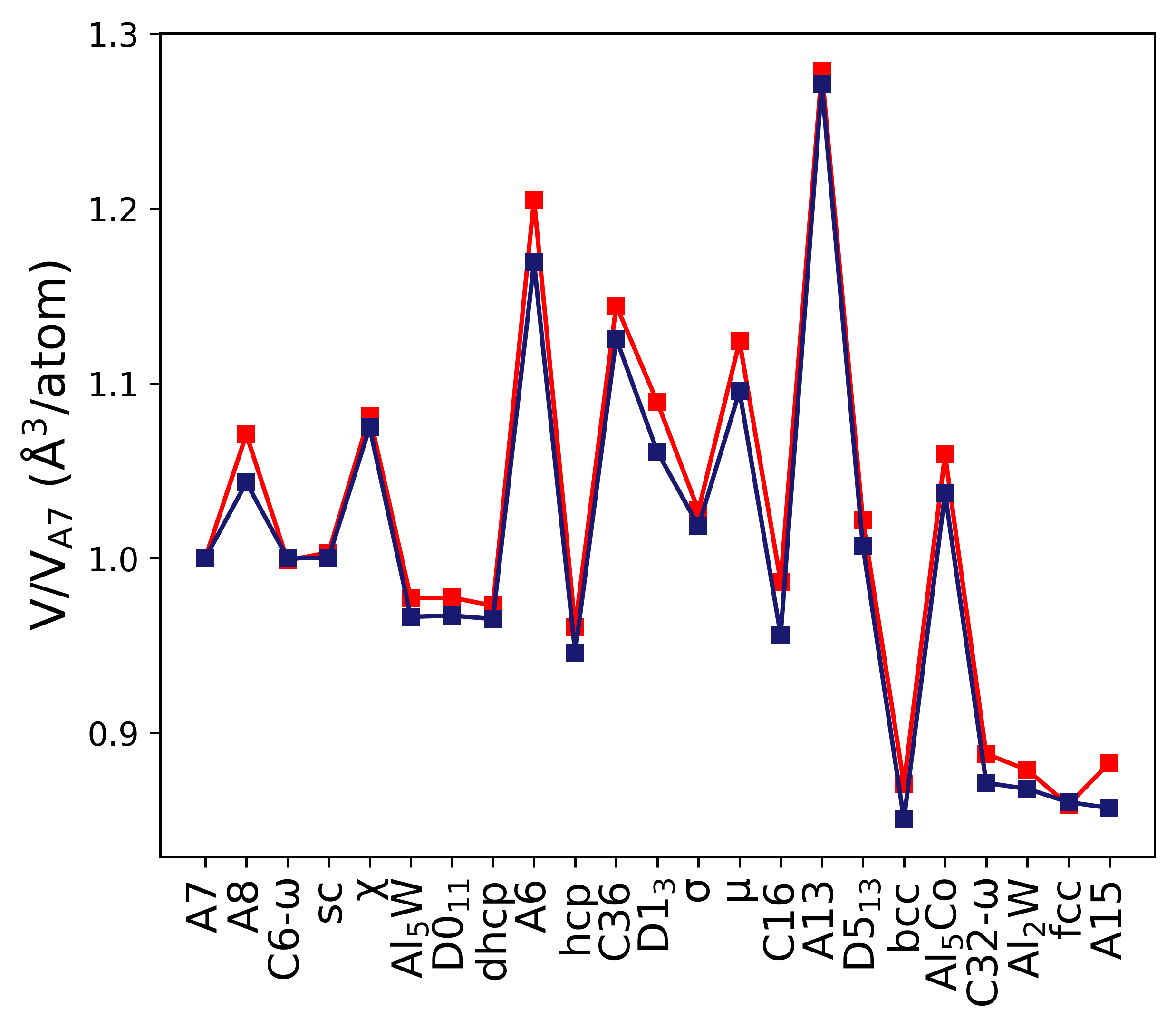}
\caption{Te (Equilibrium volumes)}
\label{fig:Te_V}
\end{subfigure}\qquad
\begin{subfigure}[b]{0.7\columnwidth}
\includegraphics[width=\textwidth]{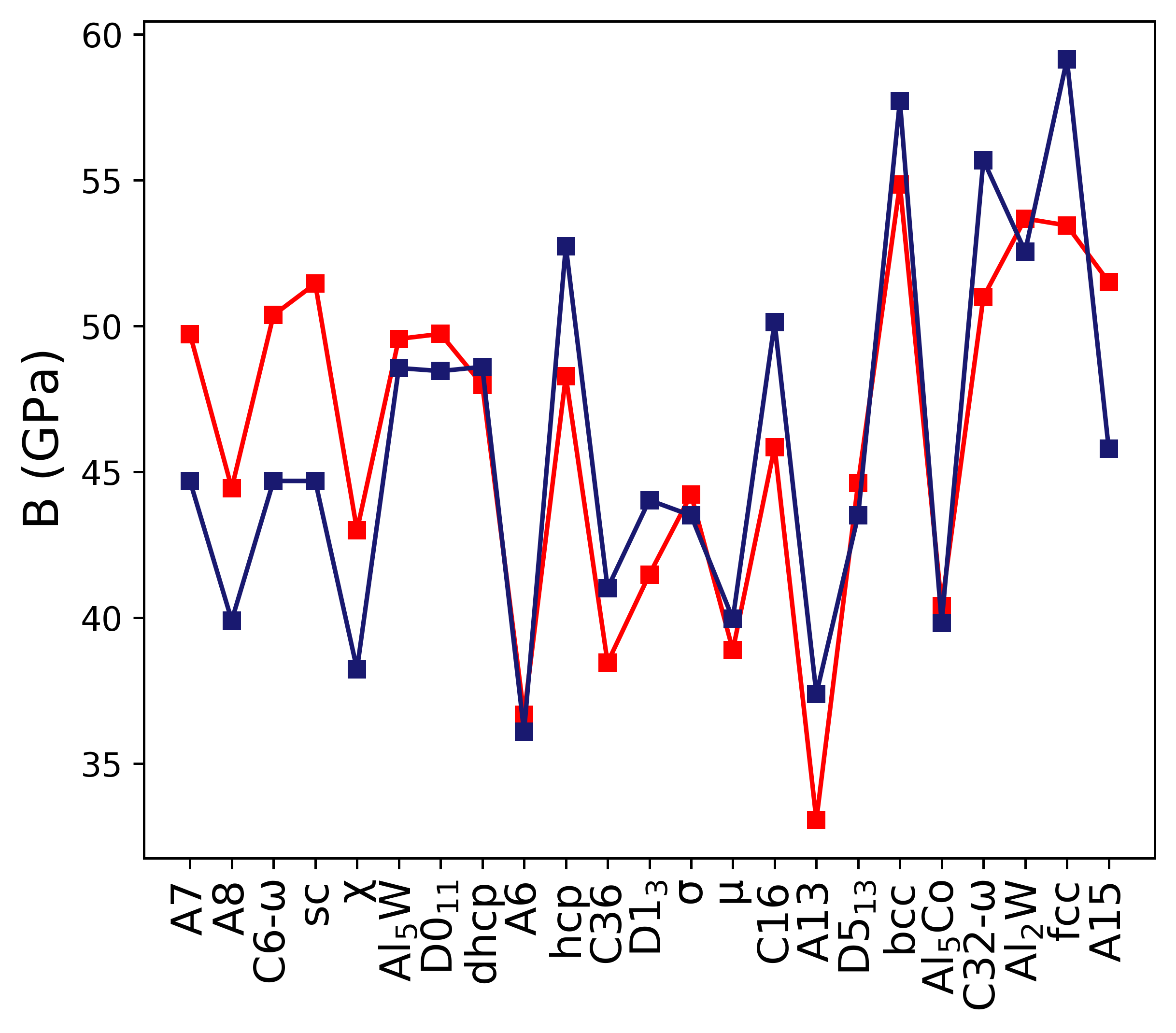}
\caption{Te (Bulk moduli)}
\label{fig:Te_B}
\end{subfigure}\qquad
\caption{Performance of the Rev-EM ACE model (blue) on prototypical crystal structures for C, Sb and Te.} 
\label{fig:Prototypes_C}
\end{figure*}

In comparison to the other elements discussed above, C is unique due to its open ground state graphite structure which is sp$^2$ hybridized, leading to large volumes per atom in spite of short first neighbor distances. The range of densities employed to generate Rev-EM, which was originally tuned for close-packed materials, proved too narrow to adequately capture the ground state structure. Rev-EM was therefore enriched with additional entropy-maximized configurations generated at lower densities, as described in Sec.\ \ref{sec:rescaling}. 

As shown in Tab.\ \ref{tab:ACE_New_to_DE_TM23}, the model for C shows higher RMSEs on the test dataset compared to the rest of the elements, reflecting the unique and complex chemistry of C. Nonetheless, this model accurately predicts the energetics, equilibrium volumes, and bulk moduli of 28 different crystal structures including BC8, $\beta$ Sn (A5), BCT5, hexagonal diamond (hex-dia) and fullerene (C60), as shown in Fig.\ \ref{fig:Prototypes_C}. In addition to the prototypes, Table \ref{tab:C_defects_ACE-RevEM_DFT} reports the performance of the model on the (111), (100), and (110) diamond (A4) surfaces. ASE \cite{ASE_Larsen_2017} was used to construct the (111) and the (100) surfaces, while the (110) reference surface structure is from Ref.\cite{Tran2016}. The formation energies are computed by allowing only the relaxation of atomic positions until forces on all atoms are less 0.01 eV/atom. No surface reconstructions were taken into account. We again emphasize that none of these structures were explicitly added to the training set, but that all structures were generated through the entropy maximization procedure. These encouraging results suggest that Rev-EM contains not only information regarding high energy structures, but also on thermodynamically-relevant low-energy configurations for a broad range of different materials.
\begin{table}[!h]
\resizebox{0.6\columnwidth}{!}{%
\begin{tabular}{c | c | c}
\hline\hline
& DFT & ACE \\
\hline
& (mJ/m$^2$) & (mJ/m$^2$) \\
\hline
111 & 13.0 & 12.54 \\
110 & 5.84 & 5.08 \\
100 & 9.09 & 8.9 \\
\hline\hline
\end{tabular}}
\caption{Surface formation energies in diamond.}
\label{tab:C_defects_ACE-RevEM_DFT} 
\end{table}

To demonstrate the dynamical stability of the C ACE model, MD simulations were performed in the NVT ensemble for 45 structures at temperatures ranging from 500 K to 10,500 K in steps of 1000 K. The structures were randomly selected from the Rev-EM dataset over densities in the range of 1.5-4.5 g/cc. Supercells of these structures were generated, resulting in cells containing anywhere between 400 to 900 atoms depending on the density. Each simulation was run successfully for a total of 150 ps with a time step of 0.25 fs, corresponding to a total of 67.5 ns of aggregate simulation time. All MD simulations were performed using the LAMMPS software package \cite{LAMMPS}.
The configurations towards the end of the simulation were analyzed by computing their coordination numbers. At 3500 K, sp$^2$ coordinated atoms make up more than 85\% of the atoms for the low and intermediate density (1.5 to 3.0 g/cc) configurations. For configurations with densities in the range of 3.0-3.3 g/cc, a mixture of sp$^2$ and sp$^3$ hybridized atoms were found with the fraction of sp$^3$ atoms being 50-70\%. At densities close to 3.5 g/cc, sp$^3$ hybridized atoms were found to constitute more than 80\% of the atoms. As the volumes of the simulation cells are held fixed, configurations with densities above 4.0 g/cc result in pressures greater than 200 GPa. At these densities and pressures, the fraction of sp$^3$ coordinated atoms is greater than 95\% even at 4500 K. Snapshots extracted from three simulations generated at different densities are reported in the Supplementary Material (see Fig.\ \textcolor{blue}{S6}). This demonstrates that the potential naturally produces very stable trajectories even in extreme conditions of temperature and pressure.

In addition, quench simulations were carried out as described in Refs. \cite{Qamar2023, Jana_2019}. 2000 atoms were randomly placed in a simple cubic cell at densities ranging from 1.5-3.5 g/cc. These supercells were melted at 10,000 K in the NVT ensemble and held for 10 ps, before quenching to 300 K by linearly decreasing the temperature at a rate of 100 K/ps, before holding the temperature constant for a further 13 ps, after which configurations were analyzed. The obtained fraction of sp$^3$ bonds are in very good agreement with the DFT results \cite{Jana_2019}, as shown in the Supplementary Material (see Fig.\ \textcolor{blue}{S7}). Both the above simulations suggest that, the current C ACE model, accurately describes the governing crystal structures at all densities.

\subsection{Comparison with existing datasets}
\label{sec:comparison_MLIAPs}

\begin{table*}[ht]
\centering
\begin{tabular}{ | c | c | c | c | c | c | }
\hline
& \diagbox{MLIAP}{Dataset} & ANI \cite{ANI_Al} & GAP \cite{fellman2024fastaccuratemachinelearnedinteratomic} & UNEP \cite{Song2024} & Rev-EM \\
\hline
\multirow{4}{*}{Al} & ANI \cite{ANI_Al} & 2.57/60.28 & 233.89/143.95 & 209.64/55.29 & 73.34/77.99 \\ 
& GAP \cite{fellman2024fastaccuratemachinelearnedinteratomic} & 76.11/307.37 & 0.76/31.32 & 92.69/96.3 & 155.76/263.75 \\
& UNEP \cite{Song2024} & 13.88/156.29 & 114.79/88.29 & 19.62/69.0 & 114.88/111.49 \\
& ACE-2 & 8.8/94.42 & 56.93/64.75 & 17.09/53.94 & 9.56/44.23\\
\hline\hline
& \diagbox{MLIAP}{Dataset} & DP \cite{Ding2024} & GAP \cite{PhysRevB.100.144105} & UNEP \cite{Song2024} & Rev-EM \\
\hline
\multirow{4}{*}{W} & DP \cite{Ding2024} & - & 236.41/410.75 & 347.23/446.5 & 1280.09/2135.85 \\ 
& GAP \cite{PhysRevB.100.144105} & - & 1.22/180.49 & 134.22/241.2 & 417.76/647.59 \\
& UNEP \cite{Song2024} & - & 61.74/370.52 & 27.77/246.57 & 520.85/795.87 \\
& ACE-2 & - & 70.87/252.13 & 44.37/263.54 & 60.95/433.6\\
\hline
\end{tabular}
\caption{Performance of several MLIAPs for Al and W on existing datasets and on the Rev-EM dataset. The RMSE values are shown as energy/force in units (meV/atom)/(meV/\AA~). Note that the entirety of the respective datasets were used to obtain these errors.}
\label{tab:Al_W_MLIAP_Rev-EM_Cross_Validation} 
\end{table*}

In order to compare the relative robustness of the potentials obtained from the the Rev-EM dataset, we tested our ACE models for Al, W, and C on multiple datasets available from literature. Configurations with energies above 10 eV/atom from the minimum energy structure and force magnitudes above 20 eV/\AA~have been removed from all the datasets for Al. Similarly, configurations with energies above 20 eV/atom from the minimum energy structure and force magnitudes above 100 eV/\AA~have been removed from all the datasets for W. The ANI \cite{ANI_Al}, GAP \cite{fellman2024fastaccuratemachinelearnedinteratomic} and Unified Neuroevolution Potential (UNEP) \cite{Song2024} models for Al were also cross-validated against all available datasets. For W, deep potential (DP) \cite{Ding2024}, GAP \cite{PhysRevB.100.144105} and UNEP \cite{Song2024} models were again tested on the entire Rev-EM dataset and cross-validated on their respected training datasets when available. The results for Al and W are shown in Tab.~\ref{tab:Al_W_MLIAP_Rev-EM_Cross_Validation}. Note that the entire datasets were used to compute the errors, which might include both training and test data when assessing a model on its own dataset. The reported values for the ACE-2 models hence differ slightly from the values in Tabs. \ref{tab:ACE_New_to_Old_DE_TM23} and \ref{tab:ACE_New_to_DE_TM23}, where test errors are reported. For the same reason, the errors for MLIAPs tested on their own datasets may also differ from those reported in the original references.

The ACE-2 model for Al performs uniformly well on the ANI, GAP and UNEP datasets, as the errors are generally comparable to the training error on the Rev-EM dataset. On the other hand, the ANI, GAP, and UNEP models have very low test/train RMSEs when assessed on their own datasets, but cross-validation to other datasets shows much higher errors, indicating that different datasets likely contain mutually exclusive regions of configuration space. In all cases, the ACE-2 model trained to Rev-EM comes in second in terms of accuracy, only behind the matched models/dataset combinations. This indicates that the Rev-EM dataset produces models that are significantly more robust than alternative approaches.
A broadly similar behavior is observed for W, where most potentials do significantly worse when tested on datasets other than the ones they were trained on, and especially when tested on the Rev-EM dataset, where almost 10-fold higher energy errors are observed when comparing other models to ACE-2. In contrast, the ACE-2 model trained to Rev-EM perform similarly when tested on different datasets, suggesting that Rev-EM is effectively a super-set of these.

\begin{table*}[ht]
\centering
\resizebox{\linewidth}{!}{
\begin{tabular}{| c | c | c | c | c | c | c |}
\hline
\diagbox{MLIAP}{Dataset} & GAP-20 \cite{10.1063/5.0005084} & GAP \cite{Bernstein2019} & SNAP \cite{PhysRevB.106.L180101} & ANI-1xnr \cite{Zhang2024} & ACE \cite{Qamar2023} & Rev-EM \\
\hline
GAP-20U \cite{10.1063/5.0091698} & 47.92/777.63 & 140.28/1113.17 & - & - & 809.48/1232.56 & 1042.11/2846.62 \\ 
GAP \cite{Bernstein2019} & 342.09/1200.09 & 95.93/589.93 & - & - & 680.93/1440.19 & 687.36/2026.89 \\
SNAP \cite{PhysRevB.106.L180101} & 318.90/1659.66 & 631.15/1152.23 & - & - & 2008.34/2115.11 & 1907.04/2322.9 \\ 
ANI-1xnr \cite{Zhang2024} & 360.34/794.76 & 243.35/679.56 & - & - & 725.98/1037.11 & 1311.81/1523.53 \\
ACE \cite{Qamar2023} & 70.11/554.07 & 68.58/772.70 & - & - & 86.67/655.43 & 675.04/5728.02 \\
ACE (this work) & 160.99/773.02 & 104.89/582.91 & - & - & 214.20/924.11 & 105.27/948.75\\
\hline
\end{tabular}
}
\caption{Performance of several MLIAPs for C on existing datasets and on the Rev-EM dataset. The RMSE values are shown as energy/force in units (meV/atom)/(meV/\AA~). Note that the entirety of the respective datasets were used to obtain these errors.}
\label{tab:C_MLIAP_Rev-EM_Cross_Validation} 
\end{table*}

Finally, Table~\ref{tab:C_MLIAP_Rev-EM_Cross_Validation} reports the performance of various MLIAPs for C: ACE \cite{Qamar2023}, SNAP \cite{PhysRevB.106.L180101}, GAP \cite{10.1063/5.0005084, 10.1063/5.0091698, Bernstein2019}, and the ANI-1xnr \cite{Zhang2024}, when tested on the Rev-EM dataset and cross-validated against the other datasets when available in the literature. In this case, configurations with energies above 10 eV/atom from the minimum energy structure and magnitude of forces above 40 eV/\AA~ have been removed from all the datasets. The energy and force RMSE for ACE (this work) is different from the values in Tab.~\ref{tab:ACE_New_to_DE_TM23} because the upper limit of force magnitude contained in the (test-only Rev-EM) dataset is increased to 40 eV/\AA~, while the training dataset had a maximum force magnitude of 25 eV/\AA~. Once again, irrespective of the underlying ML architecture and training sets, none of the MLIAPs show good transferability to the Rev-EM dataset, while the ACE model from this work shows comparable performance on all datasets.

As the cross-validation of MLIAPs on different datasets from literature is based on as-published DFT data, this includes inconsistencies due not only to different DFT settings (functionals/convergence parameters) but also due to different software packages used to generate their training datasets. To partially account for this this, we compute the RMSEs between $\Delta \mathrm{E}^\mathrm{DFT}$ and $\Delta \mathrm{E}^\mathrm{MLIAP}$ for all MLIAPs when cross-validated on other datasets. The $\Delta \mathrm{E}$ values are the energy differences from the respective minimum energy structure. However, absolute values are considered when comparing against matching MLIAPs/datasets. Forces are not affected by this correction.

\section{Discussion}\label{sec:Discussion}

The results presented above demonstrate the the problem of developing truly robust MLIAPs that are transferable across a broad range of phases and structures remains an outstanding challenge even for "simple" unary systems. Indeed, many potentials previously developed by the community, sometimes over many years of iterative improvements, do show very high, "near quantum", accuracy for configurations similar to those in the training set but significantly higher errors (sometimes by a factor 10 to 100) on novel configurations. While problems with very well defined structural spaces benefit from the very high "local" accuracy provided by conventional approaches, many other problems where it is difficult to {\em a priori} delineate the relevant structural space, e.g., for systems under extreme conditions, can be expected to strongly benefit from approaches such as the ones presented here, even at the cost of a modest decrease in local accuracy.\\
Indeed, the extreme diversity of Rev-EM makes it particularly challenging to capture using MLIAPs, especially when combined with a complex chemistry as in the case of C. As discussed above, a thorough exploration of the hyperparameter space of the MLIAPs (cutoff radius, energy reweighting, choice of the type of basis functions, etc.) is likely to significantly improve the accuracy of the ACE models presented here. In addition, simply increasing the complexity of the MLIAPs is also a promising avenue, although it incurs a corresponding increase in computational cost. For example, Qamar {\em et al.} \cite{Qamar2023} demonstrated that increasing the complexity of ACE models results in a steady decrease of the observed errors for models of C. Repeating their experiment by increasing the number of basis functions from 387 to 488, causing an increase in adjustable parameters from 1094 to 1816, led to a decrease in test energy RMSE from 101.35 meV/atom to 84.12 meV/atom and of force RMSE from 750.67 meV/\AA~ to 645.76 meV/\AA~. As no error saturation was observed in Ref.\ \cite{Qamar2023} in this range of complexity, significant further reductions in errors can be expected through a continued increase in complexity of the potentials. It therefore appears possible to largely achieve both high accuracy and broad robustness, albeit at an increased computational cost.

\section{Conclusion}\label{sec:Conclusion}
This study demonstrates that the curation of training datasets for MLIAPs can be recast as an optimization problem where the objective function is the information entropy of the features contained in the dataset. This reformulation of the problem enables the simple, systematic, and autonomous generation of ultra-diverse datasets that distil a very high amount of information in a relatively small number of small configurations. Extensive validation demonstrates that ACE potentials trained to Rev-EM are extremely robust up to large values of energies and forces, while capturing the properties of interest that are usually manually targeted during training set generation. Rev-EM is also shown to be transferable to a diverse set of other unary materials with different thermodynamics, including Be, C, Al, Sb, Te, W, Re, and Os. While capturing the extended feature space sampled by Rev-EM is extremely challenging for MLIAPs, reweighting strategies, hyperparameter optimization, and increase in complexity of the MLIAPs offer multiple venues to achieve the best of both worlds where the tradeoff between accuracy and transferability is minimized.
The initial indications provided here are therefore a promising step towards eventually meeting the ultimate objective of uniform near-quantum accuracy over an extremely broad span of conditions.

\section{Methods}\label{sec:Methods}
\subsection{Implementation details}
The results reported above were obtained using the first 55 bispectrum components (corresponding to $2J_{\mathrm{max}}=8$), which form the basis of the GAP \cite{PhysRevLett.104.136403} and SNAP \cite{THOMPSON2015316} MLIAPs. 
The MLIAPPY package in LAMMPS \cite{LAMMPS} is used to compute the features, their gradient with respect to atomic positions, and the repulsion potential $V_\mathrm{core}$, which is given by $V_{\mathrm{core}}(r) = (1+\cos(\frac{\pi r}{R_{\mathrm{core}}}))$. The function $F$ from Eq.\ \ref{eq:objective} and its gradient are computed in Python using the automatic differentiation framework JAX \cite{jax2018github}. The ASE package \cite{ASE_Larsen_2017} has been used to generate the random initial configurations. 

As discussed in Sec.\ \ref{sec:diversity}, the bispectrum components exhibit a strong density dependence. This increases the probability that a high-density candidate configuration would increase the information entropy of the dataset. In order to avoid introducing a bias, independent instances of the greedy optimization loop described in Fig. \ref{fig:workflow} are carried out in windows associated with different core exclusion radii $R_{\mathrm{core}}$. Each candidate configuration is randomly initialized with atoms between 2 and 25 at a volume that is randomly sampled within a window of volume/atom [$V_{\mathrm{start}}$, $V_{\mathrm{end}}$] and with a randomized triclinic shape. 

The volumes per atom in candidate configurations are sampled from 17 windows that span 0.45 to 3 times the equilibrium volume per atom in the ground state. For each window, the core exclusion radius $R_{\mathrm{core}}$ is set so that the exclusion volume scales with $V_{\mathrm{start}}$. 
The minimum distance allowed between atoms $R_{\mathrm{min}}$ is 90\% of the $R_{\mathrm{core}}$. Configurations that do not meet this condition are rejected. 

The parameter $K$ in Eq. \ref{eq:objective} that controls the relative importance of $V_{\mathrm{core}}$ and the entropy term is empirically tuned on-the-fly. Within the optimization loop, repeated failures of candidates $\mathcal{C}_k$ to increase the entropy $S(\mathcal{S}_k \cup \mathcal{C}_k)$ lead to a small increase of $K$, while rejections of candidates $\mathcal{C}_k$ due to violations of the minimum distance criterion cause a decrease of $K$. At every instance of volume windows, the model is boostrapped with 10 randomly generated configurations to initialize the covariance matrix and run for 25,000 iterations, both for the per-atom mode and the per-configuration mode. 

\subsection{Datasets}
\label{sec:datasets}
Rev-EM is compared to 7 other datasets available from the literature, namely Poul-Mg \cite{PhysRevB.107.104103}, Bernstein-Ti \cite{Bernstein2019}, Smith-Al \cite{ANI_Al}, Lysogorskiy-Cu \cite{PACE_implementation}, Song-W \cite{Song2024}, Owen-W \cite{Owen2024} and Karabin-W, a subset of 20,000 configuration from Ref.\ \cite{10.1063/5.0013059}. In Poul-Mg, a dataset for Mg was created by generating random crystal structures belonging to a broad range of spacegroups using the RandSPG program \cite{AVERY2017208}. These structures were further subjected to random atomic displacements and  volumetric and shear strains. In Bernstein-Ti, the authors generate an initial dataset using the ab-inito random structure search (AIRSS) method \cite{Pickard_2011} and use CUR decomposition \cite{mahoney2009cur} to sub-select diverse structures based on similarity index of configuration averaged descriptors. In Smith-Al, the dataset for Al is generated by improving the initial dataset, created randomly, by using an active learning strategy driven by molecular dynamics.  
In Lysogorskiy-Cu, the training dataset consists of expert-curated clusters, prototypical and defected structures. These structures were then subjected to random atomic displacements and high strains. In Song-W, the training dataset was constructed by including bulk, defected structures and their randomly perturbed counterparts, as well as structures from MD simulations carried out using an EAM potential at various temperatures and pressures.
This initial dataset was then augmented by active learning. The Owen-W dataset was generated by performing ab-initio molecular dynamics simulations on a supercell constructed from the ground state crystal structure in which a vacancy was introduced. The Karabin-W dataset has been generated by maximizing the entropy of the feature distribution of atoms in individual configurations. A subset of 20,000 configurations have been selected from the $\sim$ 200,000 total configurations. 

\subsection{Rescaling Rev-EM}\label{sec:rescaling}
A simple volumetric rescaling is performed on Rev-EM to generate datasets for Be, Al, W, Re and Os based on their respective equilibrium volumes obtained from the Materials Project database \cite{10.1063/1.4812323}. The ground state crystal structure of C is graphite, which exhibits short intra-planar distances (1.4 \AA), but large Voronoi volumes due to large inter-planar distances (3.2 \AA). These configurations were not adequately sampled using the density range deemed adequate for close-packed materials. This was remedied by generating additional optimization cycles with smaller $R_{\mathrm{core}}$ and larger volumes, producing $\sim$11,000 structures, which were combined with $\sim$15,000 structures sampled from the original Rev-EM. This results in a C dataset consisting of 26,269 configurations and a total of 175,935 atoms.
\begin{figure*}[h]
\begin{subfigure}[b]{\columnwidth}
\includegraphics[width=\textwidth]{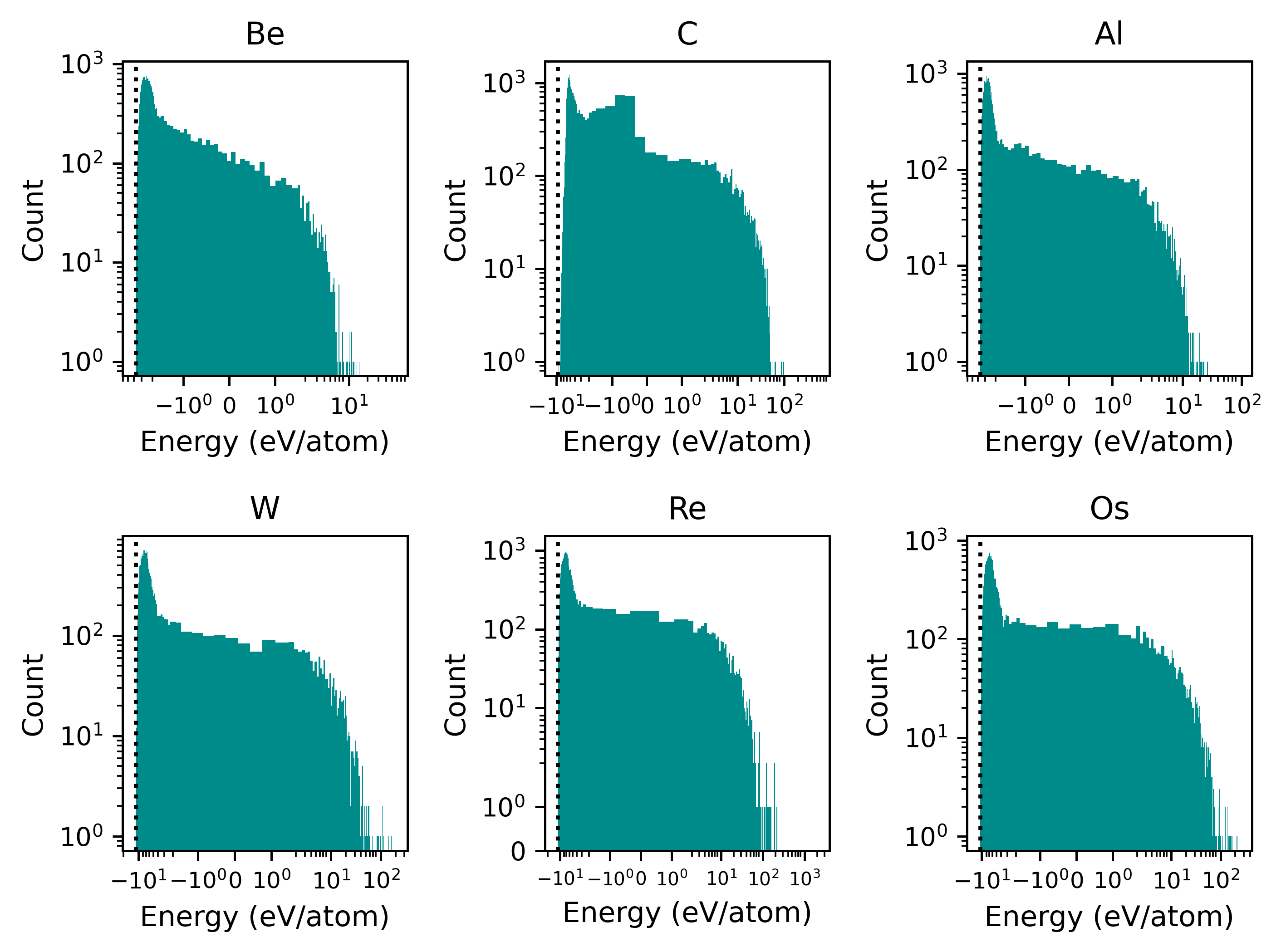}
\caption{Energy distribution}
\label{fig:energies_elements}
\end{subfigure}\qquad
\begin{subfigure}[b]{\columnwidth}
\includegraphics[width=\textwidth]{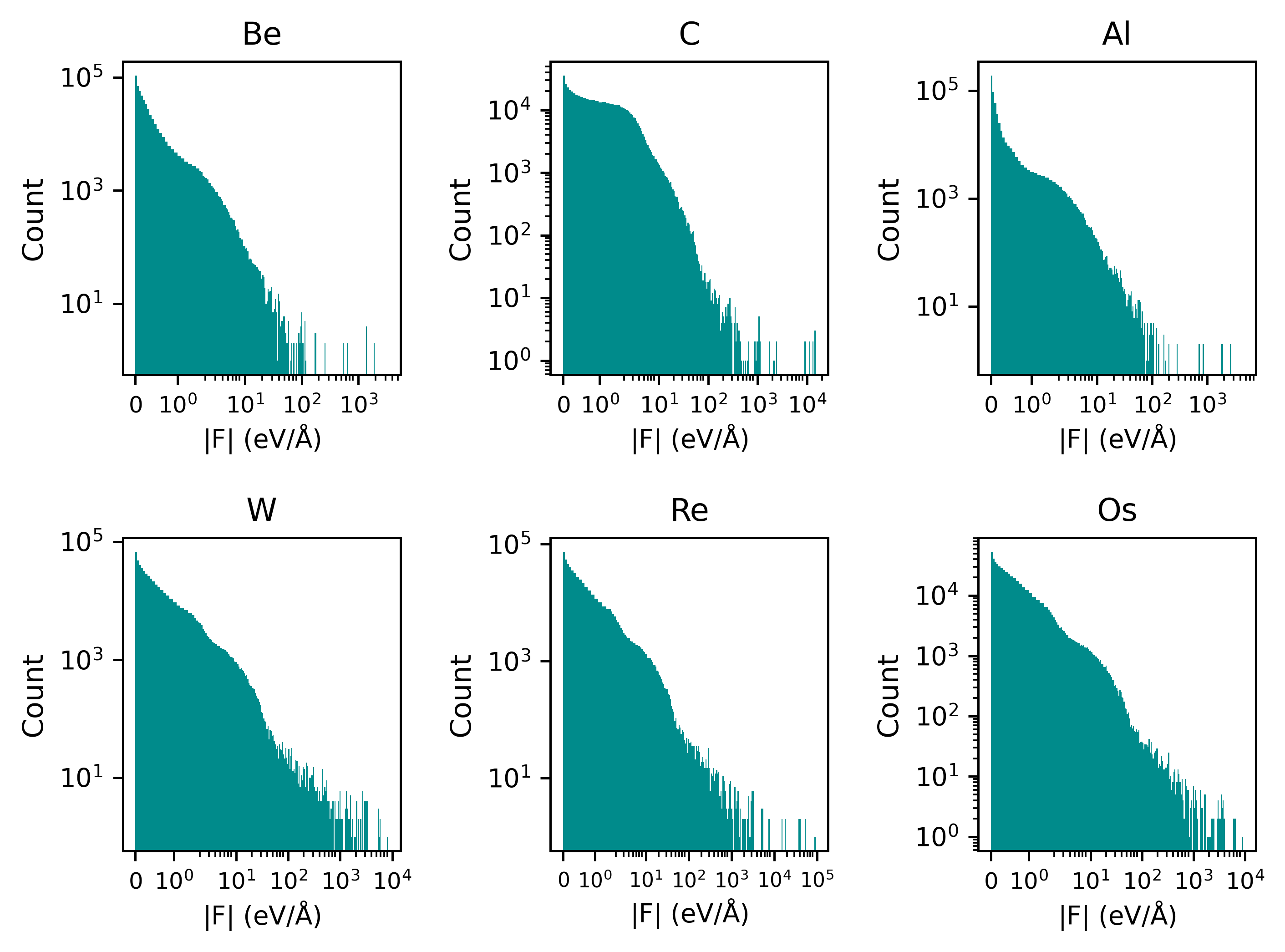}
\caption{Force distribution}
\label{fig:forces_elements}
\end{subfigure}\qquad
\caption{The energy (left) and force (right) distributions for different elements Be, C, Al, W, Re and Os. The black dotted vertical line shows the equilibrium energy of the ground state crystal structure of the respective element.}
\label{fig:rescale_datasets}
\end{figure*}

The energy and force distributions for the different instantiations of Rev-EM  are shown in Fig.\ \ref{fig:rescale_datasets}. Similar to the case of W, the distributions are extremely broad (almost 100 eV/atom in energy and $10^4$ eV/\AA~ in forces), but still concentrated at relatively low values. 

\subsection{DFT calculations}\label{sec:DFT}
Non-spin-polarized density-functional theory (DFT) \cite{PhysRev.140.A1133} calculations for the entire Rev-EM dataset were performed for Be, C, Al, W, Re, and Os using the VASP (version 6.3.2) software package \cite{KRESSE199615, PhysRevB.54.11169, PhysRevB.59.1758}. The projector augmented-wave (PAW) \cite{PhysRevB.50.17953} method was used with the generalized gradient approximation \cite{PhysRevLett.77.3865}. A plane-wave cutoff energy of 700 eV and spacing between K-points of 0.125 \AA$^{-1}$ were used for all elements. Methfessel-Paxton smearing with a width of 0.1 eV was applied for all elements except for C, where Gaussian smearing with a width of 0.05 eV was applied. 
The atoms were not allowed to relax and an electronic self-consistency energy convergence of 10$^{-6}$ eV is used. Only a very small number of calculations failed to converged, resulting in 30,348 configurations for Be and Al, 30,347 configurations for W and Os, 30,341 configurations for Re and all 26,269 configurations for C. 

The elastic constants for the ground state BCC structure for W and the energy-volume curves for all elements were calculated using the pyiron framework \cite{JANSSEN201924}. The equilibrium volume, energy, and bulk modulus were obtained by fitting a Murnaghan equation of state \cite{doi:10.1073/pnas.30.9.244} as implemented in ASE to the structures within $\pm$ 15\% of the volume corresponding to the minimum energy structure. Long range Van der Waals interactions were accounted for C for only the prototypical structures shown in Fig.\ \ref{fig:Prototypes_C} by considering the D2 \cite{D2_Grimme} corrections. These dispersion corrections were tabulated for use with LAMMPS and made available by Ref.\ \cite{Qamar2023}, which were then used with the ACE model for C. Identical corrections are applied to both the ACE models and the DFT.

\subsection{ACE parameterization}
\label{sec:ace}
The {\em pacemaker} \cite{bochkarev2022efficient, PhysRevB.99.014104, PACE_implementation} code was used for parameterization of nonlinear ACE \cite{PhysRevB.99.014104} models for the Be, C, Al, Sb, Te, W, Re and Os systems. An identical setup is used for all materials, without any explicit hyperparameter tuning. One of the material specific parameters is the cut-off radius which was set to 4.5~\AA~ for Be, 5.5~\AA~ for Al/C, 6.5~\AA~ for Sb, 7.5~\AA~ for Te and 6.0~\AA~ for W/Re/Os. Another describes the parameterization of the short-range repulsion, given for all elements in the Supplementary Materials. The parameters of the short-range core repulsion (see Ref.\ \cite{bochkarev2022efficient}) have been estimated empirically without optimization.

Finally, configurations that have energies (relative to the minimum energy in the set) and (absolute) forces above a certain threshold are removed from the reference datasets (both for training and testing). The energy thresholds are 10 eV/atom for Be/C/Al, 20 eV/atom for W/Re and 40 eV/atom for Os. The force thresholds are 20 eV/\AA~ for Be/Al, 25 eV/\AA~ for C and 100 eV/\AA~ for W/Re/Os. A subset of 10,000 structures were randomly sampled from the Rev-EM dataset out of which 8,000 configurations are used for training and the remaining 20\% were used for testing during optimization of the ACE models. The resulting number of structures and atomic environments present in the training and testing datasets for all elements are given in Tab.\ \ref{tab:Test_train_datasets}. 
\begin{table}[!h]
\resizebox{\columnwidth}{!}{%
\begin{tabular}{c | c | c }
\hline\hline
Element (Model) & Training & Test \\
\hline
& configurations/atoms & configurations/atoms \\
\hline
Be & 7,956/46,345 & 22,230/131,082 \\
C  & 7,023/46,842 & 15,943/107,435 \\
Al & 7,893/46,355 & 22,074/130,135 \\
Sb & 7,898/46,079 & 21,854/124,804 \\
Te & 7,922/46,070 & 22,178/131,100 \\ 
W  & 7,808/46,106 & 21,878/129,511 \\
Re & 7,692/45,231 & 21,589/128,297 \\
Os & 7,862/45,952 & 22,020/130,252 \\
\hline\hline
\end{tabular}}
\caption{Total number of configurations/atoms in the training and test datasets.}
\label{tab:Test_train_datasets}
\end{table}

For all elements, 10\% of the total weight is given to forces and 90\% of the total weight is given to  energies, with 75\% of weight being distributed to configurations that are less than $\Delta E$ from the minimum energy $E_{\mathrm{min}}$ in the training dataset. We used the energy-based weighting policy implemented in the {\em pacemaker} code and explained in detail in Ref.\ \cite{bochkarev2022efficient}. $\Delta E$ is 2 eV/atom for Be, Al, W, Re, Os and is 3 eV/atom for C. As explained in Sec.\ \ref{sec:tradeoff}, $\Delta E$ is tuned to explore the accuracy/robustness tradeoff for W. The normalized weights based on the energy differences to the minimum energy structure are shown in Fig.\ \textcolor{blue}{S4} for the three ACE models for W. 

The ACE models for all elements consist of 387 Bessel radial basis functions combined up to the sixth body-order, resulting in 1,094 adjustable parameters.

\backmatter

\bmhead{Data availability}
The datasets for all elements will be made available upon publication.

\bmhead{Author contributions}
AS refined the production code, generated the datasets, trained the potentials, and analyzed the results. DP developed the research idea, secured funding, and wrote the initial version of the entropy maximization code. Both authors wrote the manuscript.

\bmhead{Acknowledgments}
We acknowledge support by the Exascale Computing Project (17-SC-20-SC), a collaborative effort of the U.S. Department of Energy Office of Science and the National Nuclear Security Administration. This research used computing resources provided by the Laboratory Institutional Computing Program and by the Darwin testbed at Los Alamos National Laboratory (LANL) which is funded by the Computational Systems and Software Environments subprogram of LANL's Advanced Simulation and Computing program (NNSA/DOE). Los Alamos National Laboratory is operated by Triad National Security LLC, for the National Nuclear Security administration of the U.S. DOE under Contract No. 89233218CNA0000001.

\bmhead{Competing interests}
The authors declare no financial or non-financial competing interests.


\begin{thebibliography}{10}
\expandafter\ifx\csname url\endcsname\relax
  \def\url#1{\burl{#1}}\fi
\expandafter\ifx\csname urlprefix\endcsname\relax\def\urlprefix{URL }\fi
\providecommand{\bibinfo}[2]{#2}
\providecommand{\eprint}[2][]{\url{#2}}
\providecommand{\doi}[1]{\url{https://doi.org/#1}}
\bibcommenthead

\bibitem{PhysRevMaterials.8.013803}
\bibinfo{author}{Subramanyam, A. P.~A.}, \bibinfo{author}{Jenke, J.},
  \bibinfo{author}{Ladines, A.~N.}, \bibinfo{author}{Drautz, R.} \&
  \bibinfo{author}{Hammerschmidt, T.}
\newblock \bibinfo{title}{Parametrization protocol and refinement strategies
  for accurate and transferable analytic bond-order potentials: Application to
  {R}e}.
\newblock \emph{\bibinfo{journal}{Phys. Rev. Mater.}}
  \textbf{\bibinfo{volume}{8}}, \bibinfo{pages}{013803} (\bibinfo{year}{2024}).

\bibitem{PhysRevLett.98.146401}
\bibinfo{author}{Behler, J.} \& \bibinfo{author}{Parrinello, M.}
\newblock \bibinfo{title}{Generalized neural-network representation of
  high-dimensional potential-energy surfaces}.
\newblock \emph{\bibinfo{journal}{Phys. Rev. Lett.}}
  \textbf{\bibinfo{volume}{98}}, \bibinfo{pages}{146401}
  (\bibinfo{year}{2007}).

\bibitem{10.1063/1.4712397}
\bibinfo{author}{Jose, K. V.~J.}, \bibinfo{author}{Artrith, N.} \&
  \bibinfo{author}{Behler, J.}
\newblock \bibinfo{title}{{Construction of high-dimensional neural network
  potentials using environment-dependent atom pairs}}.
\newblock \emph{\bibinfo{journal}{The Journal of Chemical Physics}}
  \textbf{\bibinfo{volume}{136}}, \bibinfo{pages}{194111}
  (\bibinfo{year}{2012}).

\bibitem{10.1063/1.3553717}
\bibinfo{author}{Behler, J.}
\newblock \bibinfo{title}{{Atom-centered symmetry functions for constructing
  high-dimensional neural network potentials}}.
\newblock \emph{\bibinfo{journal}{The Journal of Chemical Physics}}
  \textbf{\bibinfo{volume}{134}}, \bibinfo{pages}{074106}
  (\bibinfo{year}{2011}).

\bibitem{PhysRevB.87.184115}
\bibinfo{author}{Bart\'ok, A.~P.}, \bibinfo{author}{Kondor, R.} \&
  \bibinfo{author}{Cs\'anyi, G.}
\newblock \bibinfo{title}{On representing chemical environments}.
\newblock \emph{\bibinfo{journal}{Phys. Rev. B}} \textbf{\bibinfo{volume}{87}},
  \bibinfo{pages}{184115} (\bibinfo{year}{2013}).

\bibitem{PhysRevLett.104.136403}
\bibinfo{author}{Bart\'ok, A.~P.}, \bibinfo{author}{Payne, M.~C.},
  \bibinfo{author}{Kondor, R.} \& \bibinfo{author}{Cs\'anyi, G.}
\newblock \bibinfo{title}{Gaussian approximation potentials: The accuracy of
  quantum mechanics, without the electrons}.
\newblock \emph{\bibinfo{journal}{Phys. Rev. Lett.}}
  \textbf{\bibinfo{volume}{104}}, \bibinfo{pages}{136403}
  (\bibinfo{year}{2010}).

\bibitem{THOMPSON2015316}
\bibinfo{author}{Thompson, A.}, \bibinfo{author}{Swiler, L.},
  \bibinfo{author}{Trott, C.}, \bibinfo{author}{Foiles, S.} \&
  \bibinfo{author}{Tucker, G.}
\newblock \bibinfo{title}{Spectral neighbor analysis method for automated
  generation of quantum-accurate interatomic potentials}.
\newblock \emph{\bibinfo{journal}{Journal of Computational Physics}}
  \textbf{\bibinfo{volume}{285}}, \bibinfo{pages}{316--330}
  (\bibinfo{year}{2015}).

\bibitem{10.1063/1.5017641}
\bibinfo{author}{Wood, M.~A.} \& \bibinfo{author}{Thompson, A.~P.}
\newblock \bibinfo{title}{{Extending the accuracy of the SNAP interatomic
  potential form}}.
\newblock \emph{\bibinfo{journal}{The Journal of Chemical Physics}}
  \textbf{\bibinfo{volume}{148}}, \bibinfo{pages}{241721}
  (\bibinfo{year}{2018}).

\bibitem{doi:10.1137/15M1054183}
\bibinfo{author}{Shapeev, A.~V.}
\newblock \bibinfo{title}{Moment tensor potentials: A class of systematically
  improvable interatomic potentials}.
\newblock \emph{\bibinfo{journal}{Multiscale Modeling \& Simulation}}
  \textbf{\bibinfo{volume}{14}}, \bibinfo{pages}{1153--1173}
  (\bibinfo{year}{2016}).

\bibitem{PhysRevB.99.014104}
\bibinfo{author}{Drautz, R.}
\newblock \bibinfo{title}{Atomic cluster expansion for accurate and
  transferable interatomic potentials}.
\newblock \emph{\bibinfo{journal}{Phys. Rev. B}} \textbf{\bibinfo{volume}{99}},
  \bibinfo{pages}{014104} (\bibinfo{year}{2019}).

\bibitem{fu2023forces}
\bibinfo{author}{Fu, X.} \emph{et~al.}
\newblock \bibinfo{title}{Forces are not enough: Benchmark and critical
  evaluation for machine learning force fields with molecular simulations}
  (\bibinfo{year}{2023}).

\bibitem{Stocker_2022}
\bibinfo{author}{Stocker, S.}, \bibinfo{author}{Gasteiger, J.},
  \bibinfo{author}{Becker, F.}, \bibinfo{author}{G\"unnemann, S.} \&
  \bibinfo{author}{Margraf, J.~T.}
\newblock \bibinfo{title}{How robust are modern graph neural network potentials
  in long and hot molecular dynamics simulations?}
\newblock \emph{\bibinfo{journal}{Machine Learning: Science and Technology}}
  \textbf{\bibinfo{volume}{3}}, \bibinfo{pages}{045010} (\bibinfo{year}{2022}).

\bibitem{NPJ_Montes}
\bibinfo{author}{Montes~de Oca~Zapiain, D.} \emph{et~al.}
\newblock \bibinfo{title}{{Training data selection for accuracy and
  transferability of interatomic potentials}}.
\newblock \emph{\bibinfo{journal}{npj Computational Materials}}
  \textbf{\bibinfo{volume}{8}} (\bibinfo{year}{2022}).

\bibitem{Qamar2023}
\bibinfo{author}{Qamar, M.}, \bibinfo{author}{Mrovec, M.},
  \bibinfo{author}{Lysogorskiy, Y.}, \bibinfo{author}{Bochkarev, A.} \&
  \bibinfo{author}{Drautz, R.}
\newblock \bibinfo{title}{Atomic cluster expansion for quantum-accurate
  large-scale simulations of carbon}.
\newblock \emph{\bibinfo{journal}{Journal of Chemical Theory and Computation}}
  \textbf{\bibinfo{volume}{19}}, \bibinfo{pages}{5151--5167}
  (\bibinfo{year}{2023}).

\bibitem{PhysRevB.106.L180101}
\bibinfo{author}{Willman, J.~T.} \emph{et~al.}
\newblock \bibinfo{title}{Machine learning interatomic potential for
  simulations of carbon at extreme conditions}.
\newblock \emph{\bibinfo{journal}{Phys. Rev. B}}
  \textbf{\bibinfo{volume}{106}}, \bibinfo{pages}{L180101}
  (\bibinfo{year}{2022}).

\bibitem{Zhang2024}
\bibinfo{author}{Zhang, S.} \emph{et~al.}
\newblock \bibinfo{title}{Exploring the frontiers of condensed-phase chemistry
  with a general reactive machine learning potential}.
\newblock \emph{\bibinfo{journal}{Nature Chemistry}}
  \textbf{\bibinfo{volume}{16}}, \bibinfo{pages}{727--734}
  (\bibinfo{year}{2024}).

\bibitem{10.1063/5.0091698}
\bibinfo{author}{Rowe, P.}, \bibinfo{author}{Deringer, V.~L.},
  \bibinfo{author}{Gasparotto, P.}, \bibinfo{author}{Csányi, G.} \&
  \bibinfo{author}{Michaelides, A.}
\newblock \bibinfo{title}{Erratum: “an accurate and transferable machine
  learning potential for carbon” [j. chem. phys. 153, 034702 (2020)]}.
\newblock \emph{\bibinfo{journal}{The Journal of Chemical Physics}}
  \textbf{\bibinfo{volume}{156}}, \bibinfo{pages}{159901}
  (\bibinfo{year}{2022}).

\bibitem{ANI_Al}
\bibinfo{author}{Smith, J.~S.} \emph{et~al.}
\newblock \bibinfo{title}{Automated discovery of a robust interatomic potential
  for aluminum}.
\newblock \emph{\bibinfo{journal}{Nature Communications}}
  \textbf{\bibinfo{volume}{12}} (\bibinfo{year}{2021}).

\bibitem{PhysRevMaterials.3.023804}
\bibinfo{author}{Zhang, L.}, \bibinfo{author}{Lin, D.-Y.},
  \bibinfo{author}{Wang, H.}, \bibinfo{author}{Car, R.} \& \bibinfo{author}{E,
  W.}
\newblock \bibinfo{title}{Active learning of uniformly accurate interatomic
  potentials for materials simulation}.
\newblock \emph{\bibinfo{journal}{Phys. Rev. Mater.}}
  \textbf{\bibinfo{volume}{3}}, \bibinfo{pages}{023804} (\bibinfo{year}{2019}).

\bibitem{PhysRevB.100.014105}
\bibinfo{author}{Jinnouchi, R.}, \bibinfo{author}{Karsai, F.} \&
  \bibinfo{author}{Kresse, G.}
\newblock \bibinfo{title}{On-the-fly machine learning force field generation:
  Application to melting points}.
\newblock \emph{\bibinfo{journal}{Phys. Rev. B}}
  \textbf{\bibinfo{volume}{100}}, \bibinfo{pages}{014105}
  (\bibinfo{year}{2019}).

\bibitem{10.1063/1.5020067}
\bibinfo{author}{Herr, J.~E.}, \bibinfo{author}{Yao, K.},
  \bibinfo{author}{McIntyre, R.}, \bibinfo{author}{Toth, D.~W.} \&
  \bibinfo{author}{Parkhill, J.}
\newblock \bibinfo{title}{Metadynamics for training neural network model
  chemistries: A competitive assessment}.
\newblock \emph{\bibinfo{journal}{The Journal of Chemical Physics}}
  \textbf{\bibinfo{volume}{148}}, \bibinfo{pages}{241710}
  (\bibinfo{year}{2018}).

\bibitem{vanderOord2023}
\bibinfo{author}{van~der Oord, C.}, \bibinfo{author}{Sachs, M.},
  \bibinfo{author}{Kov\'acs, D.~P.}, \bibinfo{author}{Ortner, C.} \&
  \bibinfo{author}{Cs\'anyi, G.}
\newblock \bibinfo{title}{Hyperactive learning for data-driven interatomic
  potentials}.
\newblock \emph{\bibinfo{journal}{npj Computational Materials}}
  \textbf{\bibinfo{volume}{9}}, \bibinfo{pages}{168} (\bibinfo{year}{2023}).

\bibitem{Kulichenko2023}
\bibinfo{author}{Kulichenko, M.} \emph{et~al.}
\newblock \bibinfo{title}{Uncertainty-driven dynamics for active learning of
  interatomic potentials}.
\newblock \emph{\bibinfo{journal}{Nature Computational Science}}
  \textbf{\bibinfo{volume}{3}}, \bibinfo{pages}{230--239}
  (\bibinfo{year}{2023}).

\bibitem{PhysRevB.99.064114}
\bibinfo{author}{Podryabinkin, E.~V.}, \bibinfo{author}{Tikhonov, E.~V.},
  \bibinfo{author}{Shapeev, A.~V.} \& \bibinfo{author}{Oganov, A.~R.}
\newblock \bibinfo{title}{Accelerating crystal structure prediction by
  machine-learning interatomic potentials with active learning}.
\newblock \emph{\bibinfo{journal}{Phys. Rev. B}} \textbf{\bibinfo{volume}{99}},
  \bibinfo{pages}{064114} (\bibinfo{year}{2019}).

\bibitem{Bernstein2019}
\bibinfo{author}{Bernstein, N.}, \bibinfo{author}{Cs\'anyi, G.} \&
  \bibinfo{author}{Deringer, V.~L.}
\newblock \bibinfo{title}{De novo exploration and self-guided learning of
  potential-energy surfaces}.
\newblock \emph{\bibinfo{journal}{npj Computational Materials}}
  \textbf{\bibinfo{volume}{5}}, \bibinfo{pages}{99} (\bibinfo{year}{2019}).

\bibitem{Sivaraman2020}
\bibinfo{author}{Sivaraman, G.} \emph{et~al.}
\newblock \bibinfo{title}{Machine-learned interatomic potentials by active
  learning: amorphous and liquid hafnium dioxide}.
\newblock \emph{\bibinfo{journal}{npj Computational Materials}}
  \textbf{\bibinfo{volume}{6}}, \bibinfo{pages}{104} (\bibinfo{year}{2020}).

\bibitem{Qi2024}
\bibinfo{author}{Qi, J.}, \bibinfo{author}{Ko, T.~W.}, \bibinfo{author}{Wood,
  B.~C.}, \bibinfo{author}{Pham, T.~A.} \& \bibinfo{author}{Ong, S.~P.}
\newblock \bibinfo{title}{Robust training of machine learning interatomic
  potentials with dimensionality reduction and stratified sampling}.
\newblock \emph{\bibinfo{journal}{npj Computational Materials}}
  \textbf{\bibinfo{volume}{10}}, \bibinfo{pages}{43} (\bibinfo{year}{2024}).

\bibitem{PODRYABINKIN2017171}
\bibinfo{author}{Podryabinkin, E.~V.} \& \bibinfo{author}{Shapeev, A.~V.}
\newblock \bibinfo{title}{Active learning of linearly parametrized interatomic
  potentials}.
\newblock \emph{\bibinfo{journal}{Computational Materials Science}}
  \textbf{\bibinfo{volume}{140}}, \bibinfo{pages}{171--180}
  (\bibinfo{year}{2017}).

\bibitem{10.1063/1.5005095}
\bibinfo{author}{Gubaev, K.}, \bibinfo{author}{Podryabinkin, E.~V.} \&
  \bibinfo{author}{Shapeev, A.~V.}
\newblock \bibinfo{title}{{Machine learning of molecular properties: Locality
  and active learning}}.
\newblock \emph{\bibinfo{journal}{The Journal of Chemical Physics}}
  \textbf{\bibinfo{volume}{148}}, \bibinfo{pages}{241727}
  (\bibinfo{year}{2018}).

\bibitem{GUBAEV2019148}
\bibinfo{author}{Gubaev, K.}, \bibinfo{author}{Podryabinkin, E.~V.},
  \bibinfo{author}{Hart, G.~L.} \& \bibinfo{author}{Shapeev, A.~V.}
\newblock \bibinfo{title}{Accelerating high-throughput searches for new alloys
  with active learning of interatomic potentials}.
\newblock \emph{\bibinfo{journal}{Computational Materials Science}}
  \textbf{\bibinfo{volume}{156}}, \bibinfo{pages}{148--156}
  (\bibinfo{year}{2019}).

\bibitem{PhysRevMaterials.7.043801}
\bibinfo{author}{Lysogorskiy, Y.}, \bibinfo{author}{Bochkarev, A.},
  \bibinfo{author}{Mrovec, M.} \& \bibinfo{author}{Drautz, R.}
\newblock \bibinfo{title}{Active learning strategies for atomic cluster
  expansion models}.
\newblock \emph{\bibinfo{journal}{Phys. Rev. Mater.}}
  \textbf{\bibinfo{volume}{7}}, \bibinfo{pages}{043801} (\bibinfo{year}{2023}).

\bibitem{Vandermause2020}
\bibinfo{author}{Vandermause, J.} \emph{et~al.}
\newblock \bibinfo{title}{On-the-fly active learning of interpretable bayesian
  force fields for atomistic rare events}.
\newblock \emph{\bibinfo{journal}{npj Computational Materials}}
  \textbf{\bibinfo{volume}{6}}, \bibinfo{pages}{20} (\bibinfo{year}{2020}).

\bibitem{Xie2023}
\bibinfo{author}{Xie, Y.} \emph{et~al.}
\newblock \bibinfo{title}{Uncertainty-aware molecular dynamics from bayesian
  active learning for phase transformations and thermal transport in sic}.
\newblock \emph{\bibinfo{journal}{npj Computational Materials}}
  \textbf{\bibinfo{volume}{9}}, \bibinfo{pages}{36} (\bibinfo{year}{2023}).

\bibitem{PhysRevB.107.104103}
\bibinfo{author}{Poul, M.}, \bibinfo{author}{Huber, L.},
  \bibinfo{author}{Bitzek, E.} \& \bibinfo{author}{Neugebauer, J.}
\newblock \bibinfo{title}{Systematic atomic structure datasets for machine
  learning potentials: Application to defects in magnesium}.
\newblock \emph{\bibinfo{journal}{Phys. Rev. B}}
  \textbf{\bibinfo{volume}{107}}, \bibinfo{pages}{104103}
  (\bibinfo{year}{2023}).

\bibitem{10.1063/5.0013059}
\bibinfo{author}{Karabin, M.} \& \bibinfo{author}{Perez, D.}
\newblock \bibinfo{title}{{An entropy-maximization approach to automated
  training set generation for interatomic potentials}}.
\newblock \emph{\bibinfo{journal}{The Journal of Chemical Physics}}
  \textbf{\bibinfo{volume}{153}} (\bibinfo{year}{2020}).

\bibitem{beirlant1997nonparametric}
\bibinfo{author}{Beirlant, J.}, \bibinfo{author}{Dudewicz, E.~J.},
  \bibinfo{author}{Gy{\"o}rfi, L.}, \bibinfo{author}{Van~der Meulen, E.~C.}
  \emph{et~al.}
\newblock \bibinfo{title}{Nonparametric entropy estimation: An overview}.
\newblock \emph{\bibinfo{journal}{International Journal of Mathematical and
  Statistical Sciences}} \textbf{\bibinfo{volume}{6}}, \bibinfo{pages}{17--39}
  (\bibinfo{year}{1997}).

\bibitem{Larsen_2016}
\bibinfo{author}{Larsen, P.~M.}, \bibinfo{author}{Schmidt, S.} \&
  \bibinfo{author}{Schi\o{}tz, J.}
\newblock \bibinfo{title}{Robust structural identification via polyhedral
  template matching}.
\newblock \emph{\bibinfo{journal}{Modelling and Simulation in Materials Science
  and Engineering}} \textbf{\bibinfo{volume}{24}}, \bibinfo{pages}{055007}
  (\bibinfo{year}{2016}).

\bibitem{Stukowski_2010}
\bibinfo{author}{Stukowski, A.}
\newblock \bibinfo{title}{Visualization and analysis of atomistic simulation
  data with {OVITO}-the open visualization tool}.
\newblock \emph{\bibinfo{journal}{Modelling and Simulation in Materials Science
  and Engineering}} \textbf{\bibinfo{volume}{18}}, \bibinfo{pages}{015012}
  (\bibinfo{year}{2009}).

\bibitem{PACE_implementation}
\bibinfo{author}{Lysogorskiy, Y.} \emph{et~al.}
\newblock \bibinfo{title}{{Performant implementation of the atomic cluster
  expansion (PACE) and application to copper and silicon}}.
\newblock \emph{\bibinfo{journal}{npj Computational Materials}}
  \textbf{\bibinfo{volume}{7}} (\bibinfo{year}{2021}).

\bibitem{Song2024}
\bibinfo{author}{Song, K.} \emph{et~al.}
\newblock \bibinfo{title}{General-purpose machine-learned potential for 16
  elemental metals and their alloys}.
\newblock \emph{\bibinfo{journal}{Nature Communications}}
  \textbf{\bibinfo{volume}{15}}, \bibinfo{pages}{10208} (\bibinfo{year}{2024}).

\bibitem{Owen2024}
\bibinfo{author}{Owen, C.~J.} \emph{et~al.}
\newblock \bibinfo{title}{Complexity of many-body interactions in transition
  metals via machine-learned force fields from the {TM}23 data set}.
\newblock \emph{\bibinfo{journal}{npj Computational Materials}}
  \textbf{\bibinfo{volume}{10}}, \bibinfo{pages}{92} (\bibinfo{year}{2024}).

\bibitem{MEHL2017S1}
\bibinfo{author}{Mehl, M.~J.} \emph{et~al.}
\newblock \bibinfo{title}{The {AFLOW} library of crystallographic prototypes:
  Part 1}.
\newblock \emph{\bibinfo{journal}{Computational Materials Science}}
  \textbf{\bibinfo{volume}{136}}, \bibinfo{pages}{S1--S828}
  (\bibinfo{year}{2017}).

\bibitem{HICKS2019S1}
\bibinfo{author}{Hicks, D.} \emph{et~al.}
\newblock \bibinfo{title}{The {AFLOW} library of crystallographic prototypes:
  Part 2}.
\newblock \emph{\bibinfo{journal}{Computational Materials Science}}
  \textbf{\bibinfo{volume}{161}}, \bibinfo{pages}{S1--S1011}
  (\bibinfo{year}{2019}).

\bibitem{HICKS2021110450}
\bibinfo{author}{Hicks, D.} \emph{et~al.}
\newblock \bibinfo{title}{The {AFLOW} library of crystallographic prototypes:
  Part 3}.
\newblock \emph{\bibinfo{journal}{Computational Materials Science}}
  \textbf{\bibinfo{volume}{199}}, \bibinfo{pages}{110450}
  (\bibinfo{year}{2021}).

\bibitem{Cak_2014}
\bibinfo{author}{\u{C}\'{a}k, M.}, \bibinfo{author}{Hammerschmidt, T.},
  \bibinfo{author}{Rogal, J.}, \bibinfo{author}{Vitek, V.} \&
  \bibinfo{author}{Drautz, R.}
\newblock \bibinfo{title}{Analytic bond-order potentials for the bcc refractory
  metals {Nb}, {Ta}, {Mo} and {W}}.
\newblock \emph{\bibinfo{journal}{Journal of Physics: Condensed Matter}}
  \textbf{\bibinfo{volume}{26}}, \bibinfo{pages}{195501}
  (\bibinfo{year}{2014}).

\bibitem{Paidar_1999}
\bibinfo{author}{Paidar, V.}, \bibinfo{author}{Wang, L.~G.},
  \bibinfo{author}{Sob, M.} \& \bibinfo{author}{Vitek, V.}
\newblock \bibinfo{title}{A study of the applicability of many-body central
  force potentials in {NiAl} and {TiAl}}.
\newblock \emph{\bibinfo{journal}{Modelling and Simulation in Materials Science
  and Engineering}} \textbf{\bibinfo{volume}{7}}, \bibinfo{pages}{369}
  (\bibinfo{year}{1999}).

\bibitem{PhysRevB.66.094110}
\bibinfo{author}{Luo, W.}, \bibinfo{author}{Roundy, D.},
  \bibinfo{author}{Cohen, M.~L.} \& \bibinfo{author}{Morris, J.~W.}
\newblock \bibinfo{title}{Ideal strength of bcc molybdenum and niobium}.
\newblock \emph{\bibinfo{journal}{Phys. Rev. B}} \textbf{\bibinfo{volume}{66}},
  \bibinfo{pages}{094110} (\bibinfo{year}{2002}).

\bibitem{Tran2016}
\bibinfo{author}{Tran, R.} \emph{et~al.}
\newblock \bibinfo{title}{Surface energies of elemental crystals}.
\newblock \emph{\bibinfo{journal}{Scientific Data}}
  \textbf{\bibinfo{volume}{3}}, \bibinfo{pages}{160080} (\bibinfo{year}{2016}).

\bibitem{Batzner2022}
\bibinfo{author}{Batzner, S.} \emph{et~al.}
\newblock \bibinfo{title}{E(3)-equivariant graph neural networks for
  data-efficient and accurate interatomic potentials}.
\newblock \emph{\bibinfo{journal}{Nature Communications}}
  \textbf{\bibinfo{volume}{13}}, \bibinfo{pages}{2453} (\bibinfo{year}{2022}).

\bibitem{ASE_Larsen_2017}
\bibinfo{author}{Larsen, A.~H.} \emph{et~al.}
\newblock \bibinfo{title}{The atomic simulation environment-a python library
  for working with atoms}.
\newblock \emph{\bibinfo{journal}{Journal of Physics: Condensed Matter}}
  \textbf{\bibinfo{volume}{29}}, \bibinfo{pages}{273002}
  (\bibinfo{year}{2017}).

\bibitem{LAMMPS}
\bibinfo{author}{Thompson, A.~P.} \emph{et~al.}
\newblock \bibinfo{title}{{LAMMPS} - a flexible simulation tool for
  particle-based materials modeling at the atomic, meso, and continuum scales}.
\newblock \emph{\bibinfo{journal}{Comp. Phys. Comm.}}
  \textbf{\bibinfo{volume}{271}}, \bibinfo{pages}{108171}
  (\bibinfo{year}{2022}).

\bibitem{Jana_2019}
\bibinfo{author}{Jana, R.}, \bibinfo{author}{Savio, D.},
  \bibinfo{author}{Deringer, V.~L.} \& \bibinfo{author}{Pastewka, L.}
\newblock \bibinfo{title}{Structural and elastic properties of amorphous carbon
  from simulated quenching at low rates}.
\newblock \emph{\bibinfo{journal}{Modelling and Simulation in Materials Science
  and Engineering}} \textbf{\bibinfo{volume}{27}}, \bibinfo{pages}{085009}
  (\bibinfo{year}{2019}).

\bibitem{fellman2024fastaccuratemachinelearnedinteratomic}
\bibinfo{author}{Fellman, A.}, \bibinfo{author}{Byggm\"astar, J.},
  \bibinfo{author}{Granberg, F.}, \bibinfo{author}{Nordlund, K.} \&
  \bibinfo{author}{Djurabekova, F.}
\newblock \bibinfo{title}{Fast and accurate machine-learned interatomic
  potentials for large-scale simulations of {Cu}, {Al} and {Ni}}
  (\bibinfo{year}{2024}).
\newblock \urlprefix\url{https://arxiv.org/abs/2408.15779}.
\newblock \eprint{2408.15779}.

\bibitem{Ding2024}
\bibinfo{author}{Ding, C.-J.} \emph{et~al.}
\newblock \bibinfo{title}{A deep learning interatomic potential suitable for
  simulating radiation damage in bulk tungsten}.
\newblock \emph{\bibinfo{journal}{Tungsten}} \textbf{\bibinfo{volume}{6}},
  \bibinfo{pages}{304--322} (\bibinfo{year}{2024}).

\bibitem{PhysRevB.100.144105}
\bibinfo{author}{Byggm\"astar, J.}, \bibinfo{author}{Hamedani, A.},
  \bibinfo{author}{Nordlund, K.} \& \bibinfo{author}{Djurabekova, F.}
\newblock \bibinfo{title}{Machine-learning interatomic potential for radiation
  damage and defects in tungsten}.
\newblock \emph{\bibinfo{journal}{Phys. Rev. B}}
  \textbf{\bibinfo{volume}{100}}, \bibinfo{pages}{144105}
  (\bibinfo{year}{2019}).

\bibitem{10.1063/5.0005084}
\bibinfo{author}{Rowe, P.}, \bibinfo{author}{Deringer, V.~L.},
  \bibinfo{author}{Gasparotto, P.}, \bibinfo{author}{Csányi, G.} \&
  \bibinfo{author}{Michaelides, A.}
\newblock \bibinfo{title}{An accurate and transferable machine learning
  potential for carbon}.
\newblock \emph{\bibinfo{journal}{The Journal of Chemical Physics}}
  \textbf{\bibinfo{volume}{153}}, \bibinfo{pages}{034702}
  (\bibinfo{year}{2020}).

\bibitem{jax2018github}
\bibinfo{author}{Bradbury, J.} \emph{et~al.}
\newblock \bibinfo{title}{{JAX}: composable transformations of
  {P}ython+{N}um{P}y programs} (\bibinfo{year}{2018}).
\newblock \urlprefix\url{http://github.com/google/jax}.

\bibitem{AVERY2017208}
\bibinfo{author}{Avery, P.} \& \bibinfo{author}{Zurek, E.}
\newblock \bibinfo{title}{Randspg: An open-source program for generating
  atomistic crystal structures with specific spacegroups}.
\newblock \emph{\bibinfo{journal}{Computer Physics Communications}}
  \textbf{\bibinfo{volume}{213}}, \bibinfo{pages}{208--216}
  (\bibinfo{year}{2017}).

\bibitem{Pickard_2011}
\bibinfo{author}{Pickard, C.~J.} \& \bibinfo{author}{Needs, R.~J.}
\newblock \bibinfo{title}{Ab initio random structure searching}.
\newblock \emph{\bibinfo{journal}{Journal of Physics: Condensed Matter}}
  \textbf{\bibinfo{volume}{23}}, \bibinfo{pages}{053201}
  (\bibinfo{year}{2011}).

\bibitem{mahoney2009cur}
\bibinfo{author}{Mahoney, M.~W.} \& \bibinfo{author}{Drineas, P.}
\newblock \bibinfo{title}{Cur matrix decompositions for improved data
  analysis}.
\newblock \emph{\bibinfo{journal}{Proceedings of the National Academy of
  Sciences}} \textbf{\bibinfo{volume}{106}}, \bibinfo{pages}{697--702}
  (\bibinfo{year}{2009}).

\bibitem{10.1063/1.4812323}
\bibinfo{author}{Jain, A.} \emph{et~al.}
\newblock \bibinfo{title}{Commentary: The materials project: A materials genome
  approach to accelerating materials innovation}.
\newblock \emph{\bibinfo{journal}{APL Materials}} \textbf{\bibinfo{volume}{1}},
  \bibinfo{pages}{011002} (\bibinfo{year}{2013}).

\bibitem{PhysRev.140.A1133}
\bibinfo{author}{Kohn, W.} \& \bibinfo{author}{Sham, L.~J.}
\newblock \bibinfo{title}{Self-consistent equations including exchange and
  correlation effects}.
\newblock \emph{\bibinfo{journal}{Phys. Rev.}} \textbf{\bibinfo{volume}{140}},
  \bibinfo{pages}{A1133--A1138} (\bibinfo{year}{1965}).

\bibitem{KRESSE199615}
\bibinfo{author}{Kresse, G.} \& \bibinfo{author}{Furthm\"uller, J.}
\newblock \bibinfo{title}{Efficiency of ab-initio total energy calculations for
  metals and semiconductors using a plane-wave basis set}.
\newblock \emph{\bibinfo{journal}{Computational Materials Science}}
  \textbf{\bibinfo{volume}{6}}, \bibinfo{pages}{15--50} (\bibinfo{year}{1996}).

\bibitem{PhysRevB.54.11169}
\bibinfo{author}{Kresse, G.} \& \bibinfo{author}{Furthm\"uller, J.}
\newblock \bibinfo{title}{Efficient iterative schemes for ab initio
  total-energy calculations using a plane-wave basis set}.
\newblock \emph{\bibinfo{journal}{Phys. Rev. B}} \textbf{\bibinfo{volume}{54}},
  \bibinfo{pages}{11169--11186} (\bibinfo{year}{1996}).

\bibitem{PhysRevB.59.1758}
\bibinfo{author}{Kresse, G.} \& \bibinfo{author}{Joubert, D.}
\newblock \bibinfo{title}{From ultrasoft pseudopotentials to the projector
  augmented-wave method}.
\newblock \emph{\bibinfo{journal}{Phys. Rev. B}} \textbf{\bibinfo{volume}{59}},
  \bibinfo{pages}{1758--1775} (\bibinfo{year}{1999}).

\bibitem{PhysRevB.50.17953}
\bibinfo{author}{Bl\"ochl, P.~E.}
\newblock \bibinfo{title}{Projector augmented-wave method}.
\newblock \emph{\bibinfo{journal}{Phys. Rev. B}} \textbf{\bibinfo{volume}{50}},
  \bibinfo{pages}{17953--17979} (\bibinfo{year}{1994}).

\bibitem{PhysRevLett.77.3865}
\bibinfo{author}{Perdew, J.~P.}, \bibinfo{author}{Burke, K.} \&
  \bibinfo{author}{Ernzerhof, M.}
\newblock \bibinfo{title}{Generalized gradient approximation made simple}.
\newblock \emph{\bibinfo{journal}{Phys. Rev. Lett.}}
  \textbf{\bibinfo{volume}{77}}, \bibinfo{pages}{3865--3868}
  (\bibinfo{year}{1996}).

\bibitem{JANSSEN201924}
\bibinfo{author}{Janssen, J.} \emph{et~al.}
\newblock \bibinfo{title}{pyiron: An integrated development environment for
  computational materials science}.
\newblock \emph{\bibinfo{journal}{Computational Materials Science}}
  \textbf{\bibinfo{volume}{163}}, \bibinfo{pages}{24--36}
  (\bibinfo{year}{2019}).

\bibitem{doi:10.1073/pnas.30.9.244}
\bibinfo{author}{Murnaghan, F.~D.}
\newblock \bibinfo{title}{The compressibility of media under extreme
  pressures}.
\newblock \emph{\bibinfo{journal}{Proceedings of the National Academy of
  Sciences}} \textbf{\bibinfo{volume}{30}}, \bibinfo{pages}{244--247}
  (\bibinfo{year}{1944}).

\bibitem{D2_Grimme}
\bibinfo{author}{Grimme, S.}
\newblock \bibinfo{title}{Semiempirical {GGA}-type density functional
  constructed with a long-range dispersion correction}.
\newblock \emph{\bibinfo{journal}{Journal of Computational Chemistry}}
  \textbf{\bibinfo{volume}{27}}, \bibinfo{pages}{1787--1799}
  (\bibinfo{year}{2006}).

\bibitem{bochkarev2022efficient}
\bibinfo{author}{Bochkarev, A.} \emph{et~al.}
\newblock \bibinfo{title}{Efficient parametrization of the atomic cluster
  expansion}.
\newblock \emph{\bibinfo{journal}{Phys. Rev. Mater.}}
  \textbf{\bibinfo{volume}{6}}, \bibinfo{pages}{013804} (\bibinfo{year}{2022}).

\end{thebibliography}
\end{document}


\makeatletter
\renewcommand \thesection{S\@arabic\c@section}
\renewcommand\thetable{S\@arabic\c@table}
\renewcommand \thefigure{S\@arabic\c@figure}
\makeatother

\onecolumn
\title[Article Title]{Supplementary material for: Information-entropy-driven generation of material-agnostic datasets for machine-learning interatomic potentials}

\author*[1]{\fnm{Aparna} \sur{P. A. Subramanyam}}\email{apasubramanyam@lanl.gov}
\author*[1]{\fnm{Danny} \sur{Perez}}\email{danny\_perez@lanl.gov}

\affil[1]{\orgdiv{Theoretical Division T-1}, \orgname{Los Alamos National Laboratory}, \orgaddress{\city{Los Alamos}, \postcode{87545}, \state{New Mexico}, \country{USA}}}

\maketitle
\section{Geometric characterization}\label{secA1}
\begin{figure}[h!]
\centering
\includegraphics[width=0.7\columnwidth]{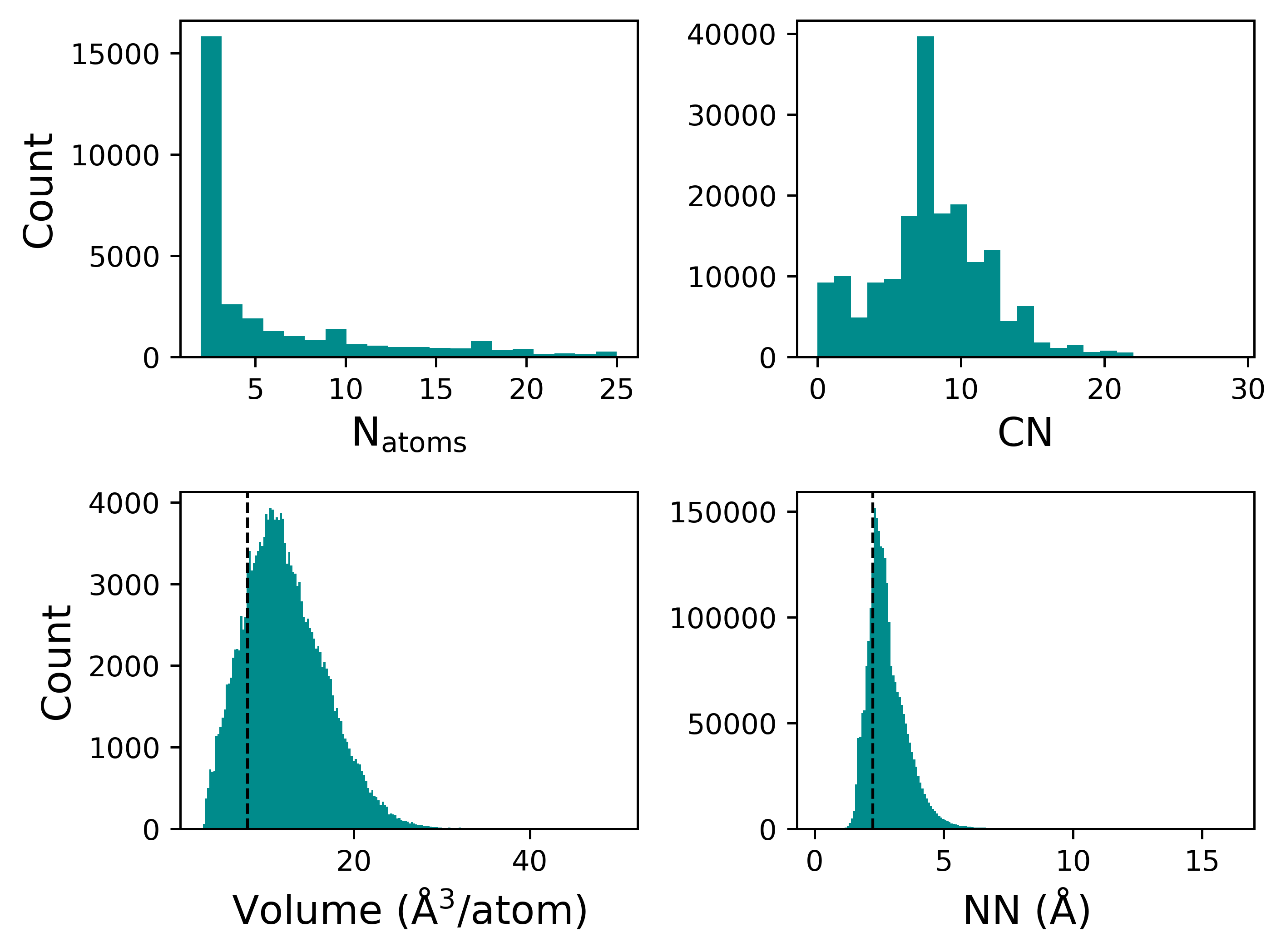}
\caption{Distribution of the number of atoms, the coordination number (CN) (generated with 120\% of the equilibrium distance of hcp-Be as cut-off distace), Voronoi volumes and distribution of nearest neighbor distances for the Rev-EM dataset instantiated for Be.}
\label{fig:characterization}
\end{figure}

The Voronoi volumes and coordination numbers, here corresponding to the Rev-EM dataset instantiated for Be, are shown in Fig.\ \ref{fig:characterization}. These quantities have been computed using the pyscal \cite{Menon2019} package. 

\section{Measure of diversity}\label{secA2}
\begin{figure}[h!]
\centering
\includegraphics[width=0.7\columnwidth]{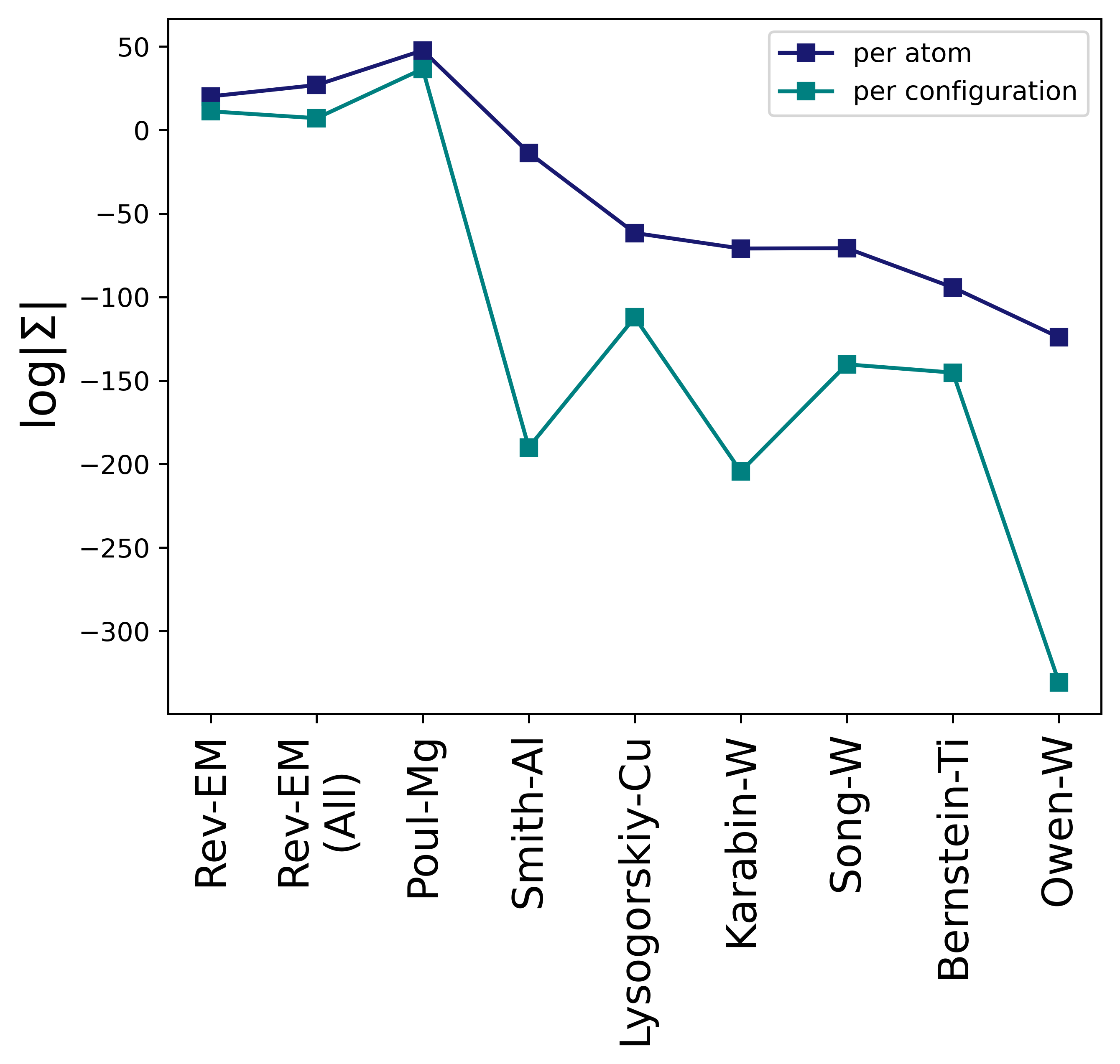}
\caption{Comparison of the information entropy of the per-configuration averaged and per-atom distribution of the bispectrum components for the Rev-EM dataset and datasets from literature. The Rev-EM dataset comprises of the structures generated in the respective mode while the Rev-EM (All) dataset is a combination of structures from both the modes.}
\label{fig:det_all}
\end{figure}
Fig. \ref{fig:det_all}, reports the information entropy of the different datasets investigated in the current work. Contrary to the equivalent figure in the main manuscript, these datasets were not uniformized to a common density range. This confers the 
 Poul-Mg dataset a very high feature entropy, due to the extremely high-density configurations included in this set.

\section{ACE model for W}
\subsection{Cold curves}
\begin{figure}[h]
\centering
\includegraphics[width=0.7\columnwidth]{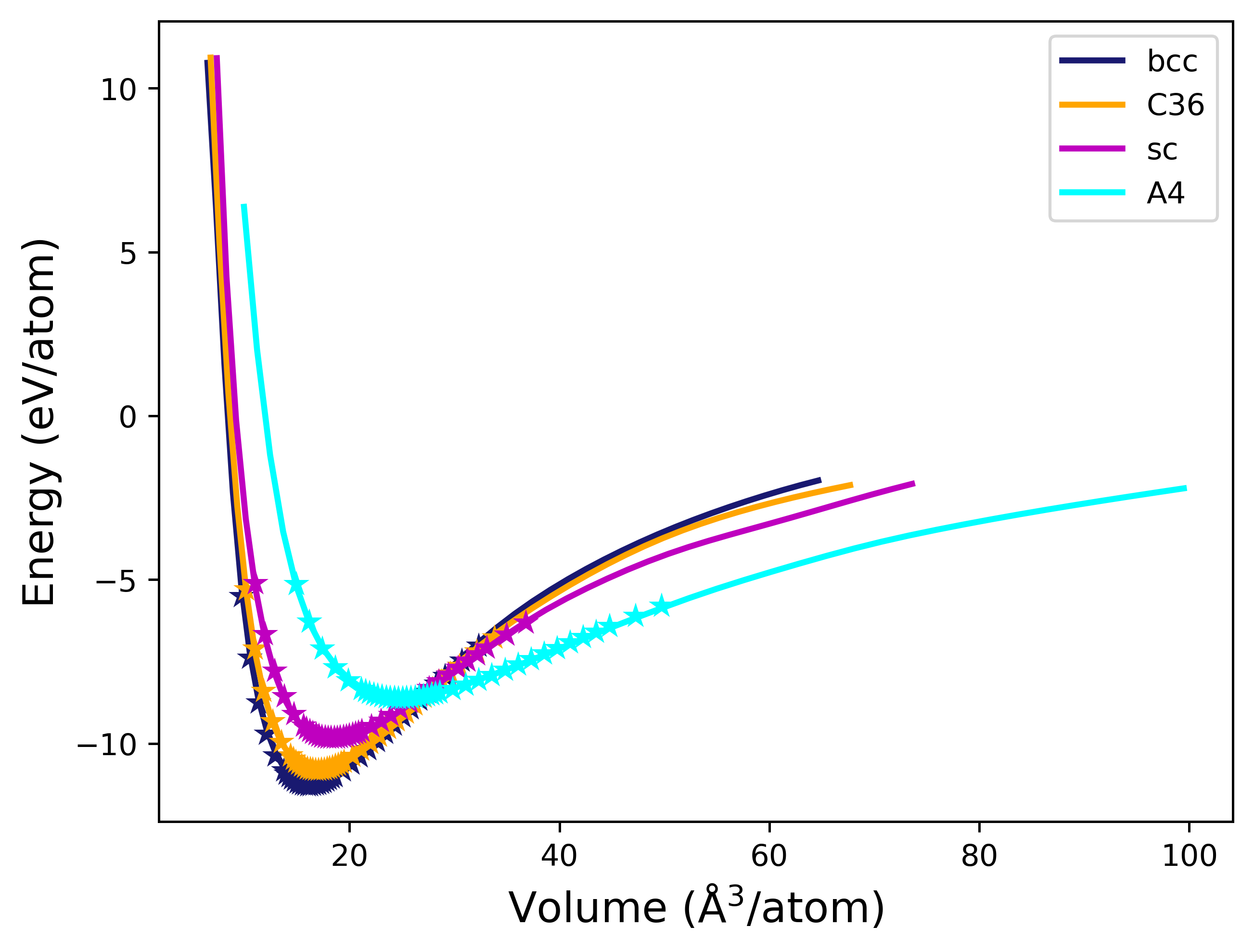}
\caption{Energy-volume curves for selected crystal structures across a wide range of volumes. DFT values are shown as stars and the ACE predictions as solid lines.}
\label{fig:W_EVsV}
\end{figure}

Relative phase stability is often a key quantity of interest in applications. The energy-volume cold curves for the ground state BCC crystal structure, the topologically-close-packed (TCP) C36 crystal structure, the diamond (A4) crystal structure and the simple cubic (sc) structure are shown in Fig. \ref{fig:W_EVsV} for the ACE-2 model. ACE-2 (solid lines) accurately reproduces the DFT energy-volume data (marked as stars) over a very broad range of densities. 

\subsection{Grain boundaries}

Grain boundary (GB) structures were generated using the aimsgb package \cite{CHENG201892}. As shown in Tab.~\ref{tab:W_GBs}, ACE-2 accurately captures the energetics of symmetric tilt GBs $\Sigma$3[1-10](112), $\Sigma$3[1-10](111), $\Sigma$5[100](013) and the twist GBs $\Sigma$3[1-10](110) and  $\Sigma$5[100](001). Due to the computational cost involved in sampling the ground state GB structure for each GB, we did not repeat DFT calculations ourselves, but rather compare with results reported in Ref.~\cite{Scheiber_2016}, keeping in mind that the different DFT settings used in both studies will introduce a possible source of discrepancy. Overall, the ACE-2 predictions are in good agreement with the DFT reference, both in terms of absolute energies and ordering between the different boundaries. 
The $\Sigma$5[100](013) GB is a special case where experiments and DFT agree that the GB structure with broken symmetry, where the two grains shift perpendicular to the GB plane along the [100] direction, is the most stable structure. Our ACE model also predicts that the structure with broken symmetry is 20.15 mJ/m$^2$ more stable than the mirror symmetric structure while DFT stabilizes the broken symmetric structure by 89 mJ/m$^2$ (Ref. \cite{Scheiber_2016}). 
\begin{table}[h!]
\centering
\begin{tabular}{c | c | c }
\hline\hline
Grain boundary & DFT & ACE \\
\hline
& (J/m$^2$) & (J/m$^2$) \\
\hline
$\Sigma$3[1-10](112) & 0.66$^{*}$  & 0.88 \\ 
$\Sigma$3[1-10](110) & 0.75$^{*}$  & 1.01 \\
$\Sigma$3[1-10](111) & 2.44$^{*}$  & 2.53 \\
$\Sigma$5[100](013) & 2.31$^{*}$  & 2.33 \\ 
$\Sigma$5[100](001) & 3.35$^{*}$  & 3.42 \\ 
\hline\hline
\end{tabular}
\caption{Formation energies of selected tilt and twist grain boundaries (GBs). The DFT values of the GB energies are from Ref.\cite{Scheiber_2016}$^{*}$.}
\label{tab:W_GBs}
\end{table}

\subsection{Energy-based weights}
\begin{figure}[h]
\centering
\includegraphics[width=0.7\columnwidth]{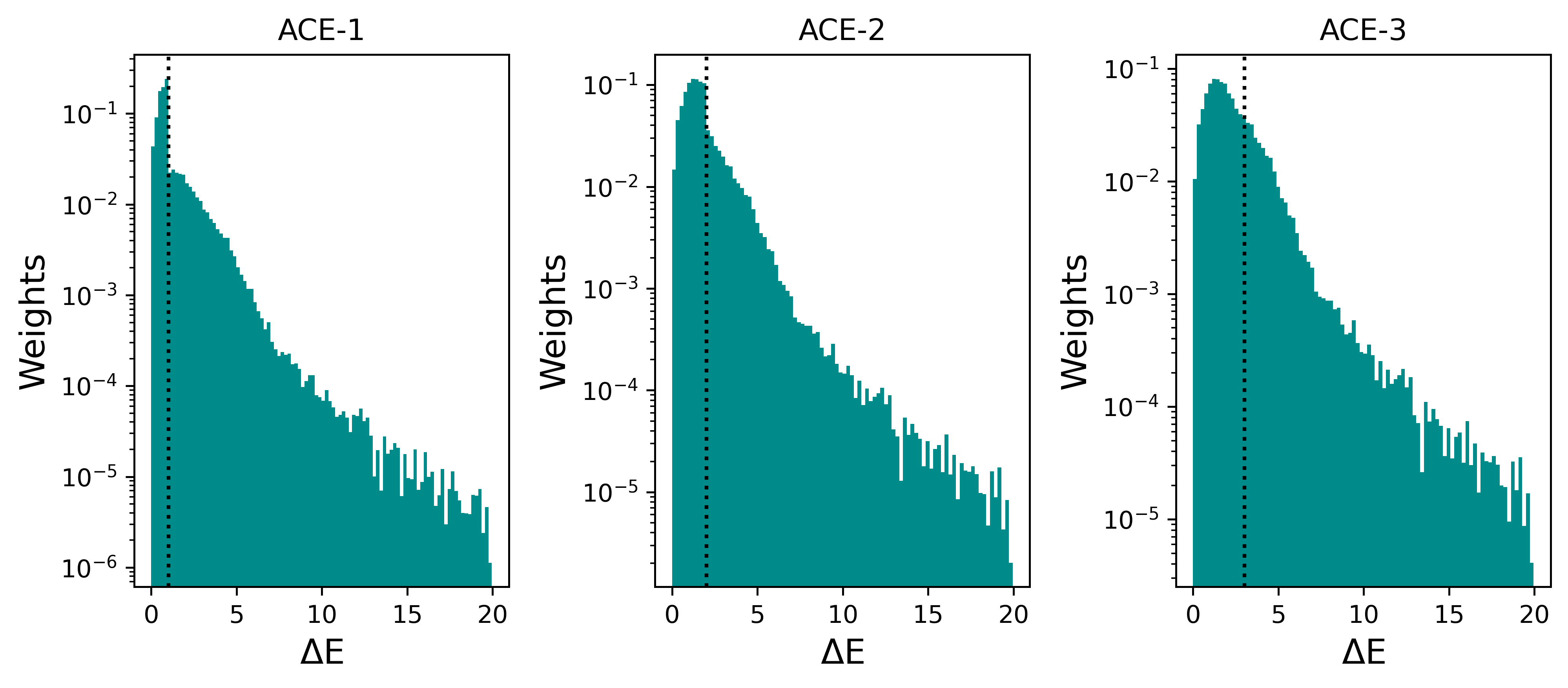}
\caption{Energy weight distribution for ACE-1, ACE-2, and ACE-3 models.} 
\label{fig:W_weights}
\end{figure}

The normalized weights based on the energy differences to the minimum energy structure are shown in Fig.\ \ref{fig:W_weights} for the three ACE models for W. 

\section{ACE model for C}
\begin{figure}[h!]
\centering
\includegraphics[width=0.65\columnwidth]{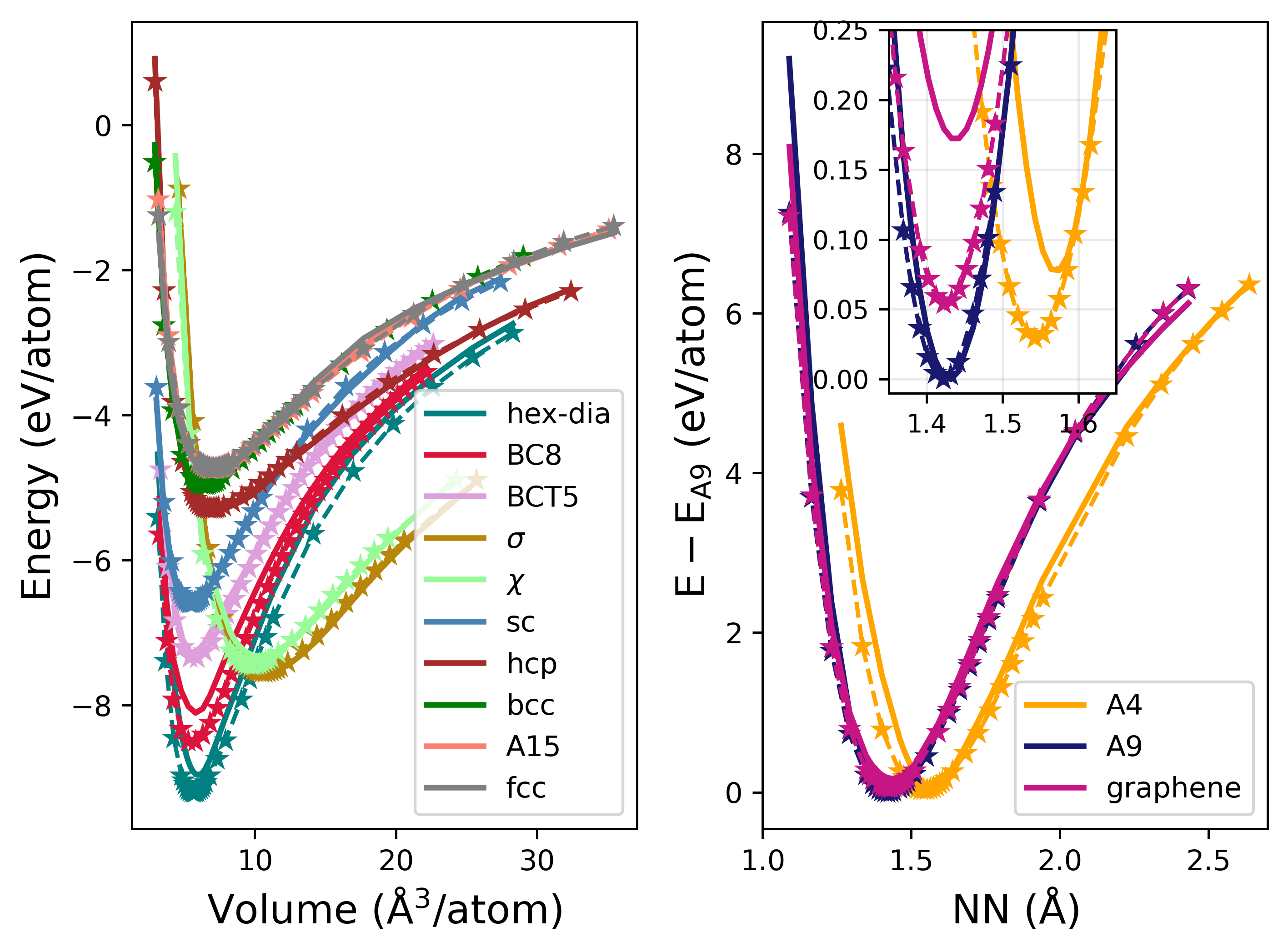} 
\caption{Energy as a function of volume for close packed prototypes (left) and as a function of nearest neighbor (NN) distance for graphite (A9), graphene and diamond (A4) structures (right) for ACE-2 (solid lines) and DFT (stars connected by dashed lines).}
\label{fig:C_EVsNN}
\end{figure}

The energy-volume curves for close packed structures and the energy-distance plots for open (A9, graphene) and the A4 structure that is close in energy to these two phases are shown in Fig.\ \ref{fig:C_EVsNN}. Both ACE and DFT energies are shown with dispersion (D2) corrections. The ACE-D2 model is close to DFT for all prototypes across the tested volume range. A9 is the most stable in both DFT and ACE followed by the A4 phase and then graphene. 

\subsection{MD simulations}
\begin{figure}[h!]
\centering
\includegraphics[width=0.85\columnwidth]{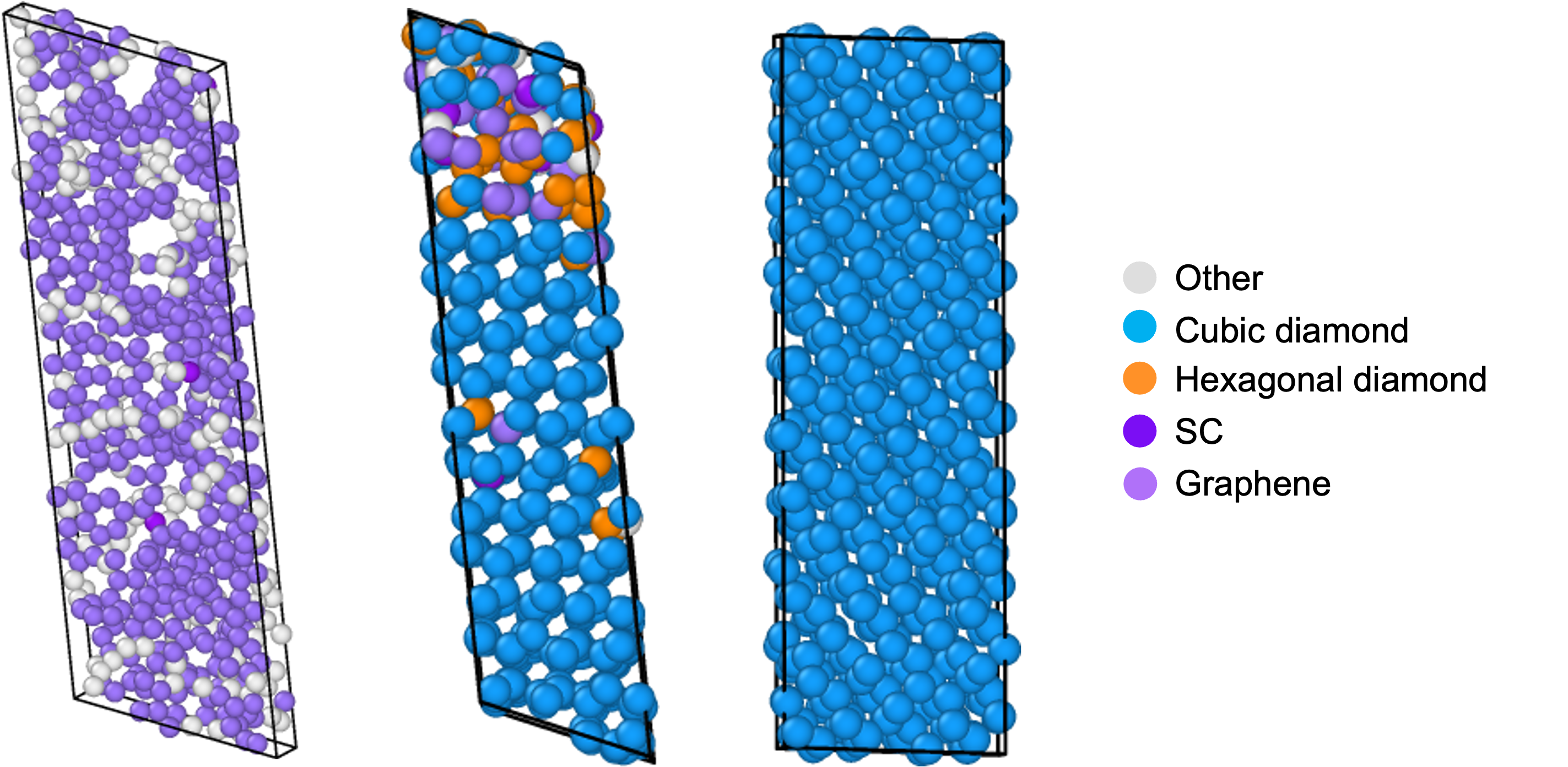} 
\caption{Snapshots of (left) 1.6 g/cc at 2500 K, (middle) 3.58 g/cc at 3500 K and (right) 4.0 g/cc supercells at 3500 K in NVT ensemble at the end of simulation time (at 150 ps). Analysis was performed using the polyhedral template matching algorithm implemented in OVITO.} 
\label{fig:MD_NVT}
\end{figure}

\begin{figure}[h!]
\centering
\includegraphics[width=0.6\columnwidth]{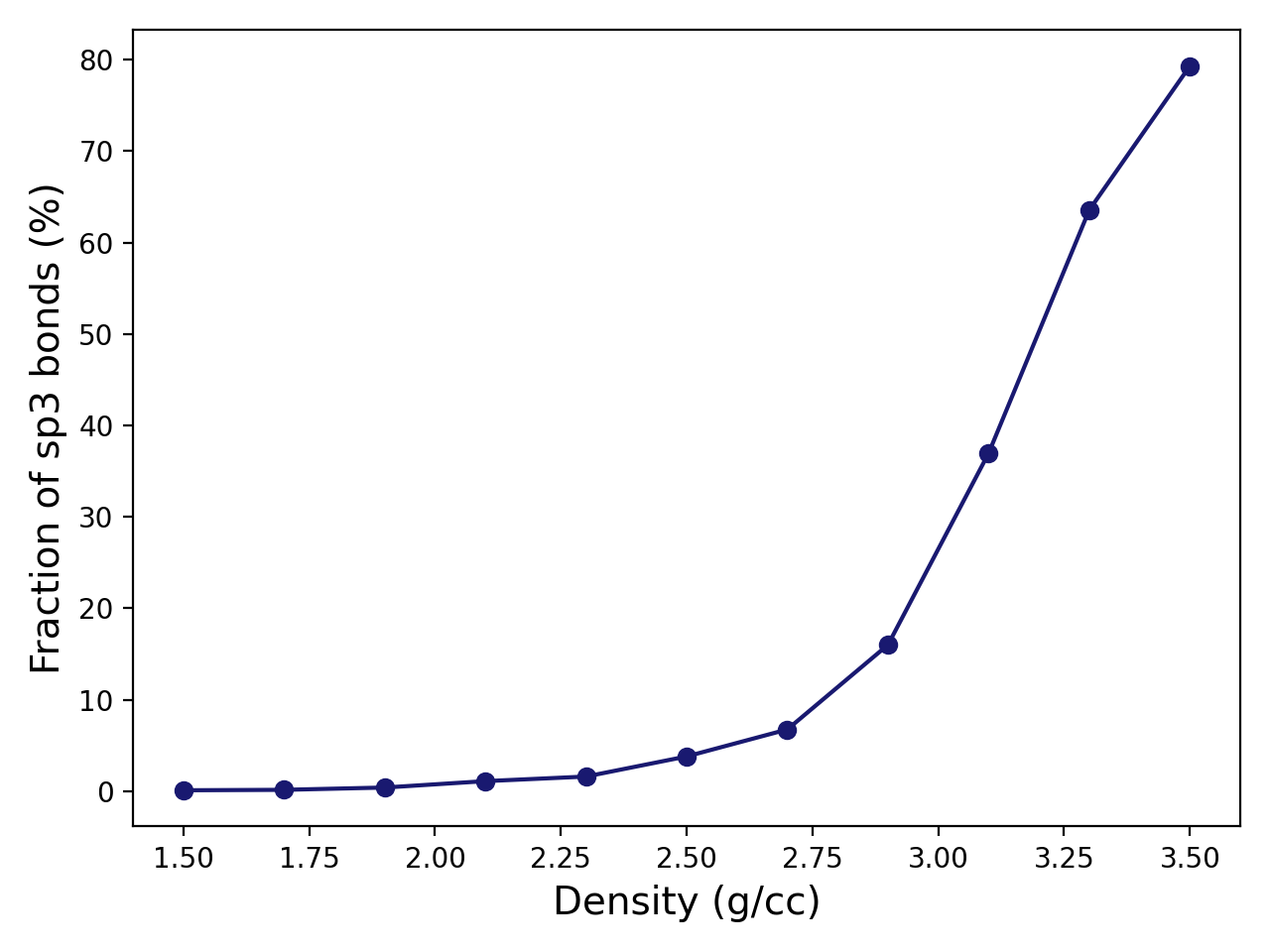} 
\caption{Density dependence of fraction of sp$^3$ bonds after quenching of amorphous-C.}
\label{fig:sp3_percent}
\end{figure}

\section{ACE models for Sb and Te}
The performance of the energy-volume cold curves for seven prototypes for Sb and eight prototypes for Te. The ground state structures A7 (for Sb) and A8 (for Te) are the only structures that are not well represented by the ACE model. ACE-2 (solid lines) accurately reproduces the DFT energy-volume data (marked as stars) over a very broad range of densities for all other prototypes.
\begin{figure}[h!]
\centering
\begin{subfigure}[b]{0.7\columnwidth}
\includegraphics[width=\textwidth]{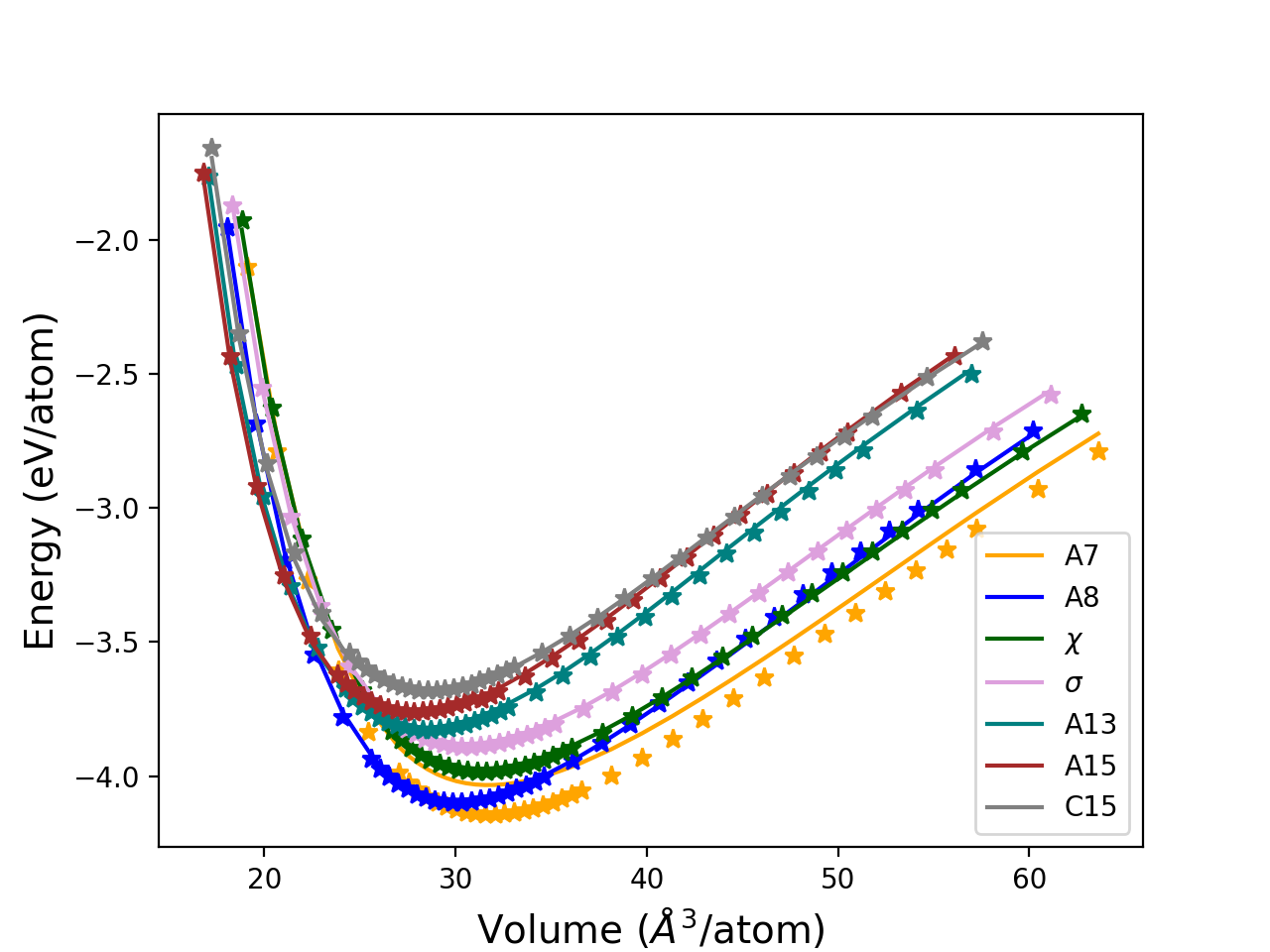}
\caption{Sb}
\label{fig:energies_Sb_Te}
\end{subfigure}\qquad
\begin{subfigure}[b]{0.7\columnwidth}
\includegraphics[width=\textwidth]{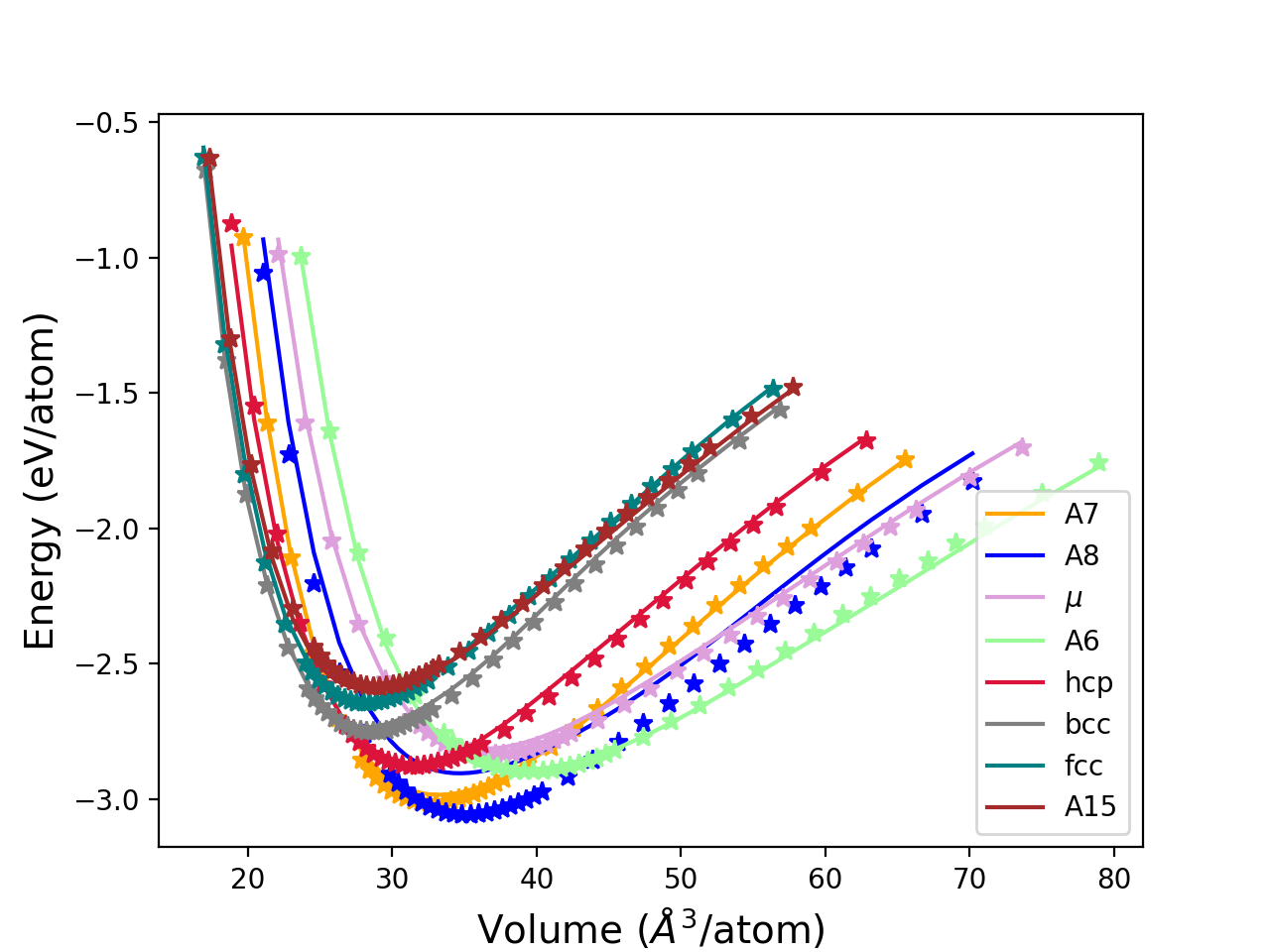}
\caption{Te}
\label{fig:forces_Sb_Te}
\end{subfigure}\qquad
\caption{The energy-volume curves for Sb (top) and Te (bottom).}
\label{fig:Sb_Te_EvsV}
\end{figure}

\section{Input files}
The complete input file used to train the ACE-2 model for Be using the {\em pacemaker} code is given below. For other elements, only the parameters that were modified are reported. 
\subsection{Be}
\begin{verbatim}
cutoff: 4.5
data:
  filename: df_Be_final_results_free_energies_set1_0d125_shifted.pckl.gzip
  cache_ref_df: False
  seed: 42
  test_size: 0.2

potential:
  deltaSplineBins: 0.001
  elements: [Be]
  embeddings:
    Be: {
     ndensity: 2,
     npot: 'FinnisSinclairShiftedScaled',
     fs_parameters: [1, 1, 1, 0.5],
     rho_core_cut: 50000,
     drho_core_cut: 27
      }
  bonds:
    BeBe: {
     rcut: 4.5,
     dcut: 0.01,
     NameOfCutoffFunction: cos,
     radbase: SBessel,
     r_in: 1.0,
     delta_in: 0.25,
     core-repulsion: [5000.0, 5.0]
     }
  functions:
    (Be): {
     nradmax_by_orders: [20, 4, 4, 2, 1],
     lmax_by_orders: [0, 3, 3, 1, 1],
     coefs_init: zero
     }

fit:
  loss: { kappa: 0.1, L1_coeffs: 0.000000001,  L2_coeffs: 0.00000000001,  w1_coeffs: 1.0, w2_coeffs: 1.0,
          w0_rad: 0.000000001, w1_rad: 0.000000001, w2_rad: 0.000000001 }
          
  weighting: {
    type: EnergyBasedWeightingPolicy,
    nfit: 10000,
    DElow: 2.0,
    DEup: 10.0,
    DFup: 20.0,
    DE: 1.0,
    DF: 1.0,
    wlow: 0.75,
    reftype: all
      }

  optimizer: BFGS
  maxiter: 800
  fit_cycles: 1
  noise_relative_sigma: 1e-4
  ladder_step: [10, 0.1]
  ladder_type: power_order

backend:
  evaluator: tensorpot
  batch_size: 1000
\end{verbatim}

\subsection{C}
\begin{verbatim}
cutoff: 5.5
rho_core_cut: 50000
drho_core_cut: 27
r_in: 0.9
delta_in: 0.2
core-repulsion: [5000.0, 5.0]
DElow: 3.0
DEup: 10.0
DFup: 25.0
\end{verbatim}

\subsection{Al}
\begin{verbatim}
cutoff: 5.5
rho_core_cut: 50000
drho_core_cut: 27
r_in: 1.0
delta_in: 0.25
core-repulsion: [5000.0, 5.0]
DElow: 2.0
DEup: 10.0
DFup: 20.0
\end{verbatim}

\subsection{W}
\begin{verbatim}
cutoff: 6.0
rho_core_cut: 20000
drho_core_cut: 270
r_in: 1.5
delta_in: 0.5
core-repulsion: [35000.0, 5.0]
DElow: 2.0
DEup: 20.0
DFup: 100.0
\end{verbatim}

\subsection{Re}
\begin{verbatim}
cutoff: 6.0
rho_core_cut: 20000
drho_core_cut: 270
r_in: 1.5
delta_in: 0.5
core-repulsion: [35000.0, 5.0]
DElow: 2.0
DEup: 20.0
DFup: 100.0
\end{verbatim}

\subsection{Os}
\begin{verbatim}
cutoff: 6.0
rho-core-cut: 20000
drho_core_cut: 270
r_in: 1.5
delta_in: 0.5
core-repulsion: [35000.0, 5.0]
DElow: 2.0
DEup: 40.0
DFup: 100.0
\end{verbatim}